\documentclass[12pt]{article}
\usepackage[usenames]{color}
\usepackage{setspace}
\usepackage{graphicx}
\usepackage{amsmath}
\usepackage{amssymb}
\usepackage{amsthm}
\usepackage{ytableau}

\usepackage{hyperref}
\hypersetup{
setpagesize=false,
 bookmarksnumbered=true,%
 bookmarksopen=true,%
 colorlinks=true,%
 linkcolor=blue,
 citecolor=blue,
}
\usepackage{bm}
\usepackage{framed,color}
\definecolor{shadecolor}{gray}{0.95}
\usepackage{fancybox}
\usepackage{ulem}
\definecolor{shadecolor}{gray}{0.95}

\usepackage{multirow}
\numberwithin{equation}{section}
\newcommand{\vev}[1]{\left\langle #1 \right\rangle}
\newcommand{\der}{\partial}
\newcommand{\Tr}{\mbox{\rm Tr}}
\newcommand{\til}[1]{\tilde{#1}}
\newcommand{\ie}{{\it i.e.}}
\newcommand{\eg}{{\it e.g.}}

\newcommand{\sumk}{\sum_{k=1}^3}
\newcommand{\sump}{\sum_{p=-1}^1}
\newcommand{\sumq}{\sum_{q=-1}^1}
\newcommand{\sumkd}{\sum_{k'=1}^3}

\newcommand{\mf}[1]{\mathfrak{#1}}

\newcommand{\zexto}{\xrightarrow{{}~{\mathbb Z}_3^{\rm (ex)}}}
\newcommand{\zex}{{\mathbb Z}_3^{\rm (ex)}}
\newcommand{\Z}[1]{{\mathbb Z}_#1}
\newcommand{\zy}{{\mathbb Z}_3^{({\rm L})}}  
\newcommand{\zpl}{\mathbb{Z}_3^{(+)}}
\newcommand{\zmi}{\mathbb{Z}_3^{(-)}}
\begin{document}
\begin{flushright}
\end{flushright}
\begin{center}
    {\LARGE\bf Grand Gauge-Higgs Unification on $T^2/{\mathbb Z}_3$
      via Diagonal Embedding Method}
\vskip 1.4cm
{\large  
Kentaro Kojima$^{a,}$\footnote{E-mail: kojima@artsci.kyushu-u.ac.jp}, 
Kazunori Takenaga$^{b,}$\footnote{E-mail: takenaga@kumamoto-hsu.ac.jp}, 
and 
Toshifumi Yamashita$^{c,}$\footnote{E-mail: tyamashi@aichi-med-u.ac.jp}
}
\\
\vskip 1.0cm
{\it $^a$ Faculty of Arts and Science, Kyushu University, Fukuoka 819-0395, Japan\\
$^b$ Faculty of Health Science, Kumamoto Health Science University, Izumi-machi, Kumamoto 861-5598, Japan\\
$^c$ Department of Physics, Aichi Medical University, Nagakute 480-1195, Japan%
}\\
\vskip 1.5cm
\begin{abstract}
    We study a novel six-dimensional gauge theory compactified on the
    $T^2/{\mathbb Z}_3$ orbifold utilizing the diagonal embedding
    method.  The bulk gauge group is $G\times G\times G$, and the
    diagonal part $G^{\rm diag}$ remains manifest in the effective
    four-dimensional theory. Further spontaneous breaking of the gauge
    symmetry occurs through the dynamics of the zero modes of the
    extra-dimensional components of the gauge field. We apply this
    setup to the $SU(5)$ grand unified theory and examine the vacuum
    structure determined by the dynamics of the zero modes. The
    phenomenologically viable models are shown, in which the unified
    symmetry $G^{\rm diag}\cong SU(5)$ is spontaneously broken down to
    $SU(3)\times SU(2)\times U(1)$ at the global minima of the
    one-loop effective potential for the zero modes. This spontaneous
    breaking provides notable features such as a realization of the
    doublet-triplet splitting without fine tuning and a prediction of
    light adjoint fields.
\end{abstract}
\end{center}
\vskip 1.0 cm
\newpage

\tableofcontents

\bigskip \bigskip

%
\section{Introduction}
%

Higher-dimensional gauge theory has been studied in the last two
decades extensively as one of the attractive possibilities for physics
beyond the standard model (SM). It is worth noting that the
higher-dimensional gauge theory can possess the dynamical mechanism
for gauge symmetry breaking via continuous Wilson line phases, called
the Hosotani mechanism~\cite{hosotani1}. It is one of the promising
approaches to understand the origin of the gauge symmetry breaking in
the electroweak theory or in the grand unified theory (GUT)~\cite{GG,
  SUSY-GUT}.  The former attempts are called gauge-Higgs
unification~\cite{hewsb,edgut}. Hence, various aspects of the
higher-dimensional gauge theory with the Hosotani mechanism have been
investigated.

The zero modes of the extra-dimensional components of the gauge field
become the dynamical degrees of freedom, which behave as scalar fields
at low energy~\cite{manton,fair}. The zero modes are closely related
with the Wilson line phases, and the quantum correction generates the
effective potential for the phases. The zero modes can acquire vacuum
expectation values (VEVs) at a minimum of the potential to induce the
gauge symmetry breaking~\cite{hosotani1}.  Interestingly, the gauge
symmetry breaking patters are definitely determined irrespective of
the detail of the dynamics in the ultraviolet region thanks to the
finiteness of the effective potential for the phases once we fix the
content of matter fields in the
theory~\cite{finite1,finite2}.\footnote{Higher-loop corrections to the
  effective potential calculated with the bare Lagrangian generally
  include divergent contributions, which originate from the loop
  integral of subdiagrams~\cite{finite2,finitesub}.  Such divergent
  contributions are canceled out with lower-order counterterms, and it
  is expected that there is no need to introduce independent
  counterterms to eliminate the divergent
  contributions~\cite{finite2}. In other words, in terms of the
  renormalized couplings, which are determined by the low-energy
  experiments and thus finite, instead of the bare couplings, the
  effective potential is free from the divergences.} One understands
the definite origin of the potential that induces the gauge symmetry
breaking.

The zero modes originally belong to the adjoint representation under
the gauge group.  Thus, it looks attractive and natural to apply the
Hosotani mechanism to the spontaneous breaking of the GUT gauge
symmetry such as $SU(5)$~\cite{GG,SUSY-GUT}.  We immediately, however,
encounter the difficulty that the existence of the scalar zero mode of
the adjoint representation tends to be incompatible with chiral
fermions, which are required in phenomenologically acceptable
models. That is, if one tries to obtain the chiral fermion, the
orbifold compactification with appropriate boundary conditions (BCs)
is a possible framework, but the scalar zero mode of the adjoint
representation is projected out for the case.  Thus, the $SU(5)$
symmetry is broken by the BCs in many higher-dimensional GUT
models~\cite{orbifoldgut}. Otherwise, an alternative direction is to
consider GUT models with higher-rank gauge groups~\cite{hrnkgut} that
are spontaneously broken by VEVs of the scalar zero modes belonging to
non-adjoint representations~\cite{hredgut}.

The diagonal embedding method~\cite{DEstring} makes the adjoint zero
mode exist and overcomes the difficulty to apply the Hosotani
mechanism to the breaking of the $SU(5)$ symmetry accompanying the
chiral fermions.  Though the method is originally invented in the
context of the heterotic string theory, it is possible to apply it to
the higher-dimensional gauge theory. In fact, we have obtained the
five-dimensional GUT models compactified on the orbifold
$S^1/{\mathbb Z}_2$, in which the $SU(5)$ gauge symmetry is broken
down to that of the SM by the Hosotani mechanism without
contradicting the existence of the chiral fermion~\cite{gghusu5}. We
call the theoretical framework the type A(djoint) grand gauge-Higgs
unification. Phenomenologically notable aspects in the type A grand
gauge-Higgs unification with $S^1/{\mathbb Z}_2$ compactification have
been investigated~\cite{gghusu5DTS,gghusu5pheno,gghusu5FT,gghuDg}.
For other types of the grand gauge-Higgs unification, referred to also
as gauge-Higgs grand unification, see Ref.~\cite{edgut}, where the
Hosotani mechanism is utilized to break the electroweak symmetry.

What is striking is that the effective potential for the phases
obtained in the diagonal embedding method maintains the desirable
nature, that is, the finiteness. 
Hence, the VEV for the zero mode can be determined by minimizing the
effective potential for the fixed matter content to induces the GUT
gauge symmetry breaking without being affected by the physics in the
ultraviolet region.  Furthermore, the diagonal embedding method can
straightforwardly be extended to the case with more complex orbifold
compactification such as $T^2/\Z{3}$.

In this paper, we shall study the gauge symmetry breaking of the
six-dimensional (6D) $SU(5)$ gauge theory compactified on the
$T^2/\Z{3}$ in the type A grand gauge-Higgs unification.  In the
counterpart in the string theory for the $\Z3$ model, the gauge
symmetry is realized at a level-3 affine Lie algebra or Kac-Moody
algebra.  We note that there is a conjecture that the generation
number is a multiple of the level~\cite{DEgen}.  Though the generation
number is just a free parameter set by hand within the field theory,
it is meaningful to construct field theoretical models that can be
considered as effective theories of the string theoretical models with
three generations.  The 6D model compactified on the $T^2/\Z{3}$
orbifold is their simplest example.  It is important and interesting
to study the $T^2/\Z{3}$ compactification from the side of quantum
gauge field theory. One can study the breaking of the $SU(5)$ gauge
symmetry by minimizing the one-loop effective potential for the
phases.  We shall determine the gauge symmetry breaking patterns
through the Hosotani mechanism for various matter contents from the
one-loop effective potential and find matter contents that result the
SM gauge symmetry.  We also discuss the phenomenological implications
such as four-dimensional (4D) chiral fermions, fermion masses, proton
decay, and so on.

This paper is organized as follows. In the next section, we introduce
the basic aspects of the orbifold $T^2/\Z{3}$. We discuss the field
theoretical realization of the diagonal embedding method focusing on
the gauge fields on the orbifold $T^2/\Z{3}$ in
Sec.~\ref{sec:gauge}. This section contains the fundamental
ingredients for studying the gauge symmetry breaking in our model. The
matter fields are introduced in Sec.~\ref{sec:mat}, where the BCs and
the mass spectrum are studied. We compute the effective potential for
the Wilson line phases in one-loop approximation and study the gauge
symmetry breaking patterns, including the breaking down to the gauge
symmetry of the SM, in Secs.~\ref{Sec:onelooppot}
and~\ref{Sec:vac}. We also discuss the phenomenological aspects of our
model in Sec.~\ref{Sec:vac}. The final section is devoted to
conclusions and discussions. Some details on the calculations are
given in the appendices.

%
\section{$T^2/{\mathbb Z}_3$ orbifold}
\label{sec_t2z3}
%

We consider the orbifold $T^2/{\mathbb Z}_3$ as the compact extra
dimensions. To deal with coordinate vectors in $T^2/{\mathbb Z}_3$, it
is convenient to use the basis vectors ${\bm e}_i$ and the metric
$g_{ij}$, which satisfies
\begin{align}
  {\bm e}_i\cdot {\bm e}_j&=g_{ij}, 
\qquad 
  {\bm e}_{i+2}=-{\bm e}_i-{\bm e}_{i+1}, \qquad {\bm e}_{i+3}={\bm e}_i,
\end{align}
where $i\in {\mathbb Z}$. Among ${\bm e}_i$, we can choose ${\bm e}_1$
and ${\bm e}_2$ as a linearly independent set.  A coordinate vector
${\bm y}$ in $T^2/{\mathbb Z}_3$ is spanned by the basis vector
as
\begin{align}\label{vecydef1}
  {\bm y}=y^i{\bm e}_i=y^1  {\bm e}_1+y^2  {\bm e}_2,
  \qquad y^i\in \mathbb R,
\end{align}
and it satisfies the following identifications:
\begin{align}
  &    {\bm y}\sim {\bm y}+2\pi R{\bm e}_1, \qquad 
    {\bm y}\sim {\bm y}+2\pi R{\bm e}_2,\label{t2vecid}
  \\
  &  {\bm y}=  y^i{\bm e}_i\sim y^i{\bm e}_{i+1}=y^1{\bm e}_2+y^2{\bm e}_3
    =-y^2{\bm e}_1+(y^1-y^2){\bm e}_2,\label{z3vecid}
\end{align}
where $R$ parametrizes the size of the compact space. Contractions
between upper and lower indices $i$ imply the summation over $i=1,2$
hereafter. By requiring that the metric $g_{ij}$ is invariant under
the transformation ${\bm e}_i\to {\bm e}_{i+1}$, we can fix it as
\begin{align}\label{t2z3met01}
  \begin{pmatrix}
      g_{11}&g_{12}
      \\g_{21}&g_{22}
  \end{pmatrix}=
                \begin{pmatrix}
                    1&-1/2\\-1/2&1
                \end{pmatrix},
\end{align}
up to an overall constant, which can be absorbed into the definition
of $R$.

The two-dimensional Cartesian coordinates, which we denote by $x^5$
and $x^6$, are related to the oblique coordinates $y^1$ and $y^2$.  We
take the basis such that $x^5=y^1$ and $x^6=0$ hold for $y^2=0$ as 
\begin{align}
  \label{cart_obli1}
  \begin{pmatrix}
      x^5\\x^6
  \end{pmatrix}
  &=
  \begin{pmatrix}
      1&\cos(2\pi/3)\\
      0&\sin(2\pi/3)
  \end{pmatrix}
  \begin{pmatrix}
      y^1\\y^2
  \end{pmatrix}=
  \begin{pmatrix}
      1&-{1/2}\\
      0&{\sqrt{3}/2}
  \end{pmatrix}
  \begin{pmatrix}
      y^1\\y^2
  \end{pmatrix}, \\
  \label{cart_obli2}
  \begin{pmatrix}
      y^1\\y^2
  \end{pmatrix}
  &=
  \begin{pmatrix}
      1&-\cot(2\pi/3)\\
      0&\csc(2\pi/3)
  \end{pmatrix}
  \begin{pmatrix}
      x^5\\x^6
  \end{pmatrix}=
  \begin{pmatrix}
      1&1/\sqrt{3}\\
      0&2/\sqrt{3}
  \end{pmatrix}
  \begin{pmatrix}
      x^5\\x^6
  \end{pmatrix},
\end{align}

\begin{figure}[]
\centering
  \includegraphics[width=7cm,clip]{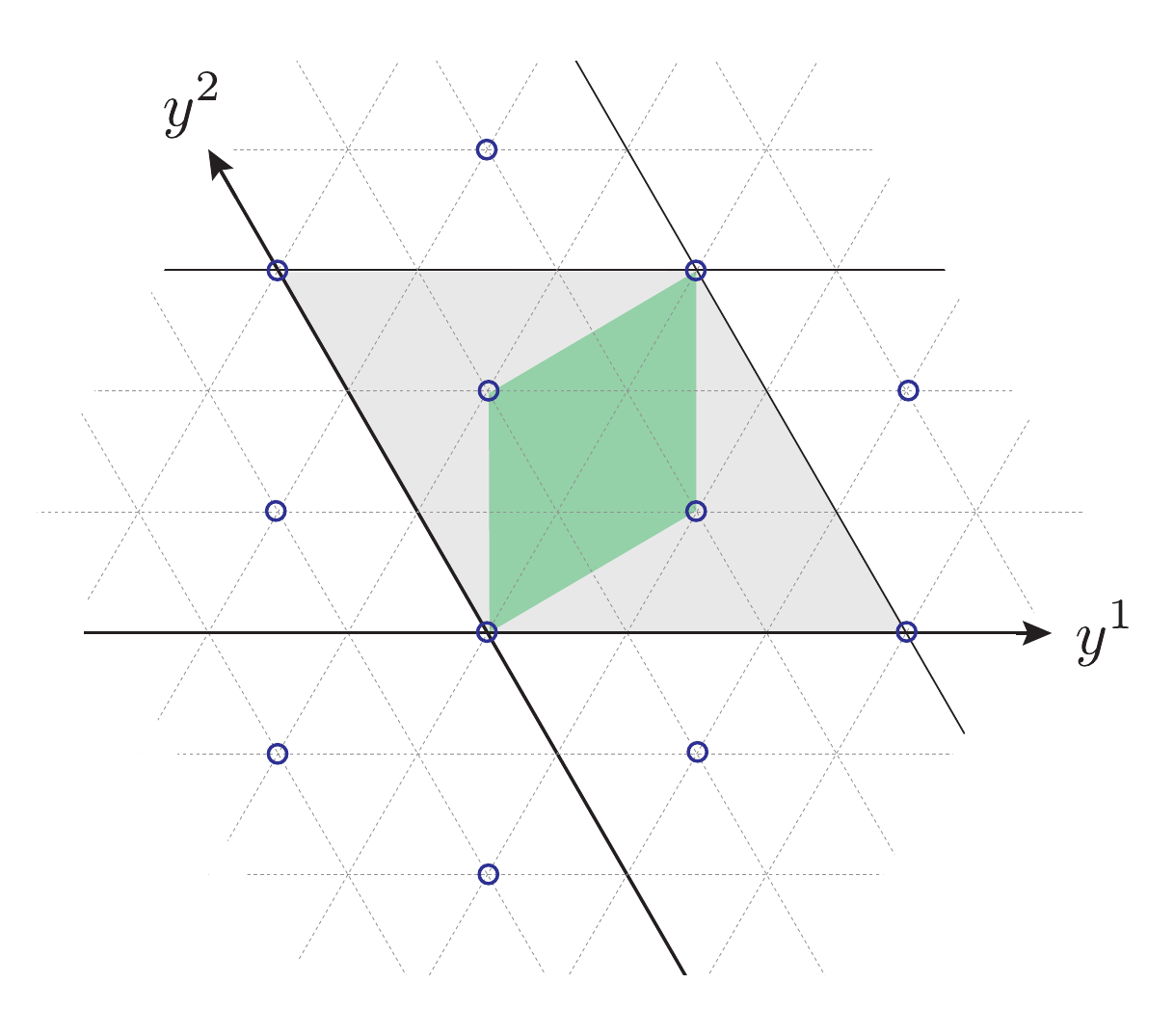} 
  \caption{The oblique coordinate system on $T^2/{\mathbb Z}_3$. The
    gray shaded region is an independent domain of the torus
    $T^2$. The green shaded region is a fundamental domain of the
    orbifold $T^2/{\mathbb Z}_3$.  The small circles correspond to the
    fixed points.}\bigskip
\label{fig_t2z3}
\end{figure}

In light of Eqs.~\eqref{t2vecid} and~\eqref{z3vecid}, let us introduce
the operators $\hat{\cal T}_j$ $(j=1,2)$ and $\hat {\cal S}_0$ that
act on the coordinates $y^i$ as
\begin{gather}
    \hat{\cal T}_1
  \begin{pmatrix}
      y^1\\ y^2
  \end{pmatrix}
  =
  \begin{pmatrix}
        \hat{\cal T}_1[y^1]\\   \hat{\cal T}_1[y^2]
  \end{pmatrix}
  =
  \begin{pmatrix}
      y^1+2\pi R\\ y^2
  \end{pmatrix},\qquad
    \hat{\cal T}_2
  \begin{pmatrix}
      y^1\\ y^2
  \end{pmatrix}
  =
  \begin{pmatrix}
        \hat{\cal T}_2[y^1]\\   \hat{\cal T}_2[y^2]
  \end{pmatrix}
  =
  \begin{pmatrix}
      y^1\\ y^2+2\pi R
  \end{pmatrix},
    \label{tis0ytrans1}\\
  \hat {\cal S}_0
  \begin{pmatrix}
      y^1\\ y^2
  \end{pmatrix}
  =
  \begin{pmatrix}
        \hat {\cal S}_0[y^1]\\   \hat {\cal S}_0[y^2]
  \end{pmatrix}
  =
  \begin{pmatrix}
      -y^2\\
y^1-y^2
\end{pmatrix}.
    \label{tis0ytrans2}
\end{gather}
The identifications in Eqs.~\eqref{t2vecid} and~\eqref{z3vecid} are
rewritten as 
\begin{align}\label{tis0def}
\hat{\cal T}_1[y^i]\sim y^i, \qquad \hat{\cal T}_2[y^i]\sim y^i, \qquad \hat{\cal S}_0[y^i]\sim y^i.
\end{align}
We can define an independent domain of the $T^2$ torus regarding the
identifications given by $\hat{\cal T}_1$ and $\hat{\cal T}_2$, where
one of the domains is shown as the gray shaded region in
Fig.~\ref{fig_t2z3}. The additional identification given by
$\hat{\cal S}_0$ defines the orbifold $T^2/{\mathbb Z}_3$, which has
the fundamental domain shown in Fig.~\ref{fig_t2z3} by the green
shaded region.

There exist fixed points on the orbifold that are invariant up to the
translations $\hat{\cal T}_1$ and $\hat{\cal T}_2$ under the discrete
rotation $\hat{\cal S}_0$. That is, the fixed points are given by the
solution to the following equation:
\begin{align}
    (\hat {\cal T}_1)^{n_1} (\hat {\cal T}_2)^{n_2} \hat {\cal S}_0
  \begin{pmatrix}
      y^1\\ y^2
  \end{pmatrix}
  =
  \begin{pmatrix}
      y^1\\ y^2
  \end{pmatrix}, \qquad {\rm where} \qquad n_1,n_2\in {\mathbb Z}. 
\end{align}
  We denote the three fixed points on the fundamental
  domain of $T^2/{\mathbb Z}_3$
  by $y^i_{{\rm f}(r)}$ $(r=0,1,2)$, which are given by
\begin{align}\label{yfp1}
  \begin{pmatrix}
   y^1_{{\rm f}(0)}\\y^2_{{\rm f}(0)}   
  \end{pmatrix}=
  \begin{pmatrix}
      0\\0
  \end{pmatrix},\qquad
  \begin{pmatrix}
   y^1_{{\rm f}(1)}\\y^2_{{\rm f}(1)}   
  \end{pmatrix}=2\pi R
  \begin{pmatrix}
      2/3\\1/3
  \end{pmatrix},\qquad
  \begin{pmatrix}
   y^1_{{\rm f}(2)}\\y^2_{{\rm f}(2)}   
  \end{pmatrix}=2\pi R
  \begin{pmatrix}
      1/3\\2/3
  \end{pmatrix}. 
\end{align}
Any other fixed points are given by the translations of
$y^i_{{\rm f}(r)}$ generated by $\hat{\cal T}_1$ and $\hat{\cal T}_2$.
In Fig.~\ref{fig_t2z3}, the fixed points are shown by the small
circles.

The operators $\hat{\cal T}_1$, $\hat{\cal T}_2$, and $\hat{\cal S}_0$
satisfy
\begin{align}
    \hat {\cal T}_1 \hat {\cal T}_2=\hat {\cal T}_2 \hat {\cal T}_1, \qquad 
  \hat {\cal S}_0 \hat {\cal T}_1=\hat {\cal T}_2 \hat {\cal S}_0.
\label{t2byt1s0}
\end{align}
Thus, $\hat{\cal T}_1$, $\hat{\cal T}_2$, and $\hat{\cal S}_0$ are not
independent to each other.  In addition, it is convenient to define
\begin{align}
  \hat {\cal S}_1\equiv \hat {\cal T}_1\hat {\cal S}_0, \qquad 
  \hat {\cal S}_2\equiv \hat {\cal T}_2 \hat {\cal T}_1\hat {\cal S}_0,
\label{defs1s2}
\end{align}
where $\hat {\cal S}_r^3=\hat {\cal I}$ for $r=0,1,2$ and
$\hat {\cal I}[y^i]=y^i$.  The operators $\hat {\cal S}_r$ give
$2\pi/3$ rotations around the fixed points $y^i_{{\rm f}(r)}$ and
satisfy $\hat {\cal S}_r[y^i_{{\rm f}(r)}]=y^i_{{\rm f}(r)}$.  Among
the above operators, we can choose $\hat{\cal T}_1$ and
$\hat{\cal S}_0$ as the independent ones, and the others
$\hat{\cal T}_2$, $\hat{\cal S}_1$, and $\hat{\cal S}_2$ can be
expressed by $\hat{\cal T}_1$ and $\hat{\cal S}_0$.

It is useful to introduce the dual basis vectors $\tilde {\bm e}^i$ as 
\begin{align}\label{dualtildeedef}
  \til {\bm e}^i\cdot {\bm e}_j=\delta_j^i, \qquad 
  \til {\bm e}^i\cdot \til {\bm e}^j=g^{ij}, \qquad g^{ik}g_{kj}=\delta_j^i, 
\end{align}
where $\delta_j^i$ is the Kronecker delta and
\begin{align}\label{invmetric1}
  \begin{pmatrix}
      g^{11}&g^{12}\\g^{21}&g^{22}
  \end{pmatrix}
                             =
 {4\over 3} \begin{pmatrix}
      1&1/2\\1/2&1
  \end{pmatrix}.
\end{align}
Note that $g^{ij}{\bm e}_j=\til {\bm e}^i$ and
$g_{ij}\til {\bm e}^j={\bm e}_i$ hold. We can introduce a dual vector
$\tilde {\bm k}$ that is spanned by the dual basis vectors as
\begin{align}
  \til {\bm k}=k_1\til {\bm e}^1+k_2\til {\bm e}^2, 
\qquad k_i\in{\mathbb R}.
\end{align}
Then, one sees $\til {\bm k}\cdot {\bm y}=k_iy^i\in {\mathbb R}$.  As
discussed in Appendix~\ref{Sec:kkexp1}, $\tilde{\bm e}^i$ is a natural
basis for a Kalzua-Klein (KK) discretized momentum, which is mapped to
a point on the lattice spanned by $\tilde{\bm e}^i$ in a
normalization.

The identification in Eq.~\eqref{z3vecid} is related to the basis
change ${\bm e}_i\to {\bm e}_{i+1}$. Under the basis change, the dual
basis vectors also change
$\tilde {\bm e}^i\to \tilde {\bm e}^{\prime\, i}$. Requiring
${\bm e}_{i+1}\cdot \tilde {\bm e}^{\prime\, i}=\delta_i^j$, we obtain
$\tilde {\bm e}^{\prime\, 1}=-\tilde {\bm e}^1+\tilde {\bm e}^2$ and
$\tilde {\bm e}^{\prime\, 2}=-\tilde {\bm e}^1$.  Thus, corresponding
to Eq.~\eqref{z3vecid}, we obtain the identification for the dual
vector as
\begin{align}
  \til{\bm k}=k_i\til {\bm e}^i\sim 
  k_1(-\til {\bm e}^1+\til {\bm e}^2)+k_2(-\til {\bm e}^1)=(-k_1-k_2)\til {\bm e}^1+k_1\til {\bm e}^2.
\end{align}
Then, the action of the operator $\hat {\cal S}_0$ on the
coordinates of dual vectors is naturally defined by
\begin{align}
  \hat {\cal S}_0[k_i]=k_{i-1}, 
  \label{kis0trans}
\end{align}
where we have also defined
\begin{align}
  k_0=-k_1-k_2, \qquad k_{i+3}=k_i,\qquad i\in {\mathbb Z}.
\end{align}
From the above, one sees
$\hat {\cal S}_0[k_i]\hat {\cal S}_0[y^i]=k_iy^i$ and
\begin{align}\label{kys_formula1}
  k_i\hat {\cal S}_0[y^i]=\hat {\cal S}_0^{-1}[k_i]y^i=k_{i+1}y^i.
\end{align}
We use Eq.~\eqref{kys_formula1} for deriving the KK expansions
of fields discussed in Appendix~\ref{Sec:kkexp1}.

%
\section{The diagonal embedding method on
  $M^4\times T^2/{\mathbb Z}_3$: gauge fields}
\label{sec:gauge}
%

\subsection{Lagrangian for gauge fields}
We start to discuss the gauge theory with the field theoretical
realization of the diagonal embedding method on
$M^4\times T^2/{\mathbb Z}_3$, where $M^4$ is the Minkowski
spacetime. In the following, we denote the 6D orthogonal coordinates
by $x^M= (x^\mu,x^5,x^6)$ ($\mu=0,1,2,3$). For the extra-dimensional
coordinates, we also use the oblique coordinates $y^1$ and $y^2$ in
Eq.~\eqref{cart_obli2} instead of $x^5$ and $x^6$.  The metric of
$M^4\times T^2/{\mathbb Z}_3$ is defined such that
$x^Mx_M=\eta_{\mu\nu}^{(4)}x^\mu x^\nu-g_{ij}y^iy^j$, where
$\eta_{\mu\nu}^{(4)}={\rm diag}(1,-1,-1,-1)$ and $g_{ij}$ is given in
Eq.~\eqref{t2z3met01}.

The action is given by the Lagrangians for the gauge fields
${\cal L}_{\rm YM}$ and the matter fields ${\cal L}_{\rm mat}$, which
will be discussed in the next section, as
\begin{align}
  S=\int_0^{2\pi R}dy^1\int_0^{2\pi R}dy^2 \sqrt{{\rm det}\, g_{ij}}
  \int d^4 x {\cal L}, \qquad {\cal L}={\cal L}_{\rm YM}+{\cal L}_{\rm mat}, 
\end{align}
where ${\rm det}\, g_{ij}=3/4$. The diagonal embedding method on the
orbifold requires that the theory respects three copies of gauge
symmetry $G$ and the global symmetry ${\zex}$ that permutes the three
copies cyclically.  Therefore, let us introduce the Lagrangian for the
gauge fields as
\begin{align}\label{lagym}
{\cal L}_{\rm YM}=  -{1\over 2}\sum_{k=1}^3{\rm Tr}\left(F_{MN}^{(k)}F^{(k)MN}\right), \quad 
  F_{MN}^{(k)}&=\der_M A_N^{(k)}-\der_N A_M^{(k)}{+}ig[A_M^{(k)},A_N^{(k)}],
\end{align}
where $g$ is the gauge coupling constant.  The gauge fields
$A_M^{(k)}$ $(k=1,2,3)$ are expanded by the generators of the gauge
symmetry as $A_M^{(k)}=A_M^{(k)a}t^{(k)}_a$, where the indices $a$ run
from 1 to the dimension of the Lie algebra of $G$, and the summation
over $a$ is implied. The operators $t^{(k)}_a$ $(k=1,2,3)$ are
representation matrices of the generators. We adopt the convention
that the matrices satisfy the following relations:
\begin{align}\label{gene_def1}
  [t^{(k)}_a,t^{(k')}_b]=if_{ab}{}^ct^{(k)}_c \delta^{kk'},
  \qquad
  \Tr[t^{(k)}_at^{(k')}_b]={1\over 2}
  \delta_{ab}\delta^{kk'}, 
\end{align}
where $f_{ab}{}^c$ is the structure constant. In Eqs.~\eqref{lagym}
and~\eqref{gene_def1}, the trace is taken over the representation
space.

The Lagrangian in Eq.~\eqref{lagym} has the gauge symmetry
$G\times G\times G$ and the global symmetry ${\zex}$.  We define the
gauge transformation of the gauge field as
\begin{align}\label{gtragauge}
  A_M^{(k)}\to \Omega^{(k)}
  \left(A_M^{(k)}-{i\over g}
  \der_M\right)\Omega^{(k)\dag}, \qquad
  \Omega^{(k)}=\exp\left({ig\alpha^{(k)a}t^{(k)}_a}\right), 
\end{align}
where $\alpha^{(k)a}(x)$ are gauge parameters. To define the global
${\zex}$ transformation of the gauge field, it is helpful to extend
the range of index $k\in\{1,2,3\}$ to $k\in {\mathbb Z}$ and to
introduce the periodicity for the index $k$, \eg,
$A_N^{(k+3)a}=A_N^{(k)a}$ and $t^{(k+3)}_a=t^{(k)}_a$. Hereafter, we
use this notation. Then, we can write the global ${\zex}$
transformation of the gauge field as follows:
\begin{align}\label{gtraz3}
  A_M^{(k)}=A_M^{(k)a}t^{(k)}_a\zexto A_M^{(k+1)a}t^{(k)}_a. 
\end{align}
Under the transformations in Eqs.~\eqref{gtragauge}
and~\eqref{gtraz3}, the Lagrangian in Eq.~\eqref{lagym} is invariant.

Using the above notation, we can define $A_M^{[p]a}$ that are the
eigenstates of the transformation in Eq.~\eqref{gtraz3} as
\begin{align}
  A_M^{[p]a}={1\over \sqrt{3}}\sum_{k=1}^3 \omega^{-kp}A_M^{(k)a},
  \qquad
  A_M^{(k)a}={1\over \sqrt{3}}\sum_{p=-1}^1 \omega^{kp}A_M^{[p]a},
  \label{apdef1}
\end{align}
where $p\in {\mathbb Z}$ and $\omega=e^{2\pi i/3}$.  From the above
definition, $A_M^{[p+3]a}=A_M^{[p]a}$ and $(A_M^{[p]a})^*=A_M^{[-p]a}$
hold. Note that $A_M^{[p]a}$ has the eigenvalue of $\omega^p$ under
the ${\zex}$ transformation in Eq.~\eqref{gtraz3}.

\subsection{Orbifold boundary conditions and residual gauge symmetries}
\label{sec:ressym}
In theories on the orbifold, field values are constrained since the
extra-dimensional coordinates obey the identifications discussed in
the previous section. To clarify the constraints, we define the
BCs~\cite{Hebecker:2001jb} for the gauge fields.  As discussed in
Sec.~\ref{sec_t2z3}, we treat $\hat {\cal T}_1$ and $\hat {\cal S}_0$
as the independent operators and define the BCs for the gauge fields
$A_M^{(k)a}(x^\mu,y^i)$ as follows:
\begin{align}\label{diagbc1}
&  A_\mu^{(k)a}(x^\mu, \hat {\cal T}_1[y^i])=A_\mu^{(k)a}(x^\mu, y^i), \qquad   A_\mu^{(k)a}(x^\mu, \hat {\cal S}_0[y^i])=A_\mu^{(k+1)a}(x^\mu, y^i),\\
&  A_{y^i}^{(k)a}(x^\mu, \hat {\cal T}_1[y^i])=A_{y^i}^{(k)a}(x^\mu, y^i), 
\qquad      A_{y^i}^{(k)a}(x^\mu, \hat {\cal S}_0[y^i])=A_{y^{i-1}}^{(k+1)a}(x^\mu, y^i),
\label{diagbc2}
\end{align}
where $A_{y^{1}}^{(k)a}=A_{5}^{(k)a}$,
$A_{y^{2}}^{(k)a}=-{1\over 2}A_{5}^{(k)a}+{\sqrt{3}\over
  2}A_{6}^{(k)a}$, and
$A_{y^{0}}^{(k)a}=-A_{y^{1}}^{(k)a}-A_{y^{2}}^{(k)a}$.  Hereafter, we
also use the notation $A_{y^{i+3}}^{(k)a}=A_{y^{i}}^{(k)a}$
$(i\in {\mathbb Z})$.

In general, BCs can nontrivially act on the representation space of
not only the discrete group $\zex$ but also the gauge group
$G$. Nevertheless, it should be emphasized that we can always take the
trivial BCs for $G$ as in Eqs.~\eqref{diagbc1} and~\eqref{diagbc2}
without loss of generality.  This is understood as follows. Although
one can introduce nontrivial transformations in the representation
space of $G$~\cite{Hebecker:2001jb,EDanom}, which we here call the
gauge twists, the nontrivial gauge twists do not affect the low-energy
physics in the present case.  For the gauge twist with respect to
$\hat {\cal S}_0$, this introduces just a difference among the bases
in the representation space of the generators $t_a^{(1)}$,
$t_a^{(2)}$, and $t_a^{(3)}$. Such difference can always be absorbed
into the redefinition of the generators $t_a^{(k)}$.  The gauge twist
with respect to $\hat {\cal T}_1$ can be absorbed by the continuous
Wilson line phases~\cite{EDanom}, which will be discussed in detail in
the next section, through the gauge transformations with the gauge
parameters depending on the extra-dimensional coordinates.  Then, if
the BCs are the same up to the gauge twist with respect to
$\hat {\cal T}_1$, these BCs are said to belong to the same
equivalence class~\cite{hosotani2,Equivclass,Equivclass6D}.  As seen
below, the vacuum is determined by a nontrivial expectation value of
the Wilson line phases.  It is known that BCs in an equivalence class
describe the same low-energy physics through the dynamics of the
Wilson line phases determined by the effective potential generated by
quantum corrections~\cite{Equivclass}.\footnote{ We introduce the
  twist for $\zex$ only associated with $\hat {\cal S}_0$ in
  Eqs.~\eqref{diagbc1} and~\eqref{diagbc2}.  One may consider a $\zex$
  twist associated with $\hat {\cal T}_1$, which cannot be absorbed by
  the Wilson line phases.}

From the BCs and Eq.~\eqref{apdef1}, it follows that
\begin{align}
\label{diagbcamu}
  &  A_\mu^{[p]a}(x^\mu, \hat {\cal T}_1[y^i])=A_\mu^{[p]a}(x^\mu, y^i), \qquad 
    A_\mu^{[p]a}(x^\mu, \hat {\cal S}_0[y^i])=\omega^pA_\mu^{[p]a}(x^\mu, y^i),\\
  &  A_{y^i}^{[p]a}(x^\mu, \hat {\cal T}_1[y^i])=A_{y^i}^{[p]a}(x^\mu, y^i), \qquad 
    A_{y^i}^{[p]a}(x^\mu, \hat {\cal S}_0[y^i])=\omega^pA_{y^{i-1}  }^{[p]a}(x^\mu, y^i).
    \label{diagbcay1y2}
\end{align}
The ${\mathbb Z}_3$ transformation of $y^i$ generated by
$\hat {\cal S}_0$ is discussed in Sec.~\ref{sec_t2z3} and is contained
in $SO(2)$ rotations that are part of the 6D Lorentz
transformation. Hence, the extra-dimensional components of the gauge
field nontrivially transform under the ${\mathbb Z}_3$ transformation,
and thus $A_{y^i}^{[p]a}$ are not the eigenstates of the BC for
$\hat {\cal S}_0$ in Eq.~\eqref{diagbcay1y2}.  We refer to the
${\mathbb Z}_3$ subgroup of the $SO(2)$ as $\zy$.  Let us denote the
eigenstates of the $\zy$ transformation by $A_{[q]}^{(k)a}$ that are
defined as
\begin{align}\label{yelltwisteq1}
  A_{[q]}^{(k)a}={1\over 3}\sum_{\ell=1}^3\omega^{-(\ell-1) q}A_{y^\ell}^{(k)a},
  \qquad A_{y^\ell}^{(k)a}=\sum_{q=1}^3\omega^{(\ell-1) q}
  A_{[q]}^{(k)a}, 
\end{align}
where $q\in {\mathbb Z}$ and the superscript of $y^\ell$ takes
$\ell=1,2,3$, whereas that of $y^i$ takes $i=1,2$ as explained in
Sec.~\ref{sec_t2z3}.  The normalization in Eq.~\eqref{yelltwisteq1} is
fixed by $A_{[\pm 1]}^{(k)a}=(A_5^{(k)a}\mp iA_6^{(k)a})/2$, which
correspond to the gauge fields $A_z^{(k)a}$ and $A_{\bar z}^{(k)a}$
associated with the complex coordinates $z=x^5+ix^6$ and
$\bar z=x^5-ix^6$.  With this definition, we find
$A_{[q+3]}^{(k)a}=A_{[q]}^{(k)a}$,
$(A_{[q]}^{(k)a})^*=A_{[-q]}^{(k)a}$, and $A_{[0]}^{(k)a}=0$.  For
fixed $k$ and $a$, there are two real degrees of freedom in
$A_{[q]}^{(k)a}$ as in $A_{y^i}^{(k)a}$.  From Eq.~\eqref{diagbc2},
the BC for $A_{[q]}^{(k)a}$ is given by
\begin{align}
  A_{[q]}^{(k)a}(x^\mu, \hat {\cal S}_0[y^i])=\omega^{-q}
  A_{[q]}^{(k+1)a}(x^\mu, y^i).
\end{align}
Namely $A_{[q]}^{(k)a}$ has the eigenvalue $\omega^{-q}$ under the
$\zy$ transformation.

From the above discussions, the eigenstates $A_{[q]}^{[p]a}$ of the
BCs are naturally defined as
\begin{align}\label{defaqpa}
  A_{[q]}^{[p]a}={1\over \sqrt{3}}\sum_{k=1}^3\omega^{-kp}A_{[q]}^{(k)a}
  ={1\over 3\sqrt{3}}\sum_{\ell=1}^3\sumk \omega^{-kp-(\ell-1)q}A_{y^\ell}^{(k)a}. 
\end{align}
Inversely, it also follows that
\begin{align}
  A_{y^\ell}^{(k)a}={1\over \sqrt{3}}\sump \sumq \omega^{(\ell-1) q+kp }A_{[q]}^{[p]a}.
\end{align}
Then, $A_{[\pm 1]}^{[p]a}$ satisfies the BCs as
\begin{align}\label{diagbcay}
  A_{[\pm 1]}^{[p]a}(x^\mu, \hat {\cal T}_1[y^i])=A_{  [\pm 1]}^{[p]a}(x^\mu, y^i), \qquad 
  A_{[\pm 1]}^{[p]a}(x^\mu, \hat {\cal S}_0[y^i])=\omega^{p\mp 1}A_{[\pm 1] }^{[p]a}(x^\mu, y^i).
\end{align}
From the BCs in Eqs.~\eqref{diagbcamu} and~\eqref{diagbcay}, it is
implied that $A_\mu^{[0]a}$, $A_{[1]}^{[1]a}$, and $A_{[-1]}^{[-1]a}$
have zero modes, which do not have ${\cal O}(1/R)$ KK masses in the 4D
effective theory.

We remind that the Lagrangian possesses $\zex$ and $\zy$
symmetries. The gauge field $A^{[p]a}_{[q]}$ has the charges
$\omega^p$ and $\omega^{-q}$ under the $\zex$ and $\zy$
transformations, respectively.  We can rearrange the two
$\mathbb{Z}_3$ symmetries as $\zpl$ and $\zmi$, under which
transformations $A^{[p]a}_{[q]}$ has the charges $\omega^{p-q}$ and
$\omega^{p+q}$, respectively. The BCs for $\hat {\cal S}_0$ introduced
in Eqs.~\eqref{diagbc1} and~\eqref{diagbc2} are regarded as the twist
for $\zpl$, and the zero mode is neutral under $\zpl$.

The BCs determine the zero modes of the gauge field. The low-energy
gauge symmetry associated with the zero modes of the 4D component of
the gauge field is referred to as the residual gauge symmetry. To
clarify the residual gauge symmetry, we focus on the covariant
derivative:
\begin{align}
\label{cd4d}
D_\mu&=
\der_\mu+i g\sumk A_\mu^{(k)a}t^{(k)}_a=  
\der_\mu+i\tilde g\sum_{p=-1}^1 A_\mu^{[p]a}t^{[-p]}_a, 
\end{align}
where we have introduced
\begin{align}\label{diggendef1}
  t^{[p]}_a&= \sumk \omega^{-kp}t^{(k)}_a,  \qquad
             t^{(k)}_a={1\over 3}\sump \omega^{kp}t^{[p]}_a,
             \qquad 
             \tilde g={g\over \sqrt{3}}.
\end{align}
The generator $t^{[p]}_a$ has a proper normalization and satisfies
\begin{align}
[t^{[p]}_a,t^{[p']}_b]=if_{ab}{}^ct^{[p+p']}_c.
  \label{tpcommutater}
\end{align}
As mentioned above, $A_\mu^{[0]a}$ have zero modes.  At a low-energy
regime, $A_\mu^{[1]a}$ and $A_\mu^{[-1]a}$ are decoupled from the
effective theory since they have no zero modes. Hence, the residual
gauge symmetry is the diagonal part of $G\times G\times G$ generated
by $t^{[0]}_a=t^{(1)}_a+t^{(2)}_a+t^{(3)}_a$. We denote this diagonal
part by $G^{\rm diag}$.  From the commutation relation in
Eq.~\eqref{tpcommutater} for $p=0$ and $p'=\pm 1$, we see that
$t^{[\pm 1]}_a$ transforms as the adjoint representation under the
residual gauge symmetry $G^{\rm diag}$.

\subsection{Wilson line phases and spontaneous symmetry breaking}
\label{sec:wilsonline}
Let us focus on the zero mode of the extra-dimensional component of
the gauge field. As discussed in the previous subsection,
$A_{[1]}^{[1]a}$ and $A_{[-1]}^{[-1]a}$ have zero modes. They carry
continuous Wilson line degrees of freedom and can develop nonzero
vacuum expectation values (VEVs) to break the gauge symmetry
spontaneously. We introduce the parametrization as
\begin{align}
  \vev{A_{[1]}^{[1]a}}\equiv {1\over R\tilde g}a_z^a, \qquad 
  \vev{A_{[-1]}^{[-1]a}}=\vev{A_{[1]}^{[1]a}}^*={1\over R\tilde g}a_z^{a*},
\label{vevspara1}
\end{align}
where $a_z^a$ is a complex parameter.  Except for the above,
$ \vev{A_{[\pm 1]}^{[p]a}}=0$ is satisfied.
We also introduce the parametrization of the VEVs of
$A_{y^\ell}^{(k)a}$ as
\begin{align}\label{tiladef2}
  \vev{A_{y^\ell}^{(k)a}}={1\over Rg}(\omega^{k+\ell-1}a_z^a+
  \bar \omega^{k+\ell-1}a_z^{a*}
  )\equiv {2\over Rg}\tilde a_{k+\ell}^a, 
\end{align}
where $\bar \omega=e^{-2\pi i/3}$. The real part of
$\omega^{k+\ell-1}a_z^a$ is equal to
$\tilde a_{k+\ell}^a$.\footnote{We have determined the normalization
  of $\tilde a_{k+\ell}^a$ in Eq.~\eqref{tiladef2} so that the Wilson
  line phase factors defined in Eq.~\eqref{wilsondef1} are invariant
  under integer shifts of $\tilde a_{k+\ell}^a$ in the $G=SU(N)$ case
  where the length of the root vectors are taken to be 1. Namely, the
  Cartan generator $H$ in the fundamental representation of the
  $SU(2)$ Lie algebra associated with a root vector is chosen as
  $H={\rm diag}(1,-1)/2$.}
From Eq.~\eqref{tiladef2}, one sees that
$\tilde a_{k+\ell}^a$ has the periodicity under the shift of its
subscript as $\tilde a_{k+\ell+3}^a =\tilde a_{k+\ell}^a$.

Let us consider the Wilson line phases defined with closed paths on
the orbifold $T^2/{\mathbb Z}_3$.  We denote the three distinct
noncontractible cycles by $C_\ell$ $(\ell=1,2,3)$. The cycle $C_1$ is
defined by the path from $y^1=0$ to $2\pi R$, while keeping $y^2=0$.
The cycle $C_2$ is defined by the path from $y^2=0$ to $2\pi R$, while
keeping $y^1=0$.  The cycle $C_3$ is defined by the path from
$-y^1-y^2=0$ to $2\pi R$, while keeping $y^1=y^2$.  By using them, we
define the Wilson line phase factors $W_\ell$ as
\begin{align}\label{wilsondef1}
  W_\ell&\equiv \exp\left(ig\sumk
       \oint_{C_\ell} dy^i
          \vev{A_{y^i}^{(k)a}}t^{(k)}_a\right)
          =\exp\left(2\pi ig R\sumk
          \vev{A_{y^\ell}^{(k)a}}t^{(k)}_a
          \right)\\
&\equiv \exp\left[i\left(
     \Theta_\ell+\Theta_\ell^\dag
     \right)
     \right],
\end{align}
where $\ell=1,2,3$, and we also define the Wilson line phases
$\Theta_\ell$ as
\begin{align}
  \Theta_\ell&={2\pi }\omega^{\ell-1} a_z^at_a^{[-1]}.
               \label{thetadef}
\end{align}
From the above, we find $\Theta_{\ell+k}=\omega^k\Theta_\ell$, which
implies $\Theta_1+\Theta_2+\Theta_3 \propto 1+\omega+\bar \omega=0$.
Let us note that the phase factors in Eq.~\eqref{wilsondef1} have
physical consequences, rather than the phases in
Eq.~\eqref{thetadef}~\cite{WuYang}.

The Wilson line phases are invariant under $\zpl$ since the phases
depend on the VEVs of $A^{[p]a}_{[p]}$ ($p=\pm 1$), which is neutral
under $\zpl$. On the other hand, $A^{[p]}_{[p]}$ has the eigenvalue
$\omega^{2p}$ under $\zmi$. One sees that the gauge fields and the
phases transform as
$\vev{A_{y^\ell}^{(k)a}}\to \vev{A_{y^{\ell+1}}^{(k+1)a}}$ and
$\tilde a_{k+\ell}^a\to \tilde a_{k+\ell-1}^a$ under $\zmi$.  This
implies the transformation law of the phase factors,
$W_\ell\to W_{\ell-1}$, under $\zmi$. Thus, the symmetry $\zmi$ is
generally broken by nontrivial VEVs of the Wilson line phases. Notice
that, if $W_1=W_2=W_3$ is satisfied, the symmetry $\zmi$ survives.
Thus, the vacuum with the alignment $W_1=W_2=W_3$ is discriminated in
view of the symmetry and is provided by
$\tilde a_\ell^a-\tilde a_{\ell+1}^a=0$ (mod 1).

The VEVs of the Wilson line phases are dynamically determined.  Thus,
we focus on the potential for the zero mode of $A_{y^i}^{(k)a}$. In
the present case, ${\cal L}_{\rm YM}$ involves the nonvanishing
potential for $A_{y^i}^{(k)a}$ at the classical level.\footnote{ In
  five-dimensional models compactified on the $S^1/{\mathbb Z}_2$
  orbifold, there is no tree-level potential only for the zero modes
  of the extra-dimensional components of the gauge field, although the
  zero modes can have tree-level potentials in supersymmetric models
  with the helps of additional scalars belonging to vector
  multiplets~\cite{nonflat5D}.}
From Eqs.~\eqref{lagym}
and~\eqref{gene_def1}, we obtain
\begin{align}
  {\cal L}_{\rm YM}\ni -{1\over 2}g^{ii'}g^{jj'}\Tr \left(
  \sumk F_{y^iy^j}^{(k)} \sumkd F_{y^{i'}y^{j'}}^{(k')}
  \right)=-{4\over 3}\Tr\left(
  \sumk F_{y^1y^2}^{(k)} \sumkd F_{y^{1}y^{2}}^{(k')}
  \right). 
\end{align}
The VEVs of the field strength tensors are written by the Wilson line
phases as
\begin{align}
  \sumk \vev{F_{y^iy^j}^{(k)}}
  =ig\left[\sumk\vev{A_{y^i}^{(k)a}}t^{(k)}_a,\sumkd\vev{A_{y^j}^{(k')a}}t^{(k')}_a  \right]
  ={i\over (2\pi R)^2g}[\Theta_i+\Theta_i^\dag,\Theta_j+\Theta_j^\dag],
\end{align}
where we have used Eq.~\eqref{thetadef}.  Therefore, the tree-level
potential for the Wilson line phases is given by
\begin{align}
  V_{\rm tree}\equiv 
  {4\over 3(2\pi R)^4g^2}\Tr({\cal F}_\Theta {\cal F}_\Theta^\dag), \quad
  {\rm where} \quad  
  {\cal F}_\Theta=[\Theta_1+\Theta_1^\dag,\Theta_2+\Theta_2^\dag]
  =(\bar \omega-\omega)[\Theta_1,\Theta_1^\dag].
\end{align}
Note that the tree-level potential is positive definite and has flat
directions.
On the flat directions,
$ [\Theta_1,\Theta_1^\dag]=0$ is satisfied, and hence the potential is
minimized as $V_{\rm tree}=0$.  In this case,
$W_1W_2W_3= e^{i\Theta_1+\Theta_2+\Theta_3}
e^{i\Theta_1^\dag+\Theta_2^\dag+\Theta_3^\dag }= 1$ is satisfied.

There are quantum corrections to the effective potential for the
phases.  As discussed above, the tree-level potential is minimized
along the flat directions.  Due to the loop factors, the quantum
corrections are generally suppressed compared to the tree-level
contribution if it is nonvanishing.  For the quadratic terms, it is
vanishing even along the nonflat direction.\footnote{ These quadratic
  terms are contained in tadpole terms of the field strength, which
  are generally generated on the fixed points~\cite{treepot6D}. In the
  present model, as long as $G$ of the bulk gauge group
  $G\times G\times G$ is semisimple, such tadpole terms are forbidden
  because $G$ remains unbroken on the fixed points.}  Thus, we
approximate that the minimum resides in the flat direction and
$[\Theta_1,\Theta_1^\dag]=0$ holds even if the quantum corrections are
incorporated. In this case, we can diagonalize $\Theta_\ell$ by
$G^{\rm diag}$ transformations without loss of generality.

The flat direction of the tree-level potential is no longer flat in
the effective potential. If some nontrivial values of the phase
degrees of freedom $\Theta_\ell+\Theta_\ell^\dag$ are determined by
the quantum corrections to the potential, the residual symmetry
$G^{\rm diag}$ is spontaneously broken to ${G_0}$, whose elements and
the Lie algebra ${\mf g}_0$ are given by
\begin{align}
  G_0=\{e^{i\alpha^at^{[0]}_a}~|~t^{[0]}_a\in {\mf g}_0,~\alpha^a\in {\mathbb R}\},
  \quad {\rm where}
  \quad {\mf g}_0=\{t^{[0]}_a~|~[t^{[0]}_a,W_j]=0~~{\rm for}~~j=1,2\}.
  \label{def_genH}
\end{align}
In this case, the zero modes of the gauge fields $A_\mu^{[0]a}$, which
is related to the broken generators corresponding to
$G^{\rm diag}/G_0$, acquire masses at low energy. This is understood
as follows. By using $y^i$-dependent gauge transformations, we can
always choose a gauge such that nontrivial VEVs of $A_{[p]}^{[p]a}$
are gauged away. After the gauge transformations, the BC for
$A_\mu^{[0]a}$ related to the translation by $\hat {\cal T}_j$ is
changed to
$A_\mu^{[0]a}(x^\mu,\hat {\cal T}_j[y^i])t^{[0]}_a
=W_jA_\mu^{[0]a}(x^\mu,y^i)t^{[0]}_aW_j^\dag$ $(j=1,2)$. Thus, the
zero modes of $A_\mu^{[0]a}$ of the broken generators are projected
out.  In this way, the spontaneous symmetry breaking can generally be
triggered by nontrivial VEVs of Wilson line phases.

Since $\Theta_\ell+\Theta_\ell^\dag$ is diagonal, we can expand them
by the elements of Cartan subalgebra ${\mf h}\in {\mf g}$, where
${\mf g}$ is the Lie algebra of $G$. We denote the generators in
${\mf h}$ by $H_{\hat a}$ $(\hat a=1,\dots,r)$, where $r$ is the rank
of ${\mf g}$.  Hence, we obtain
\begin{align}\label{diagonalwilsonline}
  \Theta_\ell+\Theta_\ell^\dag
  &=2\pi Rg\sumk
    \vev{A_{y^\ell}^{(k)\hat a}}H_{\hat a}^{(k)}
    = 2\pi\sumk\tilde a_{k+\ell}^{{\hat a}}2H_{\hat a}^{(k)}, \\\label{wilson_wl}
W_\ell
  &   =\exp({2\pi i\tilde a_{\ell+1}^{{\hat a}}2H_{\hat a}^{(1)}})\otimes
     \exp({2\pi i\tilde a_{\ell+2}^{{\hat a}}2H_{\hat a}^{(2)}})\otimes
     \exp({2\pi i\tilde a_{\ell}^{{\hat a}}2H_{\hat a}^{(3)}}).
\end{align}
To determine the VEVs of the Wilson line phases, we should evaluate
the effective potential for $\tilde a_\ell^{\hat a}$.  The quantum
corrections to the effective potential depend on the matter contents
of the theory.  Thus, we discuss bulk matter fields in the next
section, and the one-loop corrections are studied in
Sec.~\ref{Sec:onelooppot}.

%
\section{The diagonal embedding method on
  $M^4\times T^2/{\mathbb Z}_3$: bulk matter fields}
\label{sec:mat}
%

Let us start to discuss bulk matter fields. The invariance of the
Lagrangian under the ${\zex}$ transformation restricts the matter
contents of the theory.  We denote the representation of a matter
field under the bulk gauge symmetry $G\times G\times G$ by
$({\cal R}_1,{\cal R}_2,{\cal R}_3)$. In order to preserve the
${\zex}$ symmetry of the theory, matter fields should be incorporated
as the set of the representations
$({\cal R}_1,{\cal R}_2,{\cal R}_3)$,
$({\cal R}_3,{\cal R}_1,{\cal R}_2)$, and
$({\cal R}_2,{\cal R}_3,{\cal R}_1)$. We refer to this set of fields
as a ${\zex}$ {\it threefold}. However, there is an exception; if a
field belongs to the representation of
${\cal R}_1={\cal R}_2={\cal R}_3$, we can incorporate a single field
keeping ${\zex}$.  We refer to the field of the type
$({\cal R},{\cal R},{\cal R})$ as a ${\zex}$ {\it onefold}.

\subsection{Lagrangian for bulk scalar fields}
\label{sec:scalar_lag}
As the simplest example, we first discuss a threefold scalar
$\Phi_{{\cal R}}^{(k)}$ ($k =1,2,3$), which belongs to the following
representation:
\begin{align}\label{simple_threefold}
  \Phi_{{\cal R}}^{(1)}\sim ({\cal R},\bm 1,\bm 1),\qquad 
  \Phi_{{\cal R}}^{(2)}\sim (\bm 1,{\cal R},\bm 1),\qquad 
  \Phi_{{\cal R}}^{(3)}\sim (\bm 1,\bm 1,{\cal R}),
\end{align}
where $\bm 1$ means the singlet under $G$. Their components are
denoted by $(\Phi_{{\cal R}}^{(k)})_\alpha$, where $\alpha$ runs 1 to
dim(${\cal R}$).  Under the ${\zex}$ transformation, the threefold
scalar can be defined to transform as
$ (\Phi_{{\cal R}}^{(k)})_\alpha\to \omega^p(\Phi_{{\cal
    R}}^{(k+1)})_\alpha $, where $p\in \{0,\pm 1\}$.  Here, we
introduce the notation
$\Phi_{{\cal R}}^{(k+3)}\equiv \Phi_{{\cal R}}^{(k)}$
$(k\in {\mathbb Z})$ for convenience.  For a real field such as the
gauge field, the integer $p$ should be 0. For the complex scalars, the
phase factor $\omega^p$ can be absorbed by redefinitions of
$(\Phi_{{\cal R}}^{(k)})_\alpha$.  From the above definitions, one
sees that the following Lagrangian is ${\zex}$ invariant:
\begin{align}
  {\cal L}(\Phi_{{\cal R}}^{(k)})
  \equiv 
  \sum_{k=1}^3|(D_M^{(k)})_\alpha^\beta(\Phi_{{\cal R}}^{(k)})_\beta|^2,
  \qquad (D_M^{(k)})_\alpha^\beta=\der_M\delta_\alpha^\beta+igA_M^{(k)a}(T_a^{[{\cal R}]})_\alpha^\beta,
  \label{cov_der_r1}
\end{align}
where the repeated sets of upper and lower indices are summed.  The
representation matrices on ${\cal R}$ of the generators of $G$ are
denoted by $T_a^{[{\cal R}]}$.  The threefold scalar
$\Phi_{{\cal R}}^{(k)}$ transforms as ${\cal R}$ under $G^{\rm diag}$.

Next, let us discuss a more general case. Let
$\Phi_{{\cal R}_{123}}^{(k)}$ $(k =1,2,3)$ be a threefold scalar that
belongs to the following representations:
\begin{align}\label{gen_threefold}
\Phi_{{\cal R}_{123}}^{(1)}\sim   ({\cal R}_1,{\cal R}_2,{\cal R}_3),\qquad 
\Phi_{{\cal R}_{123}}^{(2)}\sim   ({\cal R}_3,{\cal R}_1,{\cal R}_2),\qquad 
\Phi_{{\cal R}_{123}}^{(3)}\sim   ({\cal R}_2,{\cal R}_3,{\cal R}_1). 
\end{align}
We denote elements of the representation matrices
$T_a^{[{\cal R}_k]}$ of the generators by
$(T_a^{[{\cal R}_k]})^{\alpha_k}_{\beta_k}$, where the indices
$\alpha_k$ and $\beta_k$
run from 1 to dim(${\cal R}_k$). The component of
$\Phi_{{\cal R}_{123}}^{(k)}$ is written as
\begin{align}\label{phir123comp1}
  (\Phi_{{\cal R}_{123}}^{(1)})_{\alpha_1\alpha_2\alpha_3}, \qquad 
  (\Phi_{{\cal R}_{123}}^{(2)})_{\alpha_3\alpha_1\alpha_2}, \qquad 
  (\Phi_{{\cal R}_{123}}^{(3)})_{\alpha_2\alpha_3\alpha_1}.
\end{align}
We introduce a convenient notations
$\Phi_{{\cal R}_{123}}^{(k+3)}\equiv \Phi_{{\cal R}_{123}}^{(k)}$
$(k\in {\mathbb Z})$ and
$(\Phi_{{\cal R}_{123}}^{(k)})_{[\alpha_1\alpha_2\alpha_3]}$, where
the latter represents the components in Eq.~\eqref{phir123comp1} all
at once, that is,
$(\Phi_{{\cal R}_{123}}^{(1)})_{[\alpha_1\alpha_2\alpha_3]}=
(\Phi_{{\cal R}_{123}}^{(1)})_{\alpha_1\alpha_2\alpha_3}$,
$(\Phi_{{\cal R}_{123}}^{(2)})_{[\alpha_1\alpha_2\alpha_3]}=
(\Phi_{{\cal R}_{123}}^{(2)})_{\alpha_3\alpha_1\alpha_2}$, and
$(\Phi_{{\cal R}_{123}}^{(3)})_{[\alpha_1\alpha_2\alpha_3]}=
(\Phi_{{\cal R}_{123}}^{(3)})_{\alpha_2\alpha_3\alpha_1}$.  The
${\zex}$ transformation law of the threefold field is defined as
\begin{align}\label{z3trphi1231}
&  
(\Phi_{{\cal R}_{123}}^{(1)})_{\alpha_1\alpha_2\alpha_3} \to
\omega^p (\Phi_{{\cal
    R}_{123}}^{(2)})_{\alpha_3\alpha_1\alpha_2}, \\\label{z3trphi1232}
&(\Phi_{{\cal R}_{123}}^{(2)})_{\alpha_3\alpha_1\alpha_2} \to
\omega^p (\Phi_{{\cal
    R}_{123}}^{(3)})_{\alpha_2\alpha_3\alpha_1}, \\\label{z3trphi1233}
&(\Phi_{{\cal R}_{123}}^{(3)})_{\alpha_2\alpha_3\alpha_1} \to
\omega^p (\Phi_{{\cal
    R}_{123}}^{(1)})_{\alpha_1\alpha_2\alpha_3},
\end{align}
which can be summarized as
\begin{align}
  (\Phi_{{\cal R}_{123}}^{(k)})_{[\alpha_1\alpha_2\alpha_3]}
  \to
  (\Phi_{{\cal R}_{123}}^{(k+1)})_{[\alpha_1\alpha_2\alpha_3]}. 
\end{align}
We note that the phase factor $\omega^p$ appearing in the above can be
absorbed by the field redefinitions for the case with complex scalars.
The ${\zex}$-invariant kinetic term for
$\Phi_{{\cal R}_{123}}^{(k)}$ is given by
\begin{align}
  {\cal L}(\Phi_{{\cal R}_{123}}^{(k)})
  &\equiv
    \sumk  |D_M^{(k)}\Phi_{{\cal R}_{123}}^{(k)}|^2\\
\notag   & =
    |(D_M^{(1)})_{\alpha_1\alpha_2\alpha_3}^{\beta_1\beta_2\beta_3}
    (\Phi_{{\cal R}_{123}}^{(1)})_{\beta_1\beta_2\beta_3}|^2
+    
    |(D_M^{(2)})_{\alpha_3\alpha_1\alpha_2}^{\beta_3\beta_1\beta_2}
    (\Phi_{{\cal R}_{123}}^{(2)})_{\beta_3\beta_1\beta_2}|^2
+    
    |(D_M^{(3)})_{\alpha_2\alpha_3\alpha_1}^{\beta_2\beta_3\beta_1}
    (\Phi_{{\cal R}_{123}}^{(3)})_{\beta_2\beta_3\beta_1}|^2,
\end{align}
where the covariant derivatives are written as follows:
\begin{align}\label{gentfcov1}
  (D_M^{(1)})_{\alpha_1\alpha_2\alpha_3}^{\beta_1\beta_2\beta_3}
  &=
    \delta^{\beta_1}_{\alpha_1}\delta^{\beta_2}_{\alpha_2}\delta^{\beta_3}_{\alpha_3}\der_M\\
&\quad\notag    +ig
    \left\{
    A_M^{(1)a}  (T_a^{[{\cal R}_{1}]})^{\beta_1}_{\alpha_1}
    \delta^{\beta_2}_{\alpha_2}\delta^{\beta_3}_{\alpha_3}
    +
    \delta^{\beta_1}_{\alpha_1}
    A_M^{(2)a}  (T_a^{[{\cal R}_{2}]})^{\beta_2}_{\alpha_2}
    \delta^{\beta_3}_{\alpha_3}
    +
    \delta^{\beta_1}_{\alpha_1}\delta^{\beta_2}_{\alpha_2}
    A_M^{(3)a}  (T_a^{[{\cal R}_{3}]})^{\beta_3}_{\alpha_3}
    \right\},  \\
    (D_M^{(2)})_{\alpha_3\alpha_1\alpha_2}^{\beta_3\beta_1\beta_2}
  &=
    \delta^{\beta_3}_{\alpha_3}\delta^{\beta_1}_{\alpha_1}\delta^{\beta_2}_{\alpha_2}\der_M\\
&\quad\notag
    +ig
    \left\{
    A_M^{(1)a}  (T_a^{[{\cal R}_{3}]})^{\beta_3}_{\alpha_3}
    \delta^{\beta_1}_{\alpha_1}\delta^{\beta_2}_{\alpha_2}
    +
    \delta^{\beta_3}_{\alpha_3}
    A_M^{(2)a}  (T_a^{[{\cal R}_{1}]})^{\beta_1}_{\alpha_1}
    \delta^{\beta_2}_{\alpha_2}
    +
    \delta^{\beta_3}_{\alpha_3}\delta^{\beta_1}_{\alpha_1}
    A_M^{(3)a}  (T_a^{[{\cal R}_{2}]})^{\beta_2}_{\alpha_2}
    \right\},\\\label{gentfcov3}
  (D_M^{(3)})_{\alpha_2\alpha_3\alpha_1}^{\beta_2\beta_3\beta_1}
  &=
    \delta^{\beta_2}_{\alpha_2}\delta^{\beta_3}_{\alpha_3}\delta^{\beta_1}_{\alpha_1}\der_M\\\notag
&\quad
    +ig
    \left\{
    A_M^{(1)a}  (T_a^{[{\cal R}_{2}]})^{\beta_2}_{\alpha_2}
    \delta^{\beta_3}_{\alpha_3}\delta^{\beta_1}_{\alpha_1}
    +
    \delta^{\beta_2}_{\alpha_2}
    A_M^{(2)a}  (T_a^{[{\cal R}_{3}]})^{\beta_3}_{\alpha_3}
    \delta^{\beta_1}_{\alpha_1}
    +
    \delta^{\beta_2}_{\alpha_2}\delta^{\beta_3}_{\alpha_3}
    A_M^{(3)a}  (T_a^{[{\cal R}_{1}]})^{\beta_1}_{\alpha_1}
    \right\}.
\end{align}
From the ${\zex}$ transformation law of the gauge field,
$A_M^{(k)a}\to A_M^{(k+1)a}$, we find that the above covariant
derivatives transform as
$(D_M^{(k)})_{[\alpha_1\alpha_2\alpha_3]}^{[\beta_1\beta_2\beta_3]}
\to
(D_M^{(k+1)})_{[\alpha_1\alpha_2\alpha_3]}^{[\beta_1\beta_2\beta_3]}$,
where we use the same notations for the indices as
$(\Phi_{{\cal R}_{123}}^{(k)})_{[\alpha_1\alpha_2\alpha_3]}$.  The
transformation law helps us to see the ${\zex}$ invariance of the
above Lagrangian.

Let us discuss the irreducible decomposition of
$\Phi_{{\cal R}_{123}}^{(k)}$ under $G^{\rm diag}$.  For any $k$,
$\Phi_{{\cal R}_{123}}^{(k)}$ transforms under $G^{\rm diag}$ as the
common reducible direct product representation
${\cal R}_1\otimes{\cal R}_2\otimes{\cal R}_3$, which can be
decomposed into the direct sum of irreducible representations
$\tilde {\cal R}_i$ $(i=1,\dots,n)$ as
\begin{align}
  {\cal R}_1\otimes{\cal R}_2\otimes{\cal R}_3=
  \tilde {\cal R}_{1}\oplus\tilde {\cal R}_{2}\oplus\dots\oplus\tilde {\cal R}_{n}.
\end{align}
This ensures that, in the representation space, there exist linear
transformations that decompose
$(\Phi_{{\cal R}_{123}}^{(k)})_{[\alpha_1\alpha_2\alpha_3]}$
into a set of irreducible representations under $G^{\rm diag}$ as
\begin{align}
 (\Phi_{{\cal R}_{123}}^{(k)})_{[\alpha_1\alpha_2\alpha_3]}
  \to (\Phi_{\tilde {\cal R}_1}^{(k)})_{\tilde \alpha_1}
  \oplus (\Phi_{\tilde {\cal R}_2}^{(k)})_{\tilde \alpha_2}
  \oplus\dots \oplus (\Phi_{\tilde {\cal R}_n}^{(k)})_{\tilde \alpha_n},
  \label{r123lin1}
\end{align}
where $\tilde \alpha_i$ runs from $1$ to
dim($\tilde {\cal R}_i$).\footnote{The linear transformations that
  give Eq.~\eqref{r123lin1} generally depend on $k$ of
  $\Phi_{{\cal R}_{123}}^{(k)}$.} Thus, we can find a basis in the
representation space such that each irreducible representation
transforms under the ${\zex}$ transformation as
\begin{align}
  (\Phi_{\tilde {\cal R}_i}^{(k)})_{\tilde \alpha_i}
 \zexto \omega^p(\Phi_{\tilde {\cal R}_i}^{(k+1)})_{\tilde \alpha_i}.
\end{align}
The above discussion means that a general threefold scalar transforms
under $G^{\rm diag}$ as a set of the threefold scalars of the type in
Eq.~\eqref{simple_threefold}; this is schematically written as
\begin{align}
&  ({\cal R}_1,{\cal R}_2,{\cal R}_3)
+
  ({\cal R}_3,{\cal R}_1,{\cal R}_2)
+
  ({\cal R}_2,{\cal R}_3,{\cal R}_1)  \sim
  \sum_{\tilde {\cal R}_i}
  \left[
  (\tilde {\cal R}_i,\bm 1,\bm 1)
  +(\bm 1,\tilde {\cal R}_i,\bm 1)
  +(\bm 1,\bm 1,\tilde {\cal R}_i)
                \right].
                \label{schemrel1}
\end{align}
We should note that the above relation for matter fields is limited
for the transformation properties under $G^{\rm diag}$, while the
couplings between the matter fields and the Wilson line phases, which
belong to $(G\times G\times G)/G^{\rm diag}$, are slightly modified
from the above. We will discuss the modification with an explicit
example in the next section.

Finally let us discuss the onefold scalar $\Phi_{{\cal R}^3}$, which
belongs to the following representation:
\begin{align}
  \Phi_{{\cal R}^3}\sim ({\cal R},{\cal R},{\cal R}).
\end{align}
The component of the onefold is denoted by
$(\Phi_{{\cal R}^3})_{\alpha_1\alpha_2\alpha_3}$, which transforms as
$(\Phi_{{\cal R}^3})_{\alpha_1\alpha_2\alpha_3}\to
\omega^p(\Phi_{{\cal R}^3})_{\alpha_3\alpha_1\alpha_2}$ under the
${\zex}$ transformation. We note that the phase factor $\omega^p$
cannot be absorbed into field redefinitions in the onefold case. If
one considers the threefold scalar of the representation
${\cal R}_1={\cal R}_2={\cal R}_3$, whose $\zex$ transformation law is
given by
$(\Phi_{{\cal R}_{123}}^{(k)})_{[\alpha_1\alpha_2\alpha_3]} \to
\omega^{p'} (\Phi_{{\cal
    R}_{123}}^{(k+1)})_{[\alpha_1\alpha_2\alpha_3]}$, the linear
combination
$\sum_{k=1}^3 \omega^{-(p-p')k}\Phi_{{\cal R}_{123}}^{(k)}$ has the
same transformation law of the onefold scalar.  The kinetic term for
the onefold scalar is given by
\begin{align}
    \label{cov_onefold}
  {\cal L}(\Phi_{{\cal R}^3})
  &\equiv 
    |(D_M)^{\beta_1\beta_2\beta_3}_{\alpha_1\alpha_2\alpha_3}(\Phi_{{\cal R}^3})_{\beta_1\beta_2\beta_3}|^2,\\    \notag
  (D_M)^{\beta_1\beta_2\beta_3}_{\alpha_1\alpha_2\alpha_3}
  &=\delta^{\beta_1}_{\alpha_1}\delta^{\beta_2}_{\alpha_2}\delta^{\beta_3}_{\alpha_3}\der_M
  \\
  & \quad +ig
    \left\{
    A_M^{(1)a}
    (T_a^{[{\cal R}]})^{\beta_1}_{\alpha_1}\delta^{\beta_2}_{\alpha_2}\delta^{\beta_3}_{\alpha_3}
    +A_M^{(2)a }
    \delta^{\beta_1}_{\alpha_1}(T_a^{[{\cal R}]})^{\beta_2}_{\alpha_2}\delta^{\beta_3}_{\alpha_3}
    +A_M^{(3)a}
    \delta^{\beta_1}_{\alpha_1}\delta^{\beta_2}_{\alpha_2}(T_a^{[{\cal R}]})^{\beta_3}_{\alpha_3}
    \right\},
\end{align}
where the operator $(D_M\Phi_{{\cal R}^3})_{\alpha_1\alpha_2\alpha_3}$
is invariant up to phase factors under the ${\zex}$ transformation.

\subsection{Lagrangian for bulk fermion fields}
Let us discuss bulk fermion fields. The notation of the fermion fields
in six dimensions is summarized in Appendix~\ref{Sec:fermion6d}. We
denote the 6D Weyl fermion with the positive and negative chiralities
by $\Psi^+$ and $\Psi^-$, respectively. Each of the 6D Weyl fermions
involves a vector-like pair of the 4D Weyl fermions, $\psi_L$ and
$\psi_R$.

Let $\Psi_{{\cal R}}^{\pm(k)}$ be a ${\zex}$ threefold 6D Weyl fermion
that belongs to the representation as in Eq.~\eqref{simple_threefold}.
Its component is denoted by $(\Psi_{{\cal R}}^{\pm(k)})_\alpha$ ,
where the subscript $\alpha$ runs from one to dim(${\cal R}$). The
${\zex}$ transformation is defined as
$(\Psi_{{\cal R}}^{\pm(k)})_\alpha\to \omega^{p^{\pm}} (\Psi_{{\cal
    R}}^{\pm (k+1)})_\alpha$, where $p^{\pm}\in \{0,\pm 1\}$. The
$\zex$-invariant kinetic term is given by
\begin{align}
  {\cal L}(\Psi_{{\cal R}}^{\pm(k)})&\equiv
  \sumk (\overline{\Psi_{{\cal R}}^{\pm (k)}})^\alpha i\Gamma^M(D_M^{(k)})_\alpha^\beta(\Psi_{{\cal R}}^{\pm(k)})_\beta , 
\end{align}
where $\Gamma^M$ is the 6D gamma matrix, given in
Appendix~\ref{Sec:fermion6d}. The covariant derivative
$(D_M^{(k)})_\alpha^\beta$ is the same form as in
Eq.~\eqref{cov_der_r1}.  Using 4D Weyl fermions
$\psi_{{\cal R},L}^{\pm(k)}$ and $\psi_{{\cal R},R}^{\pm(k)}$, we can
write
\begin{align}
  \Psi_{{\cal R}}^{+(k)}\equiv
  \begin{pmatrix}
      \psi_{{\cal R},L}^{+(k)}\\
      \psi_{{\cal R},R}^{+(k)}
  \end{pmatrix},\qquad 
  \Psi_{{\cal R}}^{-(k)}\equiv
  \begin{pmatrix}
      \psi_{{\cal R},R}^{-(k)}\\
      \psi_{{\cal R},L}^{-(k)}
  \end{pmatrix}.
\end{align}
From Eq.~\eqref{ferder61}, the Lagrangian can be rewritten by
\begin{align}
{\cal L}(\Psi_{{\cal R}}^{+(k)})&=  \sumk (\overline{\psi_{{\cal R},L}^{+(k)}},-\overline{\psi_{{\cal R},R}^{+(k)}})
        \begin{pmatrix}
           i \gamma^\mu D_\mu^{(k)} & -\overline{\tilde D}_y^{(k)}
            \\
            -\tilde D_y^{(k)}  &-i\gamma^\mu D_\mu^{(k)}
    \end{pmatrix}   
             \begin{pmatrix}
               \psi_{{\cal R},L}^{+(k)}\\\psi_{{\cal R},R}^{+(k)}
           \end{pmatrix}, \\
  {\cal L}(\Psi_{{\cal R}}^{-(k)})&=\sumk 
  (\overline{\psi_{{\cal R},R}^{-(k)}},-\overline{\psi_{{\cal R},L}^{-(k)}})
        \begin{pmatrix}
           i \gamma^\mu D_\mu^{(k)} & -\overline{\tilde D}_y^{(k)}
            \\
            -\tilde D_y^{(k)}  &-i\gamma^\mu D_\mu^{(k)}
    \end{pmatrix}   
             \begin{pmatrix}
               \psi_{{\cal R},R}^{-(k)}\\\psi_{{\cal R},L}^{-(k)}
           \end{pmatrix}, 
\end{align}
where we have defined
\begin{align}\label{deftildy1}
\tilde D_y^{(k)}=   {2\over 3}(D_{y^1}^{(k)}+ \omega D_{y^2}^{(k)}+\bar\omega D_{y^3}^{(k)}), 
  \qquad
\overline{\tilde D}_y^{(k)}=   {2\over 3}(D_{y^1}^{(k)}+ \bar \omega D_{y^2}^{(k)}+\omega D_{y^3}^{(k)}),
\end{align}
and the indices in the representation space of ${\cal R}$ are
suppressed.

For a general threefold fermion, denoted by
$\Psi_{{\cal R}_{123}}^{\pm(k)}$, whose representation is
${\cal R}_1\otimes{\cal R}_2\otimes{\cal R}_3$, we can write the
Lagrangian by using the covariant derivatives of the forms in
Eqs.~\eqref{gentfcov1}--\eqref{gentfcov3}. The irreducible
decomposition under $G^{\rm diag}$ is obtained as the scalar case,
discussed in the previous subsection.

Let us turn to deal with the onefold 6D Weyl fermions, which is
denoted by $\Psi_{{\cal R}^3}^{\pm}$. Let
$(\Psi_{{\cal R}^3}^{\pm})_{\alpha_1\alpha_2\alpha_3}$ be a component
of $\Psi_{{\cal R}^3}^{\pm}$, which is defined to transform into
$(\Psi_{{\cal R}^3}^{\pm})_{\alpha_3\alpha_1\alpha_2}$ under the
${\zex}$ transformation. The Lagrangian is written by
\begin{align}
  {\cal L}(\Psi_{{\cal R}^3}^{\pm})
  &=(\overline{\Psi_{{\cal R}^3}^{\pm}})^{\alpha_1\alpha_2\alpha_3}
    i\Gamma^M
    (D_M)^{\beta_1\beta_2\beta_3}_{\alpha_1\alpha_2\alpha_3}
    (\Psi_{{\cal R}^3}^{\pm})_{\beta_1\beta_2\beta_3},
\end{align}
where the covariant derivative is the same as in Eq.~\eqref{cov_onefold}.
Using 4D Weyl fermions $\psi_{{\cal R}^3,L}^{\pm}$ and $\psi_{{\cal R}^3,R}^{\pm}$, we can write
\begin{align}
  \Psi_{{\cal R}^3}^{+}\equiv
  \begin{pmatrix}
      \psi_{{\cal R}^3,L}^{+}\\
      \psi_{{\cal R}^3,R}^{+}
  \end{pmatrix},\qquad 
  \Psi_{{\cal R}^3}^{-}\equiv
  \begin{pmatrix}
      \psi_{{\cal R}^3,R}^{-}\\
      \psi_{{\cal R}^3,L}^{-}
  \end{pmatrix}.
\end{align}
Then the Lagrangian can be written as
\begin{align}
{\cal L}(\Psi_{{\cal R}^3}^{+})&=  \sumk (\overline{\psi_{{\cal R}^3,L}^{+}},-\overline{\psi_{{\cal R}^3,R}^{+}})
        \begin{pmatrix}
           i \gamma^\mu D_\mu & -\overline{\tilde D}_y
            \\
            -\tilde D_y  &-i\gamma^\mu D_\mu
    \end{pmatrix}   
             \begin{pmatrix}
               \psi_{{\cal R}^3,L}^{+}\\\psi_{{\cal R}^3,R}^{+}
           \end{pmatrix}, \\
  {\cal L}(\Psi_{{\cal R}^3}^{-})&=\sumk 
  (\overline{\psi_{{\cal R}^3,R}^{-}},-\overline{\psi_{{\cal R}^3,L}^{-}})
        \begin{pmatrix}
           i \gamma^\mu D_\mu & -\overline{\tilde D}_y
            \\
            -\tilde D_y  &-i\gamma^\mu D_\mu
    \end{pmatrix}   
             \begin{pmatrix}
               \psi_{{\cal R}^3,R}^{-}\\\psi_{{\cal R}^3,L}^{-}
           \end{pmatrix}, 
\end{align}
where $\tilde D_y  $ and $\overline{\tilde D}_y$ are defined as Eq.~\eqref{deftildy1}
with the covariant derivative in Eq.~\eqref{cov_onefold}, and 
the indices in the representation space are suppressed here.

In general, bulk gauge anomalies arise from 6D chiral fermions.  The
requirement of cancellations of the anomalies gives constraints on the
matter contents of
theories~\cite{Hebecker:2001jb,EDanom,edanom,Bhardwaj:2015xxa}. In our
setup, bulk anomaly cancellations can be ensured by introducing
vector-like sets of 6D Weyl fermions. There also appear 4D gauge
anomalies on the boundaries, \ie, the fixed points on
$T^2/\mathbb{Z}_3$. Such 4D anomalies depend on BCs for fermions and
will be discussed in the next subsection.

\subsection{Orbifold boundary conditions and 
  low-energy mass spectra}
\label{sec:matterbc}

We here discuss the BCs for matter fields.  First, let us see the
transformation laws of covariant derivatives under $\hat {\cal T}_1$
and $\hat {\cal S}_0$, which must be consistent with the BCs for gauge
fields. From Eqs.~\eqref{diagbc1} and~\eqref{diagbc2}, we find that
the covariant derivatives in Eqs.~\eqref{cov_der_r1}
and~\eqref{gentfcov1}--\eqref{gentfcov3} for threefolds and in
Eq.~\eqref{cov_onefold} for onefolds transform as
\begin{align}
  &  \hat {\cal T}_1[(D_{\{\mu,y^i\}}^{(k)})_\alpha^\beta]=(D_{\{\mu,y^i\}}^{(k)})_\alpha^\beta,
  &&
     \hat {\cal S}_0[(D_{\{\mu,y^i\}}^{(k)})_\alpha^\beta]=(D_{\{\mu,y^{i-1}\}}^{(k+1)})_\alpha^\beta,
  \\
    & 
      \hat {\cal T}_1[(D_{\{\mu,y^i\}}^{(k)})_{[\alpha_1\alpha_2\alpha_3]}^{[\beta_1\beta_2\beta_3]}]=
      (D_{\{\mu,y^i\}}^{(k)})_{[\alpha_1\alpha_2\alpha_3]}^{[\beta_1\beta_2\beta_3]},
    &&
       \hat {\cal S}_0[(D_{\{\mu,y^i\}}^{(k)})_{[\alpha_1\alpha_2\alpha_3]}^{[\beta_1\beta_2\beta_3]}]=
       (D_{\{\mu,y^{i-1}\}}^{(k+1)})_{[\alpha_1\alpha_2\alpha_3]}^{[\beta_1\beta_2\beta_3]},
  \\
  &  \hat {\cal T}_1[(D_{\{\mu,y^i\}})_{\alpha_1\alpha_2\alpha_3}^{\beta_1\beta_2\beta_3}]=
    (D_{\{\mu,y^i\}})_{\alpha_1\alpha_2\alpha_3}^{\beta_1\beta_2\beta_3},
  &&
     \hat {\cal S}_0[(D_{\{\mu,y^i\}})_{\alpha_1\alpha_2\alpha_3}^{\beta_1\beta_2\beta_3}]=
     (D_{\{\mu,y^{i-1}\}})_{\alpha_3\alpha_1\alpha_2}^{\beta_3\beta_1\beta_2}, 
\end{align}
where we have used the shorthand notation to show the boundary
conditions for the covariant derivative along $x^\mu$ and $y^i$ by the
subscript ${\{\mu,y^i\}}$.

The BCs for the matter fields are taken to be consistent with the
above transformations and written as
\begin{align}
&  \phi (x^\mu, \hat {\cal T}_1[y^i])=
    \omega^{p_t}   \phi(x^\mu, y^i), \qquad
  \phi (x^\mu, \hat {\cal S}_0[y^i])=
    \omega^{p_s}   \phi_{\cal S}(x^\mu, y^i), 
\end{align}
where a pair of fields $\phi$ and $\phi_{\cal S}$ represent the
scalars $\Phi$ and 6D Weyl fermions $\Psi^\pm$:
\begin{align}\notag
    (\phi,\phi_{\cal S})
  \in \bigg\{
  & \left((\Phi_{\cal R}^{(k)})_\alpha,(\Phi_{\cal R}^{(k+1)})_\alpha\right),
  &&
     \left((\Phi_{{\cal R}_{123}}^{(k)})_{[\alpha_1\alpha_2\alpha_3]},
        (\Phi_{{\cal R}_{123}}^{(k+1)})_{[\alpha_1\alpha_2\alpha_3]}\right),\\   \notag
&
   \Big((\Phi_{{\cal R}^3})_{\alpha_1\alpha_2\alpha_3},    
(\Phi_{{\cal R}^3})_{\alpha_1\alpha_2\alpha_3}\Big)  ,
  &&
    \left((\Psi_{\cal R}^{\pm(k)})_\alpha,-\tilde S_\Psi(\Psi_{\cal R}^{\pm(k+1)})_\alpha\right),\\
&   \left((\Psi_{{\cal R}_{123}}^{\pm(k)})_{[\alpha_1\alpha_2\alpha_3]},    
                  -\tilde S_\Psi(\Psi_{{\cal R}_{123}}^{\pm(k+1)})_{[\alpha_1\alpha_2\alpha_3]}\right),
  &&
    \left ((\Psi_{{\cal R}^3}^\pm)_{\alpha_1\alpha_2\alpha_3},    
     -\tilde S_\Psi(\Psi_{{\cal R}^3}^\pm)_{\alpha_3\alpha_1\alpha_2}\right)\bigg\}.
\label{bcphiphiS1}
\end{align}
The definition of $\tilde S_\Psi$ is shown in
Eq.~\eqref{tildeGam_def}, and $p_t,p_s\in \{0,\pm 1\}$ are chosen by
hand for each field. Since the 6D Weyl fermions compose of 4D Weyl
fermions, the last three pairs in Eq.~\eqref{bcphiphiS1} are
rearranged to the six pairs of the 4D Weyl fermions as
\begin{align}
  &  \left((\psi_{{\cal R},L}^{\pm(k)})_\alpha,
    \omega^{\pm 1}(\psi_{{\cal R},L}^{\pm(k+1)})_\alpha\right),
  &&  \left((\psi_{{\cal R},R}^{\pm(k)})_\alpha,
     \omega^{\mp 1}(\psi_{{\cal R},R}^{\pm(k+1)})_\alpha\right),\\
  & 
    \left((\psi_{{\cal R}_{123},L}^{\pm(k)})_{[\alpha_1\alpha_2\alpha_3]    },
    \omega^{\pm 1}(\psi_{{\cal R}_{123},L}^{\pm(k+1)})_{[\alpha_1\alpha_2\alpha_3]}\right),
  &&
     \left((\psi_{{\cal R}_{123},R}^{\pm(k)})_{[\alpha_1\alpha_2\alpha_3]},
     \omega^{\mp 1}(\psi_{{\cal R}_{123},R}^{\pm(k+1)})_{[\alpha_1\alpha_2\alpha_3]}\right),
  \\
  &  \left((\psi_{{\cal R}^3,L}^{\pm})_{\alpha_1\alpha_2\alpha_3},
    \omega^{\pm 1}(\psi_{{\cal R}^3,L}^{\pm})_{\alpha_3\alpha_1\alpha_2}\right),  
  &&  \left((\psi_{{\cal R}^3,R}^{\pm})_{\alpha_1\alpha_2\alpha_3},
     \omega^{\mp 1}(\psi_{{\cal R}^3,R}^{\pm})_{\alpha_3\alpha_1\alpha_2}\right).
\end{align}

Any BCs given above are formally written as
\begin{align}\label{tripletbc_gen_form1}
  \phi^{(k)}(x^\mu,\hat {\cal T}_1[y^i])=
  \omega^{p_t}
  \phi^{(k)}(x^\mu,y^i),\qquad 
  \phi^{(k)}(x^\mu,\hat {\cal S}_0[y^i])=
 \omega^{\tilde p_s} \phi^{(k +1)}(x^\mu,y^i),
\end{align}
where $\phi^{(k +3)}=\phi^{(k)}$ $(k \in {\mathbb Z})$ is a boson or a
4D Weyl fermion and is a component of an irreducible representation
under $G\times G\times G$. The integer $\tilde p_s$ is equal to $p_s$
for a boson and equal to $p_s\pm 1$ $(p_s\mp 1)$ for a left-handed
(right-handed) fermion with the 6D chirality $\pm$.  In most cases,
components in a set $\{\phi^{(1)},\phi^{(2)},\phi^{(3)}\}$ are not
identical and are mixed by the $\zex$ transformation; in this case we
call $\phi^{(k)}$ as a $\zex$ {\it triplet}. There is a special case,
where $\phi^{(k+1)}=\phi^{(k)}$ holds; in this case the field
$\phi^{(k)}$ is an eigenstate of the $\zex$ transformation and called
a $\zex$ {\it singlet}.  We list the sets of the form
$\{\phi^{(1)},\phi^{(2)},\phi^{(3)}\}$ as follows:
\begin{align}
&  \{(\Phi_{{\cal R}}^{(1)})_{\alpha},
      (\Phi_{{\cal R}}^{(2)})_{\alpha},
                (\Phi_{{\cal R}}^{(3)})_{\alpha}\},
  \\
&
  \{(\Phi_{{\cal R}_{123}}^{(1)})_{\alpha_1\alpha_2\alpha_3},
      (\Phi_{{\cal R}_{123}}^{(2)})_{\alpha_3\alpha_1\alpha_2},
  (\Phi_{{\cal R}_{123}}^{(3)})_{\alpha_2\alpha_3\alpha_1}\},
  \\
&
  \{(\Phi_{{\cal R}^3})_{\alpha_1\alpha_2\alpha_3},
      (\Phi_{{\cal R}^3})_{\alpha_3\alpha_1\alpha_2},
       (\Phi_{{\cal R}^3})_{\alpha_2\alpha_3\alpha_1}\},
  \\
&  \{(\Psi_{{\cal R},\{L,R\}}^{\pm (1)})_{\alpha},
      (\Psi_{{\cal R},\{L,R\}}^{\pm(2)})_{\alpha},
                (\Psi_{{\cal R},\{L,R\}}^{\pm(3)})_{\alpha}\},
  \\
&
 \{(\Psi_{{\cal R}_{123},\{L,R\}}^{\pm(1)})_{\alpha_1\alpha_2\alpha_3},
      (\Psi_{{\cal R}_{123},\{L,R\}}^{\pm(2)})_{\alpha_3\alpha_1\alpha_2},
       (\Psi_{{\cal R}_{123},\{L,R\}}^{\pm(3)})_{\alpha_2\alpha_3\alpha_1}\},
  \\
&
  \{(\Psi_{{\cal R}^3,\{L,R\}}^\pm)_{\alpha_1\alpha_2\alpha_3},
      (\Psi_{{\cal R}^3,\{L,R\}}^\pm)_{\alpha_3\alpha_1\alpha_2},
  (\Psi_{{\cal R}^3,\{L,R\}}^\pm)_{\alpha_2\alpha_3\alpha_1}\}.
\end{align}
Among them, only the onefold components with
$\alpha_1=\alpha_2=\alpha_3$ form $\zex$ singlets, and the others are
$\zex$ triplets.

Although $\zex$ singlets are eigenstates of the BCs, triplets are
not. From a $\zex$ triplet, we can define three eigenstates of the
BCs, denoted by $\phi^{[p]}$ $(p=0,\pm 1)$, as
\begin{align}
  \phi^{[p]}={1\over \sqrt{3}}\sum_{k=1}^3 \omega^{-kp}\phi^{(k)}, \qquad 
  \phi^{(k)}={1\over \sqrt{3}}\sump \omega^{k p}\phi^{[p]}.
\end{align}
Then, $\phi^{[p]}$ obeys the following BCs:
\begin{align}
  \phi^{[p]}(x^\mu,\hat {\cal T}_1[y^i])=
  \omega^{p_t}
  \phi^{[p]}(x^\mu,y^i),\qquad 
  \phi^{[p]}(x^\mu,\hat {\cal S}_0[y^i])=
  \omega^{p+\tilde p_s} \phi^{[p]}(x^\mu,y^i).
  \label{bcphip}
\end{align}
We note that $\phi^{[p]}$ is convenient to examine the KK expansions,
which are summarized in Appendix~\ref{Sec:kkexp1}, while the couplings
between matter fields and the Wilson line phases are simplified for
$\phi^{(k)}$.

From the eigenvalues of the BCs, we can find zero modes, which are
constant excitations over
the extra-dimensional space.  The zero mode can
appear as a light degree of freedom in a low-energy effective 4D
theory, where gauge symmetry $G\times G\times G$ is reduced to
$G^{\rm diag}$ as discussed in Sec.~\ref{sec:ressym}. In contrast, the
other modes have ${\cal O}(1/R)$ masses and become heavy. For $\zex$
triplets, $\phi^{[p]}$ with $p_t=p+\tilde p_s=0$ in Eq.~\eqref{bcphip}
has a zero mode. For $\zex$ singlets, fields with $p_t=\tilde p_s=0$
have zero modes.  Note that fields with $p_t=\pm 1$ do not have any
zero modes.

We discuss zero mode spectrum that arises from threefold fields in
detail.  Since threefold fields do not involve any $\zex$ singlet,
they are always organized into $\zex$ triplets.  For the case with a
threefold scalar $\Phi_{\cal R}^{(k)}$ with $p_t=0$, there appear zero
modes, contained in the triplet component $\phi^{[-p_s]}$. These zero
modes belong to the representation ${\cal R}$ under $G^{\rm
  diag}$. For the fermion case, a threefold $\Psi_{{\cal R}}^{\pm(k)}$
can be decomposed into
$\psi_{{\cal R}, L}^{\pm(k)}+\psi_{{\cal R}, R}^{\pm(k)}$. For the
case with a $\Psi_{{\cal R}}^{+(k)}$ $(\Psi_{{\cal R}}^{-(k)})$ having
$p_t=0$, a vector-like pair $\psi_{{\cal R},L}^{+(k)}$ and
$\psi_{{\cal R},R}^{+(k)}$ ($\psi_{{\cal R},R}^{-(k)}$ and
($\psi_{{\cal R},L}^{-(k)}$) has vector-like zero modes, which appear
from the triplet components $\phi^{[-p_s-1]}$ and $\phi^{[-p_s+1]}$,
respectively.  Thus, in these cases we always have vector-like fermion
zero modes, which belongs to ${\cal R}$ under $G^{\rm diag}$. Similar
discussions hold also for a more general threefold scalar
$\Phi_{{\cal R}_{123}}^{(k)}$ and fermion
$\Psi_{{\cal R}_{123}}^{\pm(k)}$.

Next, we discuss zero mode spectrum of onefold fields.  A onefold
involves both $\zex$ singlets and triplets except for the case with
the trivial representation ${\cal R}=\bm 1$. From $\zex$ singlets in a
onefold scalar $\Phi_{{\cal R}^3}$ with $p_t=0$, zero modes appear
only if $p_s=0$.  For $\zex$ triplets in $\Phi_{{\cal R}^3}$ with
$p_t=0$, zero modes appear from the component $\phi^{[-p_s]}$.  For
the fermion case, both $\psi_{{\cal R}^3, L}^{\pm}$ and
$\psi_{{\cal R}^3,R}^{\pm}$ in a onefold $\Psi_{{\cal R}^3}^\pm$ have
$\zex$ singlets. For the case with $p_t=0$, singlets in
$\psi_{{\cal R}^3,L}^{\pm}$ ($\psi_{{\cal R}^3,R}^{\pm}$) have zero
modes only if $p_s=-1$ ($p_s=1$). These zero modes of $\zex$ singlets
yield chiral fermion mass spectrum.  There also exist triplets in
$\Psi_{{\cal R}^3}^\pm$.  For $p_t=0$, zero modes appear from the
triplet components $\phi^{[-p_s\mp 1]}$ ($\phi^{[-p_s\pm 1]}$),
constructed from $\psi_{{\cal R}^3,L}^{\pm}$
($\psi_{{\cal R}^3,R}^{\pm}$). The zero modes of $\zex$ triplets
always compose vector-like pairs of 4D fermions. We note that any zero
modes belong to irreducible representations, which are contained in
the irreducible decomposition of
${\cal R}\otimes{\cal R}\otimes{\cal R}$ under $G^{{\rm diag}}$.

As an illustrative example, we consider $G=SU(N)$ and the
$N$--dimensional fundamental representation as ${\cal R}$.  In this
case, the irreducible decomposition of
${\cal R}\otimes{\cal R}\otimes{\cal R}$ is shown by the following
Young tableaux:
\begin{align}\label{youngf51}
  \ytableausetup{smalltableaux}
  \ydiagram{1}\quad \otimes\quad \ydiagram{1}
  \quad \otimes\quad  \ydiagram{1}\quad =\quad 
  \ydiagram{3}\quad \oplus\quad 
  2\times \ydiagram{2,1}\quad \oplus\quad 
  \ydiagram{1,1,1}\,.
\end{align}
One sees that $\zex$ singlets in $\Phi_{{\cal R}^3}$ and
$\Psi_{{\cal R}^3}^\pm$ always belong to the first representation on
the right-hand side of Eq.~\eqref{youngf51}.  These singlets carry $N$
out of $N^3$ degrees of freedom, and the rest $N^3-N$ degrees of
freedom form $(N^3-N)/3=N(N-1)(N+1)/3$ triplets.  For $p_t=p_s=0$,
since the $\zex$ singlets have zero modes, there appear $N+(N^3-N)/3$
degrees of freedom appear as zero modes, whose representations
correspond to the first and third terms on the right-hand side of
Eq.~\eqref{youngf51}.  On the other hand, for $p_s=\pm 1$ case with
$p_t=0$, there appear $(N^3-N)/3$ degrees of freedom as zero modes,
which transform as the representation corresponds to the second terms
on the right-hand side of Eq.~\eqref{youngf51}.  Consistently to the
above, one sees the relation
\begin{align}\notag
\sharp  \left(  \ydiagram{3}  \oplus
  \ydiagram{1,1,1}\right)
  -\sharp\left(\ydiagram{2,1}\right)
&  =\left({N\over 6}(N+1)(N+2)
  +  {N\over 6}(N-1)(N-2)\right)
  -{N\over 3}(N^2-1)\\
 & 
   =N,
   \label{Ndof1}
\end{align}
where $\sharp(*)$ is the degrees of freedom of $*$. We see that $N$ in
Eq.~\eqref{Ndof1} corresponds to the degrees of freedom of $\zex$
singlet components.

Let us examine the fermion zero modes in the $SU(N)$ case. For the
case with onefold fermions, zero mode spectrum can become chiral.  For
example, we consider the case with a onefold $\Psi_{{\cal R}^3}^+$ of
$p_t=0$.  In this case, the representations of the zero modes depend
on $p_s$, which are summarized as follows:
\begin{align}
  &  p_s=0 &&:&& (\psi_{{\cal R}^3,L}^{+})_\textrm{zero mode} \sim \ydiagram{2,1}\,,
  &&     (\psi_{{\cal R}^3,R}^{+})_\textrm{zero mode} \sim \ydiagram{2,1}\,, \\\label{chiral1}
  &  p_s=1 &&:&& (\psi_{{\cal R}^3,L}^{+})_\textrm{zero mode} \sim \ydiagram{2,1}\,,
  &&     (\psi_{{\cal R}^3,R}^{+})_\textrm{zero mode} \sim \ydiagram{3}\oplus\ydiagram{1,1,1}\,, \\\label{chiral2}
  &  p_s=-1 &&:&& (\psi_{{\cal R}^3,L}^{+})_\textrm{zero mode} \sim \ydiagram{3}\oplus\ydiagram{1,1,1}\,,
  &&     (\psi_{{\cal R}^3,R}^{+})_\textrm{zero mode} \sim \ydiagram{2,1}\,.
\end{align}
Thus, low-energy spectrum of 4D fermions is chiral for $p_s=\pm 1$, but
vector-like for $p_s=0$. A similar discussion holds for the case with
$\Psi_{{\cal R}^3}^{-}$.

Finally, we give comments on 4D gauge anomalies.  If there are fermion
zero modes, they generally contribute to the anomalies.  For threefold
fermions, their zero modes are always vector-like and do not give 4D
anomalies.  On the other hand, onefold fermions can have chiral zero
modes and thus generally generate 4D anomalies. Thus, a requirement of
the cancellation of 4D anomalies constrains the onefold fermion
contents.  In addition to the zero mode anomalies, localized anomalies
induced at the fixed points $y^i_{{\rm f}(r)}$ $(r=0,1,2)$, defined in
Eq.~\eqref{yfp1}, should also be concerned~\cite{EDanom}. The
localized contributions arise even if the fermion has $p_t=\pm 1$, in
which case there is no zero modes. In our setup, contribution to the
localized anomalies at $y^i_{{\rm f}(r)}$ can arise from a fermion
$\psi(x^\mu,y^i)$ that satisfy the BCs
$\psi(x^\mu,\hat {\cal S}_r[y^i])=\psi(x^\mu,y^i)$. One can see that
the contributions to the localized anomalies at each fixed point from
threefold fermions always cancel out since the contributions are
always vector-like. For the onefold fermion, localized anomalies
generally exist; it gives constraints on the matter content of the
theory. When the localized anomalies vanish, also the 4D anomalies
do. Conversely, the 4D anomaly cancellation does not ensure vanishing
localized anomalies.

%
\section{One-loop effective potentials for Wilson line phases in $SU(5)$ models}
\label{Sec:onelooppot}
%
In this section, we study one-loop effective potentials for the
classical background VEVs $\tilde a_{k+\ell}^a$ in
Eq.~\eqref{tiladef2}, which are related to the Wilson line phase
degrees of freedom.  As a concrete example, we focus on the case with
$G=SU(5)$. The discussion can be generalized to other gauge group
cases.

\subsection{Contributions from ${\zex}$ threefold fields}
First, we derive one-loop contributions from a ${\zex}$ threefold
scalar field to the effective potential.  The simplest example is the
threefold $\Phi_{\bm 5}^{(k)}$, which transforms under
$SU(5)\times SU(5)\times SU(5)$ as
\begin{align}
  \Phi^{(1)}_{\bm 5}\sim ({\bf 5},{\bf 1},{\bf 1}), \qquad 
  \Phi^{(2)}_{\bm 5}\sim ({\bf 1},{\bf 5},{\bf 1}), \qquad 
  \Phi^{(3)}_{\bm 5}\sim ({\bf 1},{\bf 1},{\bf 5}), 
\end{align}
where ${\bf 5}$ and ${\bf 1}$ are the fundamental and the trivial
representations of $SU(5)$, respectively. Based on the discussion in
the previous section, we define  BCs for their components,
$(\Phi^{(k)}_{\bm 5})_\alpha$ $(\alpha=1,\dots,5)$, as
\begin{align}\label{su5fndbc}
(\Phi^{(k)}_{\bm 5})_\alpha(x^\mu,\hat {\cal T}_1[y^i])=\omega^{p_t}
(\Phi^{(k)}_{\bm 5})_\alpha(x^\mu,y^i), \qquad 
(\Phi^{(k)}_{\bm 5})_\alpha(x^\mu,\hat {\cal S}_0[y^i])=\omega^{p_s}
(\Phi^{(k+1)}_{\bm 5})_\alpha(x^\mu,y^i).
\end{align}
In the following, the fundamental representation of the $SU(5)$
generators is denoted by
$(T_a^{[{\bm 5}]})_\alpha^\beta \equiv (T_a)_\alpha^\beta$.  From
Eq.~\eqref{cov_der_r1}, Lagrangian for $\Phi^{(k )}_{\bm 5}$ has the
${\zex}$-invariant form:
\begin{align}
  \label{lag_fnd}
  {\cal L}(\Phi^{(k)}_{\bm 5})
  &=
 \sum_{k=1}^3  |(D_M^{(k)})_{\alpha}^{\beta}(\Phi^{(k)}_{\bm 5})_\beta|^2,
    \qquad 
    (D_M^{(k)})_{\alpha}^{\beta}
    =\delta_\alpha^{\beta}\der_M+igA_M^{(k)a}(T_a)_\alpha^\beta.
\end{align}

To obtain the effective potential for $\tilde a_{k+\ell}^a$ in
Eq.~\eqref{tiladef2}, let us expand the Lagrangian in
Eq.~\eqref{lag_fnd} around the classical background VEVs and extract
quadratic terms of the quantum fluctuations.  As discussed in
Sec.~\ref{sec:wilsonline}, we always take a basis where the Wilson
line phases are diagonal and have the form like
Eq.~\eqref{diagonalwilsonline}.  One-loop corrections to the effective
potential for the phases can be derived through path integral over the
fluctuation $(\Phi^{(k)}_{\bm 5})_\alpha$. The quadratic terms are
written as follows:
\begin{gather}\label{quadlagsu51}
    {\cal L}^{(2)}(\Phi^{(k)}_{\bm 5}) \equiv -\sum_{k =1}^3
    (\Phi^{(k)\dag}_{\bm 5})^\alpha \left(
        \delta_{\alpha}^{\beta}\square -g^{ij}\langle{D_{y^i}^{(k
            )}}\rangle_{\alpha}^{\alpha'} \langle{D_{y^j}^{(k
            )}}\rangle_{\alpha'}^{\beta} \right)
    (\Phi^{(k)}_{\bm 5})_\beta, \\
    \langle{D_{y^i}^{(k )}}\rangle_{\alpha}^\beta \equiv
    \delta_{\alpha}^\beta\der_{y^i}+i{2\over R}\tilde a_{i+k }^{\hat
      a} (H_{\hat a})_{\alpha}^{\beta}
\label{su5fnbgcov1}                     ,
\end{gather}
where we have defined $\langle{D_{y^i}^{(k )}}\rangle_{\alpha}^\beta$
as a background covariant derivative and $\square=\der_\mu
\der^\mu$. The matrices $H_{\hat a}$ $({\hat a}=1,\dots,4)$ are the
fundamental representation of the Cartan generators of $SU(5)$, which
we can take as
\begin{align}
\begin{split}
&  H_{1}={1\over 2}{\rm diag}(1,0,0,0,-1), \qquad 
  H_{2}={1\over 2}{\rm diag}(0,1,0,0,-1), \\
&  H_{3}={1\over 2}{\rm diag}(0,0,1,0,-1), \qquad 
  H_{4}={1\over 2}{\rm diag}(0,0,0,1,-1).
\end{split}
\end{align}
Thus, the Wilson line phases in Eq.~\eqref{su5fnbgcov1} are written as 
\begin{align}
  {2}\tilde a_{i+k}^{\hat a}
  H_{\hat a}
  ={\rm diag}(\tilde a_{i+k}^1,\tilde a_{i+k}^2,\tilde a_{i+k}^3,\tilde a_{i+k}^4,\tilde a_{i+k}^5), \quad {\rm where}\quad
  \tilde a_{i+k}^5=-
  \sum_{\hat a=1}^4  \tilde a_{i+k}^{\hat a}.
  \label{tracecondsu5}
\end{align}

We now readily rewrite the quadratic Lagrangian in
Eq.~\eqref{quadlagsu51} as
\begin{align}
{\cal L}^{(2)}(\Phi^{(k)}_{\bm 5}) 
  &=-\sum_{\alpha=1}^5\sum_{k =1}^3
    (\Phi^{(k)\dag}_{\bm 5})^\alpha                       
    \left(\square +\hat M^2_{k,\alpha}
\right)
    (\Phi^{(k)}_{\bm 5})_\alpha,
\end{align}
where we have introduced the differential operator
$\hat M^2_{k,\alpha} $ as
\begin{align}\label{diff_Msq1}
\hat M^2_{k,\alpha}  
  &\equiv g^{ij}\left(-i\der_{y^i}+{\tilde a^{\alpha}_{i+k}\over R}\right)
    \left(-i\der_{y^j}+{\tilde a^{\alpha}_{j+k}\over R}\right)
  ={2\over 3}\sum_{\ell=1}^3\left(-i\der_{y^\ell}+{\tilde a^{\alpha}_{k+\ell}\over R}\right)^2. 
\end{align}
We note that the above corresponds to the operator
in Eq.~\eqref{effmassop1}. 

Based on the discussion in Sec.~\ref{sec:scalar_lag} and the BCs in
Eq.~\eqref{su5fndbc}, we see that the components
$\{(\Phi^{(1)}_{\bm 5})_\alpha,(\Phi^{(2)}_{\bm
  5})_\alpha,(\Phi^{(3)}_{\bm 5})_\alpha \}$ form a $\zex$ triplet.
In Appendix~\ref{Sec:kkexp1}, we first show the KK expansion of $\zex$
singlets in Eq.~\eqref{phi_kk_1}, and using it, we derive the
expansion of triplets in Eq.~\eqref{phik_kk_tilphik1}.  Here we
briefly provide the overview of the derivation.  From the triplet
$\phi^{(k)}$ that obeys the BCs in Eq.~\eqref{tripletbc_gen_form1}, we
can define $\phi^{[p]}$ that are eigenstates of the BCs as in
Eq.~\eqref{bcphip}. The KK expansion of $\phi^{[p]}$ yields the
corresponding KK modes $\tilde \phi^{[p]}_{N_1,N_2}$ in
Eq.~\eqref{tripletKKexp1}, where $N_i=n_i+p_t/3$ and
$n_i\in {\mathbb Z}$. From $\tilde \phi^{[p]}_{N_1,N_2}$, we can
define $\tilde \phi^{(k)}_{N_1,N_2}$ appearing in
Eq.~\eqref{phik_kk_tilphik1}. Their KK masses are given by replacing
the operator $-i\der_{y^\ell}$ in Eq.~\eqref{diff_Msq1} by $N_\ell/R$
$(N_3=-N_1-N_2)$. In the present case, the KK masses for
$(\Phi^{(k)}_{\bm 5})_\alpha$ are given by
\begin{align}\label{kkmass_fnd}
  M^2_{k,\alpha}
  ={2\over 3R^2}\sum_{\ell=1}^3\left(N_{\ell}+{\tilde a^{\alpha}_{\ell+k}}\right)^2,  \quad {\rm where}\quad N_i=n_i+p_t/3, \qquad N_3=-N_1-N_2.
\end{align}
For details, please refer to Appendix~\ref{Sec:kkexp1}.

With the above result, the 4D effective Lagrangian in
Eq.~\eqref{quadlagsu51} is rewritten by KK modes of the triplet, and
we can integrate them to obtain the effective potential. The
derivation of the potential is shown in Appendix~\ref{Sec:veffderive}.
Using the result shown in Eq.~\eqref{vpttilai}, we find that the
effective potential contribution from a real degree of freedom in
$(\Phi^{(k)}_{\bm 5})_\alpha$ is given by
\begin{align}\label{calvdef1}
  {\cal V}^{(p_t)}(\tilde a_i^{\alpha})
  &= -{\sqrt{3}\over 32\pi^7R^4}
    \sum_{w^1,w^2\in{\mathbb Z}'}
    {\cos\left(2\pi [w^1(p_t/3+\tilde a_1^{\alpha})+w^2(p_t/3+\tilde a_2^{\alpha})]\right)\over [(w^1)^2-w^1w^2+(w^2)^2]^3}, 
\end{align}
where we have used $\tilde a_i^{\alpha}$ ($i=1,2$) as the parameter of
the potential since they are taken to be the independent variables
among $\tilde a_\ell^{\alpha}$ $(\ell=1,2,3)$.  As discussed in
Appendix~\ref{Sec:veffderive}, the summation with respect to $w^1$ and
$w^2$ is taken over for all integers except for $(w^1,w^2)=(0,0)$,
which is denoted by $w^1,w^2\in {\mathbb Z}'$.  We note that the
potential in Eq.~\eqref{calvdef1} can also be naturally expressed by
the vector notation as
\begin{align}
  {\cal V}^{(p_t)}(\tilde a_i^{\alpha})
  =
-{\sqrt{3}\over 32\pi^7R^4}
    \sum_{\bm w\in\Lambda_w'}
  {\cos({2\pi \bm w\cdot \tilde{\bm{a}}^\alpha{}'})\over (|\bm w|^2)^3},
\end{align}
where we have introduced the vector $\bm w$ and the lattice
$\Lambda_w'$ as 
\begin{align}
  \Lambda_w'=\left\{
\bm w=  w^1{\bm e}_1
  +w^2 {\bm e}_2~\big|~w^1,w^2\in{\mathbb Z}'
  \right\}, 
\end{align}
and the dual vector
$\tilde{\bm{a}}^\alpha{}'=(p_t/3+\tilde a_1^{\alpha})\tilde{\bm e}^1
+(p_t/3+\tilde a_2^{\alpha})\tilde {\bm e}^2$, similar to those in
Appendix~\ref{Sec:veffderive}.

Let $\Delta V^{(p_t)}(\Phi_{\bm 5}^{(k)})$ be the
contribution to the effective potential from
$\Phi^{(k)}_{\bm 5}$ with $p_t$ defined in Eq.~\eqref{su5fndbc}. Then,
we obtain
\begin{align}\label{veff_5scalar}
\Delta V^{(p_t)}(\Phi_{\bm  5}^{(k)})&=2 \sum_{\alpha=1}^5{\cal V}^{(p_t)}(\tilde a_i^{\alpha}),
\end{align}
where the overall factor $2$ on the right-hand side arises due to the
real degrees of freedom of a complex scalar. The potential in
Eq.~\eqref{calvdef1} is manifestly invariant under integer shifts of
an arbitrarily chosen component of the Wilson line phases,
$\tilde a_i^{\alpha}\to \tilde a_i^{\alpha}\pm 1$, which preserve the
Wilson line phase factors $W_\ell$ in Eq.~\eqref{wilson_wl}.  Given
the above invariance, we relax the traceless condition imposed in
Eq.~\eqref{tracecondsu5} as
$\sum_{\hat a=1}^5 \tilde a_{i+k}^{\hat a}=0$~(mod 1) in the following
discussions.

We can generalize the above result to triplets belonging to other
representations of $SU(5)$.  Contributions to the effective potential
depend on components of weight vectors with respect to the Cartan
generators $H_{\hat a}$; for a given representation ${\cal R}$, the
representation matrix of $H_{\hat a}$ is denoted by
$H_{\hat a}^{[{\cal R}]}$. We can express eigenvalues of
$2\tilde a_{i+k}^{\hat a}H_{\hat a}^{[{\cal R}]}$ by using
$\tilde a_{i+k}^{\hat a}$. Here, let us consider a threefold scalar
$\Phi_{{\cal R}}^{(k)}$, which transforms under
$SU(5)\times SU(5)\times SU(5)$ as
\begin{align}\label{rep_gen_threefold}
  \Phi^{(1)}_{{\cal R}}\sim ({\cal R},{\bf 1},{\bf 1}), \qquad 
  \Phi^{(2)}_{{\cal R}}\sim ({\bf 1},{\cal R},{\bf 1}), \qquad 
  \Phi^{(3)}_{{\cal R}}\sim ({\bf 1},{\bf 1},{\cal R}).
\end{align}
We write the contributions to the effective potential generated by
$\Phi_{{\cal R}}^{(k)}$ as $\Delta V^{(p_t)}(\Phi_{{\cal
    R}}^{(k)})$. We find that the contributions from, \eg,
${\cal R}=\bm{10},\bm{15},\bm{24}$ cases are given by
\begin{align}
  \Delta V^{(p_t)}(\Phi_{\bm{10}}^{(k)})
  &=2 \sum_{1\leq \alpha<\beta\leq 5}{\cal V}^{(p_t)}(
  \tilde a_i^\alpha+\tilde a_i^\beta), \\
  \Delta V^{(p_t)}(\Phi_{\bm{15}}^{(k)})
  &=2 \sum_{1\leq \alpha\leq \beta\leq 5}{\cal V}^{(p_t)}(
  \tilde a_i^\alpha+\tilde a_i^\beta), \\
  \Delta V^{(p_t)}(\Phi_{\bm{24}}^{(k)})
  &=2 \sum_{1\leq \alpha\neq \beta\leq 5}{\cal V}^{(p_t)}(
  \tilde a_i^\alpha-\tilde a_i^\beta), 
\end{align}
respectively. Here, we have discarded irrelevant constants that are
independent of $\tilde a_i^\alpha$.

For general threefold scalars in Eq.~\eqref{gen_threefold}, we can
derive a differential operator as in Eq.~\eqref{diff_Msq1}.  As an
example, let us consider an
$({\cal R}_1,{\cal R}_2,{\cal R}_3)=({\bm 5},{\bm 5},{\bm 1})$
case. In this case, a component of $\Phi_{{\cal R}_{123}}^{(k)}$ has
two indices, which we denote by $\alpha_1$ and $\alpha_2$
$(\alpha_1,\alpha_2=1,\dots,5)$. Corresponding to
Eq.~\eqref{diff_Msq1}, we find the following differential operator:
\begin{align}
 \hat M_{k,\alpha_1,\alpha_2}^2=
  {2\over 3}\sum_{\ell=1}^3\left(-i\der_{y^\ell}+{\tilde a^{\alpha_1}_{k+\ell}+\tilde a^{\alpha_2}_{k+\ell+1}\over R}\right)^2.
\end{align}
From the above, we obtain a one-loop correction to the potential from
$\Phi_{{\cal R}_{123}}^{(k)}$ in a similar way to the previous cases.
The result is given by
\begin{align}
  \Delta V^{(p_t)}(\Phi_{{\cal R}_{123}}^{(k)})
  &=2 \sum_{\alpha_1,\alpha_2=1}^5{\cal V}^{(p_t)}(
 \tilde a_i^{\alpha_1}+\tilde a_{i+1}^{\alpha_2}). 
\end{align}
We note that, except for the subscripts of the phases, the potential
contribution coincides with the sum of those coming from
$\Phi_{\bm{10}}^{(k)}$ and $\Phi_{\bm{15}}^{(k)}$, which is an
explicit example of the modification explained below
Eq.~\eqref{schemrel1}.

We turn to discuss the contributions to the effective potential from
threefold fermions. The contributions mostly depend on the eigenvalues
of differential operators as in Eq.~\eqref{diff_Msq1}. Since the
covariant derivatives for bosons and fermions are the same if they
belong to the same representation of $SU(5)$, the eigenvalues are also
common for bosons and fermions. Thus, the contributions from threefold
fermions can be written by using the contributions from threefold
boson. We denote a 6D Weyl fermion $\Psi_{\cal R}^{\pm(k)}$, whose
representation is the same as in Eq.~\eqref{rep_gen_threefold}. A
contribution to the effective potential from $\Psi_{\cal R}^{\pm(k)}$
is denoted by $\Delta V^{(p_t)}(\Psi_{{\cal R}}^{(k)})$. Then, we find
$ \Delta V^{(p_t)}(\Psi_{{\cal R}}^{(k)})=-2 \Delta
V^{(p_t)}(\Phi_{{\cal R}}^{(k)})$.

\subsection{Contributions from ${\zex}$ onefold  fields }
\label{subsec:onefoldeffp1}
We start to discuss the contributions from ${\zex}$ onefolds. We first
examine a bulk matter scalar $\Phi_{\bm{5}^3}$, whose component is
written by $(\Phi_{\bm{5}^3})_{\alpha_1\alpha_2\alpha_3}$. Here, Greek
indices runs 1 to 5. The BCs can be introduced as
\begin{gather}
    (\Phi_{\bm{5}^3})_{\alpha_1\alpha_2\alpha_3}(x^\mu,\hat {\cal T}_1[y^i])=\omega^{p_t}
    (\Phi_{\bm{5}^3})_{\alpha_1\alpha_2\alpha_3}(x^\mu,y^i),\\
    (\Phi_{\bm{5}^3})_{\alpha_1\alpha_2\alpha_3}(x^\mu,\hat {\cal S}_0[y^i])=\omega^{p_s}
    (\Phi_{\bm{5}^3})_{\alpha_3\alpha_1\alpha_2}(x^\mu,y^i).
\end{gather}
The extra-dimensional component of the covariant derivative
acting on $(\Phi_{\bm{5}^3})_{\alpha_1\alpha_2\alpha_3}$
is written by 
\begin{align}
  \langle{D_{y^i}}\rangle_{\alpha_1\alpha_2\alpha_3}^{\beta_1\beta_2\beta_3} 
  &=
\delta_{\alpha_1}^{\beta_1}
    \delta_{\alpha_2}^{\beta_2}\delta_{\alpha_3}^{\beta_3}
    \left[ \der_{y^i}
    +i{2\over R}\left(\tilde a_{i+1}^{\alpha_1}
    +\tilde a_{i+2}^{\alpha_2}+\tilde  a_{i+3}^{\alpha_3}
    \right)\right],
    \label{onefoldcov_bk}
\end{align}
where the indices $\alpha_k$ $(k=1,2,3)$ are not summed on the
right-hand side in the above.

As discussed in Sec.~\ref{sec:matterbc},
$(\Phi_{\bm{5}^3})_{\alpha_1\alpha_2\alpha_3}$ contains $\zex$
triplets and singlets. The latter corresponds to the components of
$\alpha_1=\alpha_2=\alpha_3$. From Eq.~\eqref{onefoldcov_bk}, it is
clear that the singlet does not couple to the Wilson line phases since
$\tilde a_{i+1}^{\alpha} +\tilde a_{i+2}^{\alpha}+\tilde
a_{i+3}^{\alpha}=0$ holds. Thus, only $\zex$ triplets can give contribution
to the effective potential.

For a set of fixed values of $\{\alpha_1,\alpha_2,\alpha_3\}$, a
triplet is given by
\begin{align}
  \{(\Phi_{{\cal R}^3})_{\alpha_1\alpha_2\alpha_3},
  (\Phi_{{\cal R}^3})_{\alpha_3\alpha_1\alpha_2},
  (\Phi_{{\cal R}^3})_{\alpha_2\alpha_3\alpha_1}\}\equiv \{\phi^{(1)},\phi^{(2)},\phi^{(3)}\}.
\end{align}
One can see that $\phi^{(k)}$ couples to the Wilson line phases of
$\tilde a^{\alpha_1}_{i+k} +\tilde a^{\alpha_2}_{i+1+k}+ \tilde
a^{\alpha_3}_{i+2+k}$ via the covariant derivative in
Eq.~\eqref{onefoldcov_bk}. Then, as in Eq.~\eqref{kkmass_fnd}, a KK
mode of the triplet $\phi^{(k)}$ has the following KK mass:
\begin{align}
  M^2_{k,\alpha_1,\alpha_2,\alpha_3}
  ={2\over 3R^2}\sum_{\ell=1}^3\left(N_{\ell}+
{\tilde a^{\alpha_1}_{k+\ell} +\tilde
  a^{\alpha_2}_{k+\ell+1} +\tilde a^{\alpha_3}_{k+\ell+2} }\right)^2. 
\end{align}
This implies that a contribution to the effective potential from the
triplet $\phi^{(k)}$ is proportional to
$ {\cal V}^{(p_t)}({\tilde a^{\alpha_1}_{i} +\tilde a^{\alpha_2}_{i+1}
  +\tilde a^{\alpha_3}_{i+2} } ) $.

Let $\Delta V^{(p_t)}(\Phi_{{\bm 5}^3})$ be the contribution from the
onefold $\Phi_{\bm{5}^3}$. Among $5^3=125$ components of
$(\Phi_{\bm{5}^3})_{\alpha_1\alpha_2\alpha_3}$, five components are
singlets, which give constants independent of the Wilson line phases.
The remaining 120 components compose 40 triplets. We can take
summations of the contributions from the triplets as
\begin{align}
\Delta V^{(p_t)}(\Phi_{{\bm 5}^3})={2\over 3}
  \left(\sum_{\alpha_1,\alpha_2,\alpha_3=1}^5
  -\sum_{\alpha_1=\alpha_2=\alpha_3=1}^5
  \right)
  {\cal V}^{(p_t)}(
{\tilde a^{\alpha_1}_{i}
  +\tilde a^{\alpha_2}_{i+1}
  +\tilde a^{\alpha_3}_{i+2}
  }),
  \label{delVthreef1}
\end{align}
where on the right-hand side the overall factor appears since
$\Phi_{{\bm 5}^3}$ is a complex scalar, and the factor $1/3$ should be
included to correctly count 40 triplets composed of 120 components.
Let us note that the subtracted $\alpha_1=\alpha_2=\alpha_3$
contributions in Eq.~\eqref{delVthreef1} are constant, which do not
affect the vacuum structure of the potential.

Generalizations to the other representation than ${\bm 5}$ are
straightforward. For example, we find that the contribution from
$\Phi_{\bm{10}^3}$ is given by
\begin{align}\notag
\Delta V^{(p_t)}(\Phi_{\bm{10}^3})&={2\over 3}
  \left(
\left[  \prod_{i=1}^3  \sum_{1\leq \alpha_i<\beta_i\leq 5}
\right]
  -\sum_{(\alpha_1,\beta_1)=(\alpha_2,\beta_2)=(\alpha_3,\beta_3)}
  \right)\\
&\qquad \times   {\cal V}^{(p_t)}(
{\tilde a^{\alpha_1}_{i}+\tilde a^{\beta_1}_{i}
  +\tilde a^{\alpha_2}_{i+1}  +\tilde a^{\beta_2}_{i+1}
  +\tilde a^{\alpha_3}_{i+2}  +\tilde a^{\beta_3}_{i+2}
}), 
\end{align}
where the above potential consists of the contributions from
$(10^3-10)/3=330$ triplets.  As in the case of the threefold scalar,
difference between contributions from the onefold scalars and fermions
is just an overall factor. Let $\Delta V^{(p_t)}(\Psi_{{\cal R}^3})$
be the contribution to the potential from a fermion
$\Psi_{{\cal R}^3}^{\pm }$. Then, it follows that
$\Delta V^{(p_t)}(\Psi_{{\cal R}^3})= -2\Delta V^{(p_t)}(\Phi_{{\cal
    R}^3})$, where the contribution does not depend on 6D chiralities
of fermions.

%
\section{Gauge symmetry breaking patterns in $SU(5)$ models}
\label{Sec:vac}
%

\subsection{Vacuum structure and unbroken gauge symmetries}

We study the vacuum structure of the effective potential for the
Wilson line phases derived in the previous section. For simplicity, we
only consider the contributions to the potential from the gauge fields
and threefold fields of ${\Phi}_{\cal R}^{(k)}$ and
${\Psi}_{\cal R}^{\pm(k)}$ for
${\cal R}=\bm{5},\bm{10},\bm{15},\bm{24}$.  First, for a chosen
${\cal R}$, we numerically find VEVs of the Wilson line phases at a
global minimum of a contribution
$\Delta V^{(p_t)}({\Phi}_{\cal R}^{(k)})$ or
$\Delta V^{(p_t)}({\Psi}_{\cal R}^{(k)})$.  In the fundamental
representation of $SU(5)$, the Wilson line phase factors $W_\ell$ in
Eq.~\eqref{wilson_wl} take the following form:
\begin{align}
  \begin{split}
      W_\ell&=
      {\rm diag}(e^{2\pi i\tilde a_{\ell+1}^1},e^{2\pi i\tilde a_{\ell+1}^2},e^{2\pi i\tilde a_{\ell+1}^3},e^{2\pi i\tilde a_{\ell+1}^4},e^{2\pi i\tilde a_{\ell+1}^5})\\
      &\qquad \otimes
      {\rm diag}(e^{2\pi i\tilde a_{\ell+2}^1},e^{2\pi i\tilde a_{\ell+2}^2},e^{2\pi i\tilde a_{\ell+2}^3},e^{2\pi i\tilde a_{\ell+2}^4},e^{2\pi i\tilde a_{\ell+2}^5})\\
      & \qquad\qquad\otimes {\rm diag}(e^{2\pi i\tilde
        a_{\ell}^1},e^{2\pi i\tilde a_{\ell}^2},e^{2\pi i\tilde
        a_{\ell}^3},e^{2\pi i\tilde a_{\ell}^4},e^{2\pi i\tilde
        a_{\ell}^5}).
  \end{split}
           \label{Wellsu51}
\end{align}
Thus, if a VEV at a minimum is determined, we can find a gauge
symmetry breaking pattern through Eqs.~\eqref{def_genH}
and~\eqref{Wellsu51}.

Before starting to show results, let us mention that there are
degenerate vacua in potentials for the Wilson line phases.  The
degeneracy is related to the invariance of the potential under some
transformations of the Wilson line phases.  As we mentioned below
Eq.~\eqref{veff_5scalar}, ${\cal V}^{(p_t)}(\tilde a_i^{\alpha})$ in
Eq.~\eqref{calvdef1} is invariant under an integer shift
$\tilde a_i^{\alpha}\to \tilde a_i^{\alpha}\pm 1$.  Thus, effective
potentials for the Wilson line phases generally have degeneracy
related to the integer shift invariance.  This is due to the phase
property of $\tilde a_i^{\alpha}$.  In addition, from
Eq.~\eqref{calvdef1}, we see that a simultaneous change of the overall
sign of the VEVs as $\tilde a_i^{\alpha}\to -\tilde a_i^{\alpha}$ for
$i=1,2$ and $\alpha=1$--$5$ does not change the potentials for $p_t=0$
cases.  This leads to a degeneracy in the potentials.  On the other
hand, the contributions to the potentials from fields with $p_t=1$ and
$-1$ are related to each other by the overall sign change of the
phases, \ie,
${\cal V}^{(-1)}(\tilde a_i^{\alpha})={\cal V}^{(1)}(-\tilde
a_i^{\alpha})$, which is shown from Eq.~\eqref{calvdef1}. The
potentials are invariant under the permutation of the index $\alpha$,
which can be regarded as a basis change in the representation
space. The exchange of $\tilde a_1^{\alpha}$ and $\tilde a_2^{\alpha}$
also does not change the potentials.  Finally, the potential
contributions from adjoint matter fields are invariant under the
${\mathbb Z}_5$ transformation, which is the center subgroup of
$SU(5)$, with $\tilde a_i^{\alpha}+n_i/5$ ($i=1,2$), where
$n_i\in {\mathbb Z}$.

Concerning the above degeneracy, in the following, we show
representatives of VEVs at a degenerate global minimum.  In
Table~\ref{tab_ALLcom}, we show the values of $\tilde a_i^{\alpha}$ at
a global minimum of each contribution of
$\Delta V^{(p_t)}({\Phi}_{\cal R}^{(k)})$ and
$\Delta V^{(p_t)}({\Psi}_{\cal R}^{(k)})$ for $p_t=0,1$ and
${\cal R}=\bm{5},\bm{10},\bm{15},\bm{24}$.  As noted below
Eq.~\eqref{veff_5scalar}, the traceless condition holds modulo 1.  We
also show the unbroken gauge symmetry $G_0$ at the
minimum.  We don't give explicit results of $p_t=-1$ cases since they
are obtained from the ones of $p_t=1$ cases through the relation
${\cal V}^{(-1)}(\tilde a_i^{\alpha})={\cal V}^{(1)}(-\tilde
a_i^{\alpha})$ explained above.

\begin{table}[t]
    \centering
    \begin{tabular}{c|c|ll|c}\hline\hline
     ${\cal R}$& potential & $(\tilde a_1^1,\tilde a_1^2,\tilde a_1^3,\tilde a_1^4,\tilde a_1^5)$
      & $(\tilde a_2^1,\tilde a_2^2,\tilde a_2^3,\tilde a_2^4,\tilde a_2^5)$ & $G_0$ \\\hline\hline
\multirow{6}{*}{$\bm{5}$}&     $\Delta V^{(0)}({\Phi}_{\bm{5}}^{(k)})$&$(0,0,0,0,0)$&$(0,0,0,0,0)$& $SU(5)$ \\\cline{2-5}
&\multirow{2}{*}{$\Delta V^{(1)}({\Phi}_{\bm{5}}^{(k)})$}&$(3,3,3,3,3)/5$&$(3,3,3,3,3)/5$& $SU(5)$ \\
&                &$(3,3,3,3,3)/5$&$(4,4,4,4,4)/5$& $SU(5)$  \\\cline{2-5}
&      $\Delta V^{(0)}({\Psi}_{\bm{5}}^{(k)})$&$(2,1,1,1,1)/3$&$(2,1,1,1,1)/3$& $SU(4)\times U(1)$ \\\cline{2-5}
&      \multirow{2}{*}{$\Delta V^{(1)}({\Psi}_{\bm{5}}^{(k)})$}&$(0,0,0,0,0)$&$(0,0,0,0,0)$& $SU(5)$ \\
&                &$(1,1,1,0,0)/3$&$(1,1,1,0,0)/3$& $SU(3)\times SU(2)\times U(1)$ \\
      \hline\hline
\multirow{7}{*}{$\bm{10}$}&      $\Delta V^{(0)}({\Phi}_{\bm{10}}^{(k)})$&$(0,0,0,0,0)$&$(0,0,0,0,0)$& $SU(5)$ \\\cline{2-5}
&      \multirow{2}{*}{$\Delta V^{(1)}({\Phi}_{\bm{10}}^{(k)})$}&$(2,2,2,2,2)/5$&$(4,4,4,4,4)/5$& $SU(5)$ \\
&                &$(4,4,4,4,4)/5$&$(4,4,4,4,4)/5$& $SU(5)$  \\\cline{2-5}
&      \multirow{2}{*}{$\Delta V^{(0)}({\Psi}_{\bm{10}}^{(k)})$}&$(1,1,1,1,1)/5$&$(1,1,1,1,1)/5$& $SU(5)$ \\
&      &$(1,1,1,1,1)/5$&$(3,3,3,3,3)/5$& $SU(5)$ \\\cline{2-5}
&      \multirow{2}{*}{$\Delta V^{(1)}({\Psi}_{\bm{10}}^{(k)})$}&$(0,0,0,0,0)$&$(0,0,0,0,0)$& $SU(5)$ \\
&                &$(1,2,2,2,2)/3$&$(1,2,2,2,2)/3$& $SU(4)\times U(1)$ \\
      \hline\hline
\multirow{6}{*}{$\bm{15}$}&      $\Delta V^{(0)}({\Phi}_{\bm{15}}^{(k)})$&$(0,0,0,0,0)$&$(0,0,0,0,0)$& $SU(5)$ \\\cline{2-5}
&      \multirow{2}{*}{$\Delta V^{(1)}({\Phi}_{\bm{15}}^{(k)})$}&$(2,2,2,2,2)/5$&$(4,4,4,4,4)/5$& $SU(5)$ \\
&                &$(4,4,4,4,4)/5$&$(4,4,4,4,4)/5$& $SU(5)$  \\\cline{2-5}
&      \multirow{2}{*}{$\Delta V^{(0)}({\Psi}_{\bm{15}}^{(k)})$}&$(2,1,1,1,1)/6$&$(2,1,1,1,1)/6$& $SU(4)\times U(1)$ \\
&                &$(2,1,1,1,1)/6$&$(1,2,2,2,2)/3$& $SU(4)\times U(1)$ \\\cline{2-5}
&      $\Delta V^{(1)}({\Psi}_{\bm{15}}^{(k)})$&$(0,0,0,0,0)$&$(0,0,0,0,0)$& $SU(5)$ \\
      \hline\hline
\multirow{4}{*}{$\bm{24}$}&      $\Delta V^{(0)}({\Phi}_{\bm{24}}^{(k)})$&$(0,0,0,0,0)$&$(0,0,0,0,0)$& $SU(5)$ \\\cline{2-5}
&      $\Delta V^{(1)}({\Phi}_{\bm{24}}^{(k)})$&$(1,1,2,2,0)/3$&$(1,1,2,2,0)/3$ & $SU(2)^2\times U(1)^2$ 
      \\\cline{2-5}
&      $\Delta V^{(0)}({\Psi}_{\bm{24}}^{(k)})$&$(1/3,2/3,0,v_x,1-v_x)$&$(1/3,2/3,v_x,1-v_x,0)$& $U(1)^4$ \\\cline{2-5}
&      $\Delta V^{(1)}({\Psi}_{\bm{24}}^{(k)})$&$(0,0,0,0,0)$&$(0,0,0,0,0)$& $SU(5)$ \\
      \hline\hline                                         
    \end{tabular}
    \caption{The values of $\tilde a_i^{\alpha}$ at a global minimum
      of the 
      contributions $\Delta V^{(p_t)}(\phi)$, where $\phi$ is
      ${\Phi}_{\cal R}^{(k)}$ or ${\Psi}_{\cal R}^{\pm(k)}$ for
      ${\cal R}=\bm{5},\bm{10},\bm{15},\bm{24}$. We also show the
      unbroken gauge symmetry $G_0$ at the minimum. The
      constant $v_x=0.24796$ is used.}\bigskip
    \label{tab_ALLcom}
\end{table}

The gauge field also generates the contribution to the effective
potential, which is equal to
$2\Delta V^{(0)}({\Phi}_{\bm{24}}^{(k)})$, whose minimum respects
$SU(5)$ symmetry. Thus, we need bulk matter fields in the theory to
obtain the SM gauge symmetry
$G_{\rm SM}\equiv SU(3)\times SU(2)\times U(1)$ at a vacuum. Let us
remark that
$2\Delta V^{(0)}({\Phi}_{\bm{24}}^{(k)})+\Delta
V^{(0)}({\Psi}_{\bm{24}}^{(k)})=0$ at the one-loop level, and the
contribution $\Delta V^{(1)}(\Psi_{\bm 5}^{(k)})$ has degenerate
global minima with $G_{\rm SM}$ and $SU(5)$, as seen in
Table~\ref{tab_ALLcom}. Thus, we easily find matter contents that
ensure $G_{\rm SM}$ at a minimum and have no bulk and boundary
anomalies. We show two examples in Table~\ref{tab_matcon}. We refer to
the bulk matter contents shown in the left and right tables as case~(i) and (ii), respectively.  The case~(i) consists of a $p_t=0$
adjoint threefold fermion with positive chirality and 10 sets of the
$p_t=1$ fundamental threefold fermion with negative chirality.  The
case~(ii) consists of a $p_t=0$ adjoint threefold fermion with
positive chirality, the 16 sets of the $p_t=1$ fundamental threefold
fermion with negative chirality, and the 2 sets of the $p_t=0$
antisymmetric (10-dimensional) representation threefold fermion with
negative chirality.\footnote{The same $SU(5)$ representations of the
  fermionic sector are found in one of the supersymmetric models
  in~\cite{Bhardwaj:2015xxa}.}  In both cases, one sees that there are
no anomalies.  In addition, the potential contributions from the gauge field
and an adjoint fermion field cancel out. For the case~(i), the sum of
the effective potential contributions is proportional to
$\Delta V^{(1)}({\Psi}_{\bm{5}}^{(k)})$, in which $SU(5)$ and
$G_{\rm SM}$ vacua are degenerate. For the case~(ii), we numerically
find that, at the global minima of the effective potential, the values
of the Wilson line take
\begin{align}
  \tilde a_1^{\alpha}=\tilde a_2^{\alpha}=(1,1,1,0,0)/3,
  \label{smvacw}
\end{align}
and the symmetry $SU(5)$ is broken down to $G_{\rm SM}$. 
We note that on this vacuum $\tilde a_3^{\alpha}=(-2,-2,-2,0,0)/3$ and
$\tilde a_\ell^{\alpha}-\tilde a_{\ell+1}^{\alpha}=0$ (mod 1) are obtained. Thus,
this vacuum respects the symmetry
$\zpl\times \zmi \cong \zex \times \zy$, as discussed in
Sec.~\ref{sec:wilsonline}.

\begin{table}[t]
    \centering
    \begin{tabular}{ccc}
\hline\hline      \multicolumn{3}{c}{case (i)}\\
      bulk matter & $p_t$ & flavor\\\hline
      $\Psi_{\bm{24}}^{+(k)}$&0&1\\
      $\Psi_{\bm{5}}^{-(k)}$&1&10\\
      \hline\hline
    \end{tabular} \hspace{2cm}
    \begin{tabular}{ccc}
\hline\hline      \multicolumn{3}{c}{case (ii)}\\
      bulk matter & $p_t$ & flavor\\\hline
      $\Psi_{\bm{24}}^{+(k)}$&0&1\\
      $\Psi_{\bm{5}}^{-(k)}$&1&16\\
      $\Psi_{\bm{10}}^{-(k)}$&0&2\\
      \hline\hline
    \end{tabular} \qquad    
    \caption{Examples of bulk matter contents. We refer to the matter
      contents of the left (right) table as the case (i) ((ii)).}\bigskip
    \label{tab_matcon}
\end{table}

\subsection{Phenomenological implications}

On the vacuum shown in Eq.~(\ref{smvacw}), interestingly, the
so-called doublet-triplet splitting among the Higgs fields in the {\bf
  5} representation can be realized, similarly in the $S^1/\Z2$
case~\cite{gghusu5DTS}.

If we introduce a {\bf5} threefold scalar with $p_t=0$, its triplet
component gets contribution from the Wilson line phases to become
massive, while its doublet component does not and contains a massless
mode.  We note that on this vacuum, the $\Z3^{\rm(ex)}$ symmetry
remains unbroken, even though the zero modes of the extra-dimensional
components of the gauge fields which develop nonvanishing VEV,
$A_{y^i}$, have nontrivial charges of the $\Z3^{\rm(ex)}$ symmetry.
This means that the tadpole term of the zero mode of $A_{y^i}$ is
absent even in the higher-loop corrections to allow the vacuum to be a
(local) minimum without a fine tuning.  In addition, the effective
theory around the TeV scale would have a $\Z3$ symmetry, though a
soft-breaking term of the $\Z3$ symmetry may be introduced as in the
$S^1/\Z2$ case~\cite{gghusu5pheno}.

Of course, there would be large radiative corrections to the scalar
masses in non-supersymmetric (non-SUSY) models, and thus we impose the
SUSY in following.  In the SUSY limit, however, the contributions from
the fermions and the bosons to the effective potential are canceled
out.  Thus, the actual effective potential strongly depends on the
SUSY breaking.  In addition, when there is a hierarchy between the
SUSY-breaking scale and the compactification scale, the effective
potential suffers from the large logarithms, and we need to treat the
renormalization group equations.  In this way, the analysis in the
previous subsection can not be applied directly.  Nevertheless, it
provides a hope that the vacuum tends to be realized in a sizable
parameter region, besides a proof of existence.

Concerning the vacuum selection, we have proposed an interesting
scenario in Ref.~\cite{gghusu5FT}, which may be applied also to the
present case.  In the reference, we have calculated the effective
potential at a finite temperature and found that there are models
where the desired vacuum (in the $S^1/\Z2$ case) is the global minimum
at high temperature. Thus, if the universe started with very high
temperature of order the Planck scale, the vacuum would be selected
around the temperature of order the compactification/GUT scale, before
the inflation.  Then, it is natural to expect that the vacuum does not
move so much until the reheating and has been selected.

An outstanding prediction of the SUSY version is the existence of
light adjoint chiral supermultiplets of masses around the
SUSY-breaking scale, which would be a TeV scale.  This is understood
as follows.  The zero modes of $A_{y^i}$ are massless at the tree
level and receive masses through radiative corrections that are
suppressed by the SUSY-breaking scale.  Since the mass differences
among components in a single supermultiplet are at most of the
SUSY-breaking scale, the masses of their SUSY partners are also at
most of the scale.  Some collider phenomenology of them in the
$S^1/\Z2$ case was studied in Ref.~\cite{gghusu5pheno}.  In
Ref.~\cite{gghuDg}, another attractive possibility to regard the
adjoint chiral supermultiplets as those introduced in the Dirac
gaugino scenario~\cite{DiracGaugino} is studied to show that the
so-called goldstone gauginos~\cite{GoldstoneGaugino} are naturally
realized.  Similar analyses in the present case are desirable.

An unfavorable point of this prediction is that the light adjoint
chiral supermultiplets ruin the success of the gauge coupling
unification in the minimal SUSY $SU(5)$ model~\cite{SUSY-GUT}.  This
is because the adjoint multiplets give contributions of
$\Delta b_i^{adj}=(0,2,3)$ for the beta function coefficients to that
of the minimal supersymmetric SM (MSSM), $b_i^{\rm MSSM}=(33/5,1,-3)$.
It is possible, however, to recover the gauge coupling unification,
for example by introducing additional multiplets that give further
correction of $\Delta b_i^{add}=(3+n,1+n,n)$~\cite{gghusu5DTS}.

It is notable that an example with $n=0$ is naturally realized in the
present case, for instance by adding one {\bf5} and two {\bf10}
threefold hypermultiplets with $p_t=0$.  This is because the above
{\bf5} ({\bf10}) hypermultiplet contains a zero mode vector-like pairs
of the component with the SM charge $({\bf1},{\bf2})_{-1/2}$
($({\bf1},{\bf1})_{1}$).  We note that it is in contrast to the
$S^1/\Z2$ case, where the pair with $({\bf1},{\bf1})_{1}$ can not be
realized separately.  This difference would bring significant effects
on the phenomenology as the quantum corrections to the colored
particles are not so enhanced in contrast to the $n=1$ case where the
color $SU(3)$ symmetry is asymptotic nonfree (though still
perturbative around the GUT scale)~\cite{gghusu5pheno}.

Next, we discuss the matter sector.  As shown in
Sec.~\ref{sec:matterbc}, the zero modes of the threefold fermions are
vector-like, and those of the onefold fermions may be chiral but the
possible representations are restricted.  Then, the simplest way to
realize the chiral fermion in the SM is to put them on the fixed
points.  Though there are still several possibilities to put the
fermions on the three fixed points, we consider here only the case all
the SM fermions are put on a common fixed point, for simplicity.

In contrast to the usual gauge-Higgs unification models where the SM
Higgs field is unified into a gauge field, the SM Higgs field is
introduced as a {\bf5} field in our scenario, and its Yukawa coupling
can be set by hand on the fixed point.  The flavor structure of the
Yukawa couplings is similar to usual 4D models and it might be set by
hand or a flavor symmetry may be introduced.  A difference from the
usual 4D models is the $SU(5)$ breaking effect, which is carried only
by $A_{y^i}$, and thus bulk fermions are necessary as messenger of the
$SU(5)$ breaking, to solve the wrong GUT relation among the Yukawa
couplings.

Finally, we comment on the $\mu$ problem and the proton decay.  If we
put a {\bf5} threefold hypermultiplet with negative chirality and
$p_t=0$, the zero modes are a vector-like pair of the doublet chiral
supermultiplet with the $\Z3$-charge $+1$.  When these are identified
with $H_u$ and $H_d$ of the MSSM, the matter chiral supermultiplet
${\bf 10}_i$ and ${\bf{\bar5}}_i$ where the index $i$ denotes the
generation should have the $\Z3$ charge $+1$ to allow the Yukawa
couplings.  These $\Z3$ charge forbids the dimension 5 operator for
the proton decay, ${\bf 10}_i{\bf 10}_j{\bf 10}_k{\bf{\bar5}}_l$, and,
at the same time, the $\mu$ term in the MSSM.  We suppose the SUSY breaking sector
breaks the $\Z3$ symmetry softly to solve the $\mu$ problem.  Though
this $\Z3$ breaking may generate the dimension 5 proton decay
operator, its contribution to the proton decay is quite suppressed.
Then, the proton decay via the dimension 6 operators mediated by the
gauge field becomes dominant.  In the 6D spacetime, the sum of the
contributions from the KK gauge boson is (logarithmically)
divergent~\cite{LogDivInDim6PD}, when all the fermion fields are put
on a single fixed point.  Though the summation should be cut off at
some point as the 6D theory is also an effective theory, this process
is enhanced, besides the effect of the enhanced coupling of the KK
gauge field and the boundary fermions by a factor $\sqrt3$ shown in
Eq.~(\ref{diggendef1}).  Meanwhile, it also has a suppression factor.
It is possible that the dominant element of the SM fermion may come
from the "messenger field" instead of the boundary fields.  In case
that the origins of the dominant modes of the components of each
$SU(5)$ multiplet are different, the gauge interactions do not connect
them.  These points should be studied in a future work.

%
\section{Conclusions and discussions}
\label{Sec:xxx}
%
We have formulated a field theoretical realization of the diagonal
embedding method in the gauge theory compactified on the
$T^2/{\mathbb Z}_3$ orbifold. The original bulk gauge group of the
theory is $G\times G\times G$, and a global $\zex$ transformation
permutes them. Through the BCs, only the diagonal part of the gauge
group $G^{\rm diag}$, which is isomorphic to $G$, remains manifest at
a low-energy effective theory. The 4D effective theory contains the
zero mode of the extra-dimensional component of the gauge field, which
belongs to the adjoint representation of $G^{\rm diag}$. The
continuous Wilson line phase degrees of freedom, \ie, the zero mode
along the flat direction of the tree-level potential for the
extra-dimensional gauge fields, can acquire VEVs that further
spontaneously break the gauge symmetry $G^{\rm diag}$. Thus, the
theory possesses rich vacuum structure.  We have shown a
parametrization of the VEVs and the Wilson line phases, which are
required to clarify the symmetry breaking patterns.

We have also discussed the bulk scalar and fermion fields in our
setup. The representations of these bulk matter fields under the gauge
group are restricted to be the $\zex$ threefold or onefold to keep the
$\zex$ invariance of the Lagrangian. We have examined the possible BCs
for the matter fields and the KK mass spectrum.  The onefold fermions
can have 4D chiral fermions as their zero modes, although the
threefold ones always have vector-like 4D fermion zero modes.  A
particular feature is that the representations of the chiral zero
modes under the gauge group are restricted due to the diagonal
embedding method, as shown in Eqs.~\eqref{chiral1}
and~\eqref{chiral2}.

We have studied the $SU(5)$ type A grand gauge-Higgs unification model
compactified on $T^2/{\mathbb Z}_3$ with the diagonal embedding method
as an explicit application. We have derived the one-loop contributions
to the effective potential for the zero modes of extra-dimensional
gauge fields.  We have examined the vacuum structure of the effective
potential and discussed the symmetry breaking patterns related to the
bulk matter contents. Our analysis has shown that the $SU(5)$ symmetry
is broken down to $SU(3)\times SU(2)\times U(1)$ at the global minima
of the effective potential with the specific bulk matter
contents. Thus, the type A grand gauge-Higgs unification model on
$T^2/{\mathbb Z}_3$ is viable for explaining the spontaneous GUT
breaking.

In the present analysis, we utilize the dual lattice technique, which
is just a Fourier transformation.  It is actually useful to analyze
the KK expansion in the $T^2/{\mathbb Z}_3$ model, which is the
minimal $\Z{3}$ orbifold model and may be regarded as an effective
theory of the heterotic string theory with an adjoint scalar zero mode
and with three generations.  In addition, this technique can be
applied to more general orbifold models, for instance in a
ten-dimensional spacetime, straightforwardly.  It is also possible to
treat more general gauge symmetry than $SU(5)$ considered in this
article, such as $SO(10)$, $E_6$ and $E_8$.  These generalizations
would be attractive future works.

Finally, we have discussed the phenomenological implications focusing
on the GUT breaking vacuum.  A notable feature of this spontaneous GUT
breaking is to provide a solution to the doublet-triplet splitting
problem in GUT models. In addition, the vacuum is characterized by the
enhancement of a ${\mathbb Z}_3$ symmetry and is implied to be stable
against higher-loop quantum corrections. With a SUSY extension, the
light three chiral supermultiplets, which are adjoint representations
under $SU(3)$, $SU(2)$, or $U(1)$, are predicted to appear around the
SUSY-breaking scale. The unification of the three gauge couplings in
the SM can be consistently explained with the vanishing beta function
coefficient of the color $SU(3)$ at the one-loop order. We have also
given discussions about the SM matter sector and proton decay,
although detailed examinations are left for future studies.

\bigskip \bigskip
\begin{center}
    {\bf Acknowledgement}
\end{center}
The authors would like to thank Sosei Tsuchiya for collaborations at
the early stage of this work and also thank Yoshiharu Kawamura for
helpful discussions.

%
\bigskip \bigskip
\appendix
%

%
\section{Kalzua-Klein expansions on $M^4\times T^2/{\mathbb Z}_3$}
\label{Sec:kkexp1}
%
In this appendix, we discuss the KK expansion on
$M^4\times T^2/{\mathbb Z}_3$. In the following, we regard
$\hat {\cal T}_{1}$ and $\hat {\cal S}_{0}$ as the independent
operators among $\hat {\cal T}_{1,2}$ and $\hat {\cal S}_{0,1,2}$
defined in Sec.~\ref{sec_t2z3}.

\subsection{${\zex}$ singlet fields}
We first discuss the KK expansion of ${\zex}$ singlet fields.  Let
$\phi(x^\mu,y^i)$ be a ${\zex}$ singlet field that obeys the BCs as
\begin{align}
  \phi(x^\mu,\hat {\cal T}_1[y^i])=\omega^{p_t}\phi(x^\mu,y^i), \qquad 
  \phi(x^\mu,\hat {\cal S}_0[y^i])=\omega^{p_s}\phi(x^\mu,y^i), 
  \label{scalar_kk01}
\end{align}
where $p_t,p_s\in \{0,\pm 1\}$, which are consistent with
$\hat {\cal S}_r^3=\hat {\cal I}$ ($r=0,1,2$).

To examine the KK expansion, we introduce the orthonormalized
eigenfunction under the translation
$\bar y^i\to \hat {\cal T}_j[\bar y^i]=\bar y^i+\delta^i_j$
$(i,j=1,2)$, where $\bar y^i=y^i/(2\pi R)$, as
\begin{align}\label{eigenfdef1}
  &  f(\bar y^i(n_i+\alpha_i))={1\over 2\pi R(\det g_{ij})^{1/4}}e^{2\pi i\bar y^i(n_i+\alpha_i)}, \qquad n_i\in{\mathbb Z}, \quad \alpha_i\in{\mathbb R}.
\end{align}
One sees the eigenfunction satisfies
\begin{gather}\label{app_eigenfdef1}
    f(\hat {\cal T}_j[\bar y^i](n_i+\alpha_i))=e^{2\pi i
      \alpha_j}f(\bar y^i(n_i+\alpha_i)),
    \\
    \iint_{T^2} d^2y f^*(\bar y^i(n_i+\alpha_i)) f(\bar
    y^j(n_j'+\alpha_j))=\delta_{n_1n_1'}\delta_{n_2n_2'}\equiv
    \delta_{n_in_i'}^{(2)},
    \label{orthonorm_f1}
\end{gather}
where we have defined
\begin{align}
\iint_{T^2} d^2y\equiv     \int_0^{2\pi R}dy^1    \int_0^{2\pi R}dy^2
  \sqrt{\det g_{ij}}={3\over 4}\int_0^{2\pi R}dy^1    \int_0^{2\pi R}dy^2.
\end{align}
As explained in Sec.~\ref{sec_t2z3}, a pair of the same upper and
lower indices, such as $i$ in Eq.~\eqref{app_eigenfdef1}, is always
contracted as $\bar y^in_i=\bar y^1n_1+ \bar y^2n_2$.

Notice that $f(\bar y^i(n_i+\alpha_i))$ is not an eigenfunction of the
${\mathbb Z}_3$ transformation generated by $\hat {\cal S}_0$ defined
in Eq.~\eqref{tis0ytrans2}. Using Eq.~\eqref{kys_formula1}, we see
that the transformation of the function is
\begin{align}\label{z3transoff}
    f(\hat {\cal S}_0[\bar y^i](n_i+\alpha_i))
  =  f(\bar y^i\hat {\cal S}_0^{-1}[n_i+\alpha_i])
  =f(\bar y^i(n_{i+1}+\alpha_{i+1})), 
\end{align}
where we have defined $n_3=-n_1-n_2$, $\alpha_3=-\alpha_1-\alpha_2$,
$n_{i+3}=n_i$, and $\alpha_{i+3}=\alpha_i$ for $i\in {\mathbb Z}$.
The eigenfunction of both the transformations $\hat {\cal T}_{1}$ and
$\hat {\cal S}_{0}$ is given by
\begin{align}\label{t2z3eigenfunc1} 
  &  \tilde f^{[p]}(\bar y^iN_{i})
    = {1\over \sqrt{3}}\sumk \omega^{-kp}
    f(\bar y^iN_{i+k}), \qquad 
    p\in {\mathbb Z},
\end{align}
where 
\begin{align}
  &    N_{1}\equiv n_{1}+{p_t\over 3}, \quad
    N_{2}\equiv n_{2}+{p_t\over 3}, \quad
    N_{3}\equiv -n_{1}-n_2-{2p_t\over 3}, \quad
    N_{i+3}\equiv N_i.
\end{align}
Conversely, we also obtain the relation as
\begin{align}
  f(\bar y^iN_{i+k})={1\over \sqrt{3}}\sump \omega^{kp}\tilde f^{[p]}(\bar y^iN_{i}).
\end{align}
From Eq.~\eqref{t2z3eigenfunc1}, one confirms
\begin{align}
&  \tilde f^{[p_s]}({\cal T}_j[\bar y^i]N_i)
  =\omega^{p_t}\tilde f^{[p_s]}(\bar y^iN_i), \qquad 
    \tilde f^{[p_s]}({\cal S}_0[\bar y^i]N_i)
  =\omega^{p_s}\tilde f^{[p_s]}(\bar y^iN_i), 
  \label{ftilde_s0eigen}
\end{align}
where the eigenvalues are exactly the same as in
Eq.~\eqref{scalar_kk01}. Thus, a ${\zex}$ singlet field with the BCs
in Eq.~\eqref{scalar_kk01} is expanded by
$\tilde f^{[p_s]}(\bar y^iN_{i})$.\footnote{The eigenfunctions
  $\tilde f^{[p_s]}(\bar y^iN_{i})$ given by linear combinations of
  the exponential functions in Eq.~\eqref{eigenfdef1} are analogues to
  $\sin(yn/R)$ or $\cos(yn/R)$ $(n\in{\mathbb Z})$ in
  $S^1/{\mathbb Z}_2$ orbifold models.}

\begin{figure}[]
\centering
  \includegraphics[width=16cm,clip]{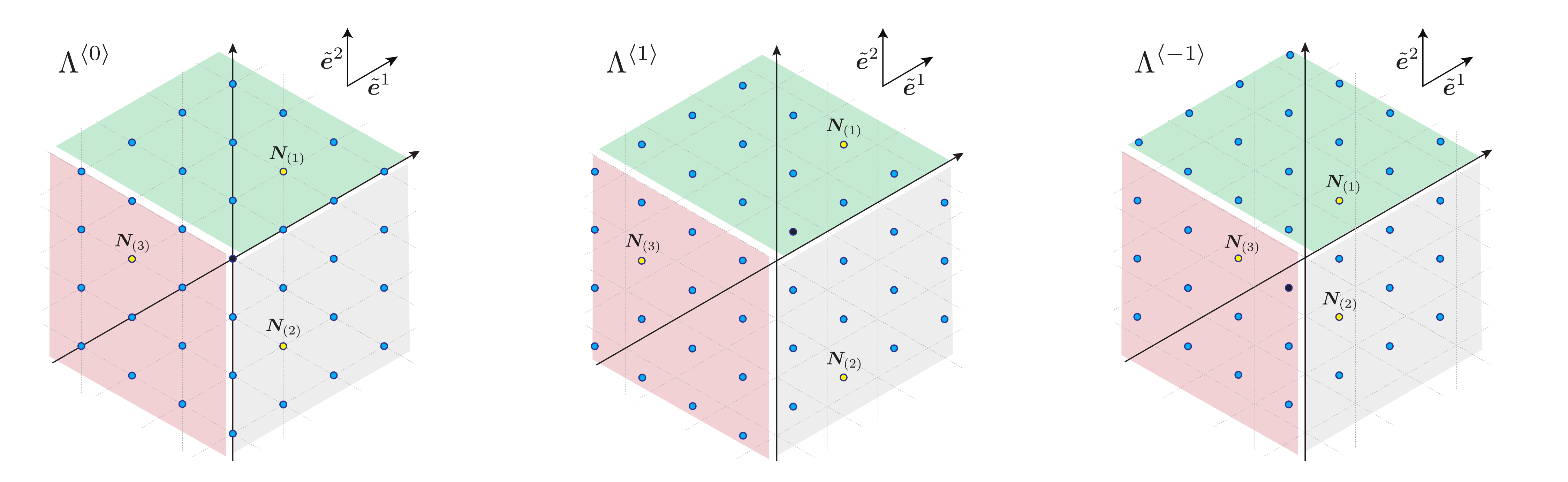} 
  \caption{The dual lattice $\Lambda^{\langle p_t\rangle}$ defined in
    Eq.~\eqref{def_dlat1} for $p_t=0,1,-1$. The black dots show the
    point with $n_1=n_2=0$.  As an example, we denote $\bm N_{(\ell)}$
    for $n_1=n_2=1$ by the yellow dots.  Dots on the green, gray, and
    red shaded regions belong to the sublattice
    $\Lambda^{\langle p_t\rangle}_{(1)}$
    $\Lambda^{\langle p_t\rangle}_{(2)}$, and
    $\Lambda^{\langle p_t\rangle}_{(3)}$ defined in
    Eq.~\eqref{def_sublattice}, respectively.}\bigskip
\label{fig_duallattice}
\end{figure}

The eigenfunctions in the set $\{\tilde f^{[p]}(\bar y^iN_{i})\}$ for
$n_i\in {\mathbb Z}$ are neither completely independent nor
orthonormalized.\footnote{This is similar to the fact that
  $\sin(yn/R)$ and $\sin(-yn/R)$ is not linearly independent even if
  $n\neq -n$ is satisfied.  If one considers
  $m,m'\in{\mathbb Z}_{\geq 0}$, $\sin(ym/R)$ and $\sin(ym'/R)$ are
  linearly independent for $m\neq m'$.}
From the right of
Eq.~\eqref{ftilde_s0eigen} and Eq.~\eqref{kys_formula1}, we find
\begin{align}\label{tildef_dependent}
\tilde f^{[p]}({\cal S}_0^\ell[\bar y^i]N_{i}) = \tilde f^{[p]}(\bar y^iN_{i+\ell})
  =  \omega^{\ell p}      
  \tilde f^{[p]}(\bar y^iN_{i}),
\end{align}
which implies a linear dependency
$\tilde f^{[p]}(\bar y^iN_{i+\ell})\propto \tilde f^{[p]}(\bar
y^iN_{i})$ related to the ${\mathbb Z}_3$ transformation. As seen
below, there is no additional linear dependencies except for the
above.  To handle the eigenfunctions, it is convenient to introduce
the normalized momentum lattice, which corresponds to the possible
momentum values on $T^2$ in a normalization and is expanded by the
dual basis vector in Eq.~\eqref{dualtildeedef} as
\begin{align}\label{def_dlat1}
  \Lambda^{\langle p_t\rangle}=\left\{
  N_1\tilde {\bm e}^1
  +N_2\tilde {\bm e}^2
  =
  \left(n_1+{p_t\over 3}\right)\tilde {\bm e}^1
  +  \left(n_2+{p_t\over 3}\right)\tilde {\bm e}^2~\big|~n_1,n_2\in{\mathbb Z}
  \right\}.
\end{align}
We hereafter refer to $\Lambda^{\langle p_t\rangle}$ as the dual
lattice. In Fig.~\ref{fig_duallattice}, the dual lattice with
$p_t=0,\pm 1$ is illustrated. Since we can relate $N_{\ell}$ and
$N_{\ell+1}$ appearing in $\tilde f^{[p]}(\bar y^iN_{i+\ell-1})$ to a
point
$\bm N_{(\ell)}\equiv N_\ell\tilde {\bm e}^1 + N_{\ell+1}\tilde {\bm
  e}^2$ on $\Lambda^{\langle p_t\rangle}$, we regard that there is a
corresponding eigenfunction on each point on the lattice.  Note that
$\bm N_{(1)}$, $\bm N_{(2)}$, and $\bm N_{(3)}$ are not identical
points on $\Lambda^{\langle p_t\rangle}$, except for the case of
$(N_1,N_2)= (0,0)$.  These points are related to each other by the
${\mathbb Z}_3$ transformation generated by $\hat {\cal S}_0$ as found
in Eq.~\eqref{z3transoff} and identified to the positions of vertices
of the equilateral triangle, whose center is located at the origin.
From the above observation, we can divide
$\Lambda^{\langle p_t\rangle}$ into the sublattice as\footnote{The
  decomposition in Eq.~\eqref{decop_lat1} with $p_t=0$ corresponds to
  ${\mathbb Z}={\mathbb Z}_{>0}+{\mathbb Z}_{<0}+\{0\}$ in
  $S^1/{\mathbb Z}_2$ orbifold models.}
\begin{align}\label{decop_lat1}
  \Lambda^{\langle p_t\rangle}&=\Lambda^{\langle p_t\rangle}_{(1)}+\Lambda^{\langle p_t\rangle}_{(2)}+\Lambda^{\langle p_t\rangle}_{(3)}
                                +\Lambda_{(0)}\delta_{p_t0},
\end{align}
where
\begin{align}\label{def_sublattice}
  \Lambda^{\langle p_t\rangle}_{(\ell)}=
  \left\{
  \bm N_{(\ell)}=  N_\ell\tilde {\bm e}^1
  +  N_{\ell+1}\tilde {\bm e}^2~\big|~N_i=n_i+{p_t\over 3},~n_1,n_2\in{\mathbb Z,
  ~N_2\geq 0,~N_1> -N_2}
  \right\}, 
\end{align}
for $\ell=1,2,3$ and $\Lambda_{(0)}$ is the origin.  If
$\tilde f^{[p]}(\bar y^iN_{i})$ corresponds to a point on
$\Lambda^{\langle p_t\rangle}_{(\ell)}$, then the dependent function
$\tilde f^{[p]}(\bar y^iN_{i+1})$ corresponds to a point on
$\Lambda^{\langle p_t\rangle}_{(\ell+1)}$. Thus, we see that the set
of the eigenfunctions \{$\tilde f^{[p]}(\bar y^iN_{i})$\} defined on a
sublattice $\Lambda^{\langle p_t\rangle}_{(\ell)}$ are linearly
independent.

In the following, we use
\begin{align}
\bm N&=    N_1\tilde {\bm e}^1
  +N_2\tilde {\bm e}^2
  =
  \left(n_1+{p_t\over 3}\right)\tilde {\bm e}^1
  +  \left(n_2+{p_t\over 3}\right)\tilde {\bm e}^2,\\
\bm N'&=    N'_1\tilde {\bm e}^1
  +N'_2\tilde {\bm e}^2
  =
  \left(n'_1+{p_t\over 3}\right)\tilde {\bm e}^1
  +  \left(n'_2+{p_t\over 3}\right)\tilde {\bm e}^2. 
\end{align}
With the help of
$\delta^{(2)}_{n_{i+k}n_{i+k'}'} =\delta_{kk'}\delta_{n_in_i'}^{(2)}$,
one can derive the orthonormal relation
\begin{align}\label{orthnormal_tildef}
  \iint_{T^2} d^2y  
  \tilde f^{[p]*}(\bar y^iN_i)
  \tilde f^{[p']}(\bar y^iN_i{}')
  =\delta_{n_in_i'}^{(2)}\delta_{pp'},
  \qquad {\rm for}~~\bm N,\bm N'\in  \Lambda^{\langle p_t\rangle}_{(\ell)},
\end{align}
for a fixed $\ell$, from the definition of
$\tilde f^{[p]}(\bar y^iN_i)$ in Eq.~\eqref{t2z3eigenfunc1} and the
relation in Eq.~\eqref{orthonorm_f1}.

Using the eigenfunction in Eq.~\eqref{t2z3eigenfunc1}, we define the
KK expansion of the singlet field in Eq.~\eqref{scalar_kk01} as
follows:
\begin{align}
  \phi(x^\mu,y^i)
  &=\sum_{\bm N \in \Lambda_{(\ell)}^{\langle p_t\rangle}}
  \tilde \phi_{N_1,N_2}(x^\mu)
  \tilde f^{[p_s]}(\bar y^iN_i)
+{\delta_{p_t0}\delta_{p_s0}\over 2\pi R(\det g_{ij})^{1/4}}\tilde \phi_{0,0}(x^\mu),
  \label{phi_kk_1} 
\end{align}
where we refer to $\tilde \phi_{N_1,N_2}(x^\mu)$ as the KK mode.  The
zero mode in the last term exists only for the case with
$(p_t,p_s)=(0,0)$.  We define the first term in Eq.~\eqref{phi_kk_1}
to be independent of a choice of $\ell$ in
$\Lambda_{(\ell)}^{\langle p_t\rangle}$. Thus, we can substitute
$\Lambda_{(\ell\mp 1)}^{\langle p_t\rangle}$ for
$\Lambda_{(\ell)}^{\langle p_t\rangle}$ in Eq.~\eqref{phi_kk_1} and
use Eq.~\eqref{tildef_dependent} to get
\begin{align}
  \sum_{\bm N \in \Lambda^{\langle p_t\rangle}_{(\ell)}}
  \tilde \phi_{N_1,N_2}(x^\mu)
  \tilde f^{[p_s]}(\bar y^iN_i)
&=    \sum_{\bm N \in \Lambda^{\langle p_t\rangle}_{(\ell \mp 1)}}
 \tilde \phi_{N_1,N_2}(x^\mu)
\omega^{\mp p_s}  \tilde f^{[p_s]}(\bar y^iN_{i\pm 1})\\
&=  \sum_{\bm N \in \Lambda^{\langle p_t\rangle}_{(\ell)}}
  \omega^{\mp p_s} \tilde \phi_{N_{1\mp 1},N_{2\mp 1}}(x^\mu)
  \tilde f^{[p_s]}(\bar y^iN_i).
\end{align}
This implies the constraint on the KK mode
$\tilde \phi_{N_1,N_2}(x^\mu)$ as
\begin{align}
  \label{kkmode_dependent_relations}
\tilde \phi_{N_{1\pm \ell},N_{2\pm \ell}}(x^\mu)
=\omega^{\mp \ell p_s} 
  \tilde \phi_{N_1,N_2}(x^\mu).
\end{align}

Let us derive the effective 4D Lagrangian for the singlet field in
Eq.~\eqref{scalar_kk01} from the KK expansion~\eqref{phi_kk_1}.  As an
example, we treat $\phi(x^\mu,y^i)$ as a scalar and consider the 6D
canonical kinetic term.\footnote{The $\zex$ singlet does not couple to
  the Wilson line phases in our model as discussed in
  Sec.~\ref{subsec:onefoldeffp1}, although the $\zex$ triplets
  generally couple to the Wilson line phases as
  Eq.~\eqref{effmassop1}.}
From the definitions of the
eigenfunctions in Eqs.~\eqref{eigenfdef1} and~\eqref{t2z3eigenfunc1},
we find the following relations:
\begin{align}
  g^{jk}    \der_{y^j} \der_{y^k}   \tilde  f^{[p_s]}(\bar y^iN_i)
={2\over 3}\sum_{\ell=1}^3(\der_{y^\ell})^2\tilde  f^{[p_s]}(\bar y^iN_i)
    =-{2\over 3R^2}\sum_{\ell=1}^3(N_{\ell})^2\tilde  f^{[p_s]}(\bar y^iN_i). 
\end{align}
Using them, we obtain the effective 4D Lagrangian
${\cal L}_{\rm eff}^{\rm singlet}$ for $\tilde \phi_{N_1,N_2}(x^\mu)$:
\begin{align}
  {\cal L}_{\rm eff}^{\rm singlet}
     &= -
\iint_{T^2}  d^2y 
       \phi^{*}(x^\mu,y^i)
       (\square-g^{jk}\der_{y^j}\der_{y^k})\phi(x^\mu,y^i)
  \\
     &=-\sum_{n_i\in \Lambda^{\langle p_t\rangle}_{(\ell)}}
       \tilde \phi^{*}_{N_1,N_2}
       \left\{\square+{2\over 3R^2}\sum_{k=1}^3(N_{k})^2\right\}
       \tilde \phi_{N_1,N_2} -\tilde \phi^{*}_{0,0}\square\tilde \phi_{0,0}
       \delta_{p_t0}\delta_{p_s0}, 
\end{align}
where $\tilde \phi_{N_1,N_2} (x^\mu)$ is an independent field for
$N_i\in \Lambda^{\langle p_t\rangle}_{(\ell)}$ with a fixed $\ell$.
Thus, the above KK mode is a canonically normalized 4D field with the
following KK mass squared:
\begin{align}
  {2\over 3R^2}\sum_{k=1}^3(N_{k})^2
  &={2\over 3R^2}
  \left\{
  \left(n_1+{p_t\over 3}\right)^2
  +\left(n_2+{p_t\over 3}\right)^2
  + \left(-n_1-n_2-{2p_t\over 3}\right)^2
  \right\}\\
  &={4\over 3R^2}
  \left\{
    \left(n_1+{p_t\over 3}\right)^2
    +\left(n_1+{p_t\over 3}\right)\left(n_2+{p_t\over 3}\right)
  +\left(n_2+{p_t\over 3}\right)^2
    \right\},
    \label{ap:kkmassZ3sing}
\end{align}
which is the squared norm of the vector $\bm N/R$ on the
momentum lattice spanned by $\tilde {\bm e}^i/R$.

\subsection{${\zex}$ triplet fields}
\label{sec:kktriplet}
We start to study KK expansion of a ${\zex}$ triplet field
$\phi^{(k)}$ that obeys the BCs as
\begin{align}\label{bcfortripapp1}
  \phi^{(k)}(\hat {\cal T}_1[y^i])
  &=\omega^{p_t}\phi^{(k)}(y^i),\quad 
  \phi^{(k)}(\hat {\cal S}_0[y^i])
  =\omega^{p_s}\phi^{(k+1)}(y^i), 
\end{align}
where we have suppressed the coordinate $x^\mu$ in the above for
shorthand notations. Let us define $\phi^{[p]}(y^i)$ as
\begin{align}\label{triplettilphidef1}
  \phi^{[p]}(y^i)= {1\over \sqrt{3}}\sumk \omega^{-kp}
  \phi^{(k)}(y^i), 
  \qquad 
  \phi^{(k)}(y^i)
  = {1\over \sqrt{3}}\sump \omega^{kp}
  \phi^{[p]}(y^i).
\end{align}
Then, $\phi^{[p]}$ becomes an eigenstate of the BCs: 
\begin{align}
  \phi^{[p]}(\hat {\cal T}_1[y^i])
  &=\omega^{p_t}\phi^{[p]}(y^i),\qquad 
  \phi^{[p]}(\hat {\cal S}_0[y^i])
  =\omega^{p_s+p}\phi^{[p]}(y^i), 
\end{align}
as the ${\zex}$ singlet in Eq.~\eqref{scalar_kk01} but $p_s$ is
substituted by $p_s+p$.  Thus, from a similar discussion deriving
Eq.~\eqref{phi_kk_1}, we define the KK expansion as
\begin{align}\label{tripletKKexp1}
  \phi^{[p]}(y^i)
  &=\sum_{\bm N \in \Lambda^{\langle p_t\rangle}_{(\ell)}}
    \tilde  \phi^{[p]}_{N_1,N_2}
    \tilde f^{[p_s+p]}(\bar y^iN_i)
    +{\delta_{p_t0}\delta_{p_s+p\,0}\over 2\pi R(\det g_{ij})^{1/4}}\tilde \phi_{0,0}^{[-p_s]}.
\end{align}
Using the above, we obtain the KK expansion of $\phi^{(k)}(y^i)$ as 
\begin{align}\label{phip_kk_tripletdef1}
  \phi^{(k)}(y^i)
  &={1\over \sqrt{3}}
    \sum_{\bm N \in \Lambda^{\langle p_t\rangle}_{(\ell)}}
    \sump \omega^{kp}
    \tilde  \phi^{[p]}_{N_1,N_2}
    \tilde f^{[p_s+p]}(\bar y^iN_i)
    +{{\omega^{-kp_s}}\delta_{p_t0}\over 2\pi R(\det g_{ij})^{1/4}}{\tilde \phi_{0,0}^{[-p_s]}
   \over \sqrt{3}}. 
\end{align}
Note that, as Eq.~\eqref{kkmode_dependent_relations}, the KK mode
$\tilde \phi^{[p]}_{N_1,N_2}$ satisfy the following constraint:
\begin{align}\label{tilphipconst1}
  \tilde  \phi^{[p]}_{N_{1\pm \ell},N_{2\pm \ell}}=\omega^{\mp \ell(p_s+p)}\tilde  \phi^{[p]}_{N_1,N_2}.
\end{align}

In view of Eq.~\eqref{triplettilphidef1}, it is natural to define the
KK mode $\tilde \phi^{(k)}_{N_1,N_2}$ as
\begin{align}\label{tripletKKmodedef1}
  \tilde \phi^{(k)}_{N_1,N_2}\equiv {1\over \sqrt{3}}\sump \omega^{kp}\tilde  \phi^{[p]}_{N_1,N_2}, \qquad
  \tilde   \phi^{[p]}_{N_1,N_2}
  ={1\over \sqrt{3}}\sumk \omega^{-kp}\tilde \phi^{(k)}_{N_1,N_2}.
\end{align}
As seen below, $\tilde \phi^{(k)}_{N_1,N_2}$ is a basis that
diagonalize contributions from the Wilson line phases to KK masses.
Combining Eq.~\eqref{phip_kk_tripletdef1} and the second equation
in~\eqref{tripletKKmodedef1}, we can expand $\phi^{(k)}(y^i)$ by
$\tilde \phi^{(k)}_{N_1,N_2}$. To see this, we use the formula
\begin{align}
  {1\over \sqrt{3}}\sump \omega^{(k-k')p}\tilde f^{[p_s+p]}(\bar y^iN_i)
  =\omega^{-(k-k')p_s}f(\bar y^iN_{i+k-k'}),
\end{align}
derived from Eq.~\eqref{t2z3eigenfunc1}. Using it, we obtain
\begin{align}
  \phi^{(k)}(y^i)
  &={1\over \sqrt{3}}
    \sum_{\bm N \in \Lambda^{\langle p_t\rangle}_{(\ell)}}
    \sumkd 
    \omega^{-k'p_s}
    \tilde \phi^{(k-k')}_{N_1,N_2}
    f(\bar y^iN_{i+k'})
        +{{\omega^{-kp_s}}\delta_{p_t0}\over 2\pi R(\det g_{ij})^{1/4}}{\tilde \phi_{0,0}^{[-p_s]}
    \over \sqrt{3}}.
    \label{phik_kk_tilphik1}
\end{align}
From Eq.~\eqref{tilphipconst1}, we find the constraint on
$\tilde \phi^{(k)}_{N_1,N_2}$ is written as
\begin{align}\label{eq:cophik}
  \tilde \phi ^{(k)}_{N_{1\pm \ell},N_{2\pm \ell}}
  =
  \omega^{\mp \ell p_s}\tilde \phi^{(k\mp \ell)}_{N_1,N_2}.
\end{align}

Let us derive the 4D effective Lagrangian for the triplet scalar
defined in in Eq.~\eqref{bcfortripapp1}. We consider the 6D kinetic
term:
\begin{align}\label{app_tripLag1}
  {\cal L}_{\rm eff}^{\rm triplet}
  &=-\iint_{T^2}d^2y
    \sumk
    \phi^{(k)*}(y^i)
    (\square+\hat M^2_k)\phi^{(k)}(y^i),
\end{align}
where $\hat M^2_k$ is a differential operator including Wilson line
phases as in Eq.~\eqref{diff_Msq1} and is defined by
\begin{align}\label{effmassop1}
\hat M^2_k=
  g^{ij}
  \left(-i\der_{y^i}+{\tilde a_{i+k}\over R}\right)
  \left(-i\der_{y^j}+{\tilde a_{j+k}\over R}\right)
  =  {2\over 3}\sum_{\ell=1}^3\left(-i\der_{y^\ell}+{\tilde a_{\ell+k}\over R}\right)^2.
\end{align}
Using the KK expansion in Eq.~\eqref{phik_kk_tilphik1} and the definition of
the eigenfunction in Eq.~\eqref{eigenfdef1}, we obtain the effective
4D Lagrangian for $\tilde \phi^{(k)}_{N_1,N_2}$ and the zero mode
$\tilde \phi_{0,0}^{[-p_s]}$ as
\begin{align}\notag 
  {\cal L}_{\rm eff}^{\rm triplet}
  &=-
    \sum_{\bm N \in \Lambda^{\langle p_t\rangle}_{(\ell')}}
    \sumk
    \tilde \phi^{(k)*}_{N_1,N_2}
    \left\{\square +
     {2\over 3R^2}\sum_{\ell=1}^3(N_{\ell}+\tilde a_{\ell+k})^2
    \right\}
    \tilde \phi^{(k)}_{N_1,N_2}
\\
  &\qquad\qquad     -\delta_{p_t0}\tilde \phi^{[-p_s]*}_{0,0}
    \left(\square+{2\over 3R^2} \sum_{\ell=1}^3\tilde a_{\ell}^2\right)\tilde \phi_{0,0}^{[-p_s]},
    \label{z3triplet_quad_term_res}
\end{align}
where $\tilde \phi^{(k)}_{N_1,N_2}$ is an independent and a
canonically normalized fields for
$N_i\in \Lambda^{\langle p_t\rangle}_{(\ell')}$ with a fixed $\ell'$.
With the help of Eq.~\eqref{eq:cophik}, we can rewrite
Eq.~\eqref{z3triplet_quad_term_res} with the summation over the dual
lattice $\Lambda^{\langle p_t\rangle}$ as
\begin{align}
  {\cal L}_{\rm eff}^{\rm triplet}
  &=-
    \sum_{\bm N \in \Lambda^{\langle p_t\rangle}}
    \tilde \phi^{(0)*}_{N_1,N_2}
    \left\{\square +
     {2\over 3R^2}\sum_{\ell=1}^3(N_{\ell}+\tilde a_{\ell})^2
    \right\}
    \tilde \phi^{(0)}_{N_1,N_2},
    \label{z3triplet_lag_wLamall}
\end{align}
where we have defined
$\tilde \phi_{0,0,}^{(0)}\equiv \tilde \phi_{0,0}^{[-p_s]}$. Although
we choose $k=0$ of $\tilde \phi^{(k)}_{N_1,N_2}$ as a representative
in the above, a similar expression holds for any choice of $k$.  The
KK mass in Eq.~\eqref{z3triplet_lag_wLamall} is again the squared norm
of the vector as the case in Eq.~\eqref{ap:kkmassZ3sing}, but the
vector $\bm N$ is shifted by the Wilson line phases as
$\bm N+\tilde a_i\tilde {\bm e}^i$.

The KK mass spectrum in Eq.~\eqref{z3triplet_lag_wLamall} is a similar
one in models with $T^2$ compactification.  This situation is shared
with the $S^1/{\mathbb Z}_2$ model with the diagonal embedding method,
where the resulting KK mass spectrum is a similar one in models with
$S^1$ compactification.

%
\section{Calculation of effective potentials on 
  $M^4\times T^2/{\mathbb Z}_3$}
\label{Sec:veffderive}
%
We derive contributions from a ${\zex}$ triplet field to the effective
potential obtained from the Lagrangian in Eq.~\eqref{app_tripLag1}.
For later convenience, we denote the KK mass in
Eq.~\eqref{z3triplet_lag_wLamall} by
\begin{align} \label{eff4lagtrip1}
  M^2(\tilde a_i)\equiv 
  {2\over 3R^2}\sum_{\ell=1}^3(N_\ell+\tilde a_{\ell})^2
      ={4\over 3R^2}
    \left\{
    \left(N_1+\tilde a_1\right)^2
    +\left(N_1+\tilde a_1\right)\left(N_2+\tilde a_2\right)
  +\left(N_2+\tilde a_2\right)^2
    \right\}. 
\end{align}
In addition, we define $\tilde {\bm{a}}=\tilde a_i\tilde{\bm e}^i$.
Then, we can write $M^2(\tilde a_i)=|\bm N+\tilde{\bm{a}}|^2/R^2$.
Since Eq.~\eqref{app_tripLag1} is rewritten as
Eq.~\eqref{z3triplet_lag_wLamall}, by performing the path integration
of the KK modes $\tilde \phi^{(0)}_{N_1,N_2}$, we obtain the following
contribution to the effective potential:
\begin{align}
  \Delta V^{(p_t)}\equiv {\hat N_{\rm deg}\over 2}
  \sum_{\bm N \in \Lambda^{\langle p_t\rangle}}
  \int{d^4p_{\rm E}\over (2\pi)^4}  
  \ln\left(p_{\rm E}^2+
  M^2(\tilde a_i) \right), 
\end{align}
where $\hat N_{\rm deg}=2$ for the case of a complex scalar, and the
square of an Euclidean four-momentum is denoted by $p_{\rm E}^2$.

To deal with the divergent momentum integral, we use the zeta function
regularization and introduce
\begin{align}
  \zeta(s)\equiv
  \sum_{\bm N \in \Lambda^{\langle p_t\rangle}}  
  \int{d^4p_{\rm E}\over (2\pi)^4}
  \left(p_{\rm E}^2+ M^2(\tilde a_i) \right)^{-s}. 
\end{align}
Then, the contributions to the potential is rewritten as 
\begin{align}
  \Delta V^{(p_t)}&=-{\hat N_{\rm deg}\over 2}\lim_{s\to 0}{d\over ds}\zeta(s).
\end{align}
A straightforward calculation shows
\begin{align}
  \zeta(s)
  &={1\over 8\pi^2}  \sum_{\bm N \in \Lambda^{\langle p_t\rangle}}  
  \int_0^\infty d\bar p \bar p^3\int_0^\infty dt {t^{s-1}\over \Gamma(s)}e^{-[\bar p^2+M^2(\tilde a_i)]t}\\
  &={s\over 16\pi^2}  \sum_{\bm N \in \Lambda^{\langle p_t\rangle}}  
\int_0^\infty dt  t^{-3}e^{-M^2(\tilde a_i)t}+{\cal O}(s^2), 
\end{align}
where $\Gamma(s)$ is the Gamma function, and $|s|\ll 1$ is implied. Thus, we get
\begin{align}
  {d\over ds}\zeta(s)=
  {1\over 16\pi^2}  \sum_{\bm N \in \Lambda^{\langle p_t\rangle}}  
  \int_0^\infty  dt t^{-3}e^{-M^2(\tilde a_i)t}
  +{\cal O}(s),
  \label{app_efpint1}
\end{align}
where the singularity associated with $t\to 0$ corresponds to the
ultraviolet divergence in the integral.

The ${\cal O}(s^0)$ term in Eq.~\eqref{app_efpint1} is evaluated by
using the Poisson resummation formula, which is derived in the next
section.  In Eq.~\eqref{poissonDdim}, we set $D=2$ and
\begin{align}
  d_i={p_t\over 3}+\tilde a_i,\qquad 
  A_{ij}={\pi R^2\over t}g_{ij},\qquad 
  (A^{-1})^{ij}={t\over \pi R^2}g^{ij},
\qquad \det{A}={3\pi^2R^4\over 4t^2},
\end{align}
where $g_{ij}$ and $g^{ij}$ are the metric given in
Sec.~\ref{sec_t2z3}. Let us introduce
the vector $\bm w$ and the lattice $\Lambda_w$ expanded by $\bm e_i$, which is associated with the metric $g_{ij}$, as 
\begin{align} 
  \Lambda_w=\left\{
\bm w=  w^1{\bm e}_1
  +w^2 {\bm e}_2~\big|~w^1,w^2\in{\mathbb Z}
  \right\}.
\end{align}
Then, we obtain
\begin{align}
  \sum_{\bm N \in \Lambda^{\langle p_t\rangle}}  
  e^{-M^2(\tilde a_i)t}
=  \sum_{\bm N \in   \Lambda^{\langle p_t\rangle}}  
  e^{-{t\over R^2}|\bm N+\tilde{\bm {a}}|^2}
  ={\sqrt{3}\pi R^2\over 2\sqrt{t^2}}
    \sum_{\bm w\in \Lambda_w}
e^{-{\pi^2R^2\over t} |\bm w|^2}e^{2\pi i \bm w\cdot\tilde{\bm {a}}'}, 
\end{align}
where we have defined
$\tilde {\bm {a}}'\equiv (p_t/3+\tilde a_1)\tilde{\bm
  e}^1+(p_t/3+\tilde a_2)\tilde{\bm e}^2$ and
\begin{align}
  |\bm w|^2=(w^1)^2+(w^2)^2-w^1w^2, \qquad
  \bm w\cdot\tilde{\bm{a}}'=(p_t/3+\tilde a_1)w^1+(p_t/3+\tilde a_2)w^2.
\end{align}
Thus, we can replace the summation over $\Lambda^{\langle p_t\rangle}$
by the summation over $\Lambda_w$. The integers $w^i$ are often called
winding numbers. Let us consider a continuum path on the covering
space of $T^2/{\mathbb Z}_3$, where the separation between the
endpoints of the path corresponds to the vector $2\pi R\bm w$. In this
case, such continuum path represents a noncontractible cycle on
$T^2/{\mathbb Z}_3$, whose winding number along the $\bm e_i$
direction is given by $w^i$, except for the case of $\bm w=0$. This
implies that the summation over the possible momentum states, \ie, KK
modes, in the evaluation of the effective potential is replaced by
performing the summation over the possible winding numbers. Notice
that the term with $\bm w=0$ in the summation represents a local
effect and is independent of the nonlocal Wilson line $\tilde a_i$. To
deal with it, we define $\Lambda_w'=\Lambda_w\backslash\{\bm w=0\}$
and write
$\sum_{\bm w\in \Lambda_w} F(\bm w)=F(0)+\sum_{\bm w\in
  \Lambda_w'}F(\bm w)$ for a function $F(\bm w)$.

From the above, we find
\begin{align}
  \lim_{s\to 0}{d\over ds}\zeta(s)
  &={\sqrt{3}\over 16\pi^7R^4}
    \sum_{\bm w\in\Lambda_w'}
    {e^{2\pi i\bm w\cdot \tilde{\bm{a}}'}\over (|\bm w|^2)^3}+({\rm const.})
    ={\sqrt{3}\over 16\pi^7R^4}
    \sum_{\bm w\in\Lambda_w'}
      {\cos({2\pi \bm w\cdot \tilde{\bm{a}}'})\over (|\bm w|^2)^3}+({\rm const.}).
\end{align}
In the last equation, we have used that $|\bm w|^2$ is symmetric under
$w^i\to -w^i$. We have separated the irrelevant constant term with
$\bm w=0$ in the above. 
In this paper, we discard the constant 
term in the effective potential. The summation over all integers for
$w^1$ and $w^2$ except for $(w^1,w^2)=(0,0)$ is denoted by
$w^1,w^2\in {\mathbb Z}'$. Finally, we obtain
\begin{align}
  \Delta V(\tilde a_i)&=\hat N_{\rm deg}{\cal V}^{(p_t)}(\tilde a_i),\\
  {\cal V}^{(p_t)}(\tilde a_i)&\equiv -{\sqrt{3}\over 32\pi^7R^4}
                                \sum_{w^1,w^2\in{\mathbb Z}'}
                                {\cos\left(2\pi [w^1(p_t/3+\tilde a_1)+w^2(p_t/3+\tilde a_2)]\right)\over [(w^1)^2-w^1w^2+(w^2)^2]^3}.
                                \label{vpttilai}
\end{align}

%
\section{The Poisson resummation formula in $D$ dimensions}
\label{Sec:aaa}
%

Let us consider the summation including a matrix $A^{-1}$, which is
the inverse of a symmetric $D\times D$ matrix $A$, as
\begin{align}
  I_D=\sum_{n_1,\dots,n_D\in {\mathbb Z}}e^{-\pi(n_i+d_i)(A^{-1})^{ij}(n_j+d_j)},
\end{align}
where $(A^{-1})^{ij}$ are elements of $A^{-1}$, and the indices $i,j$
run from 1 to $D$. Since $I_D$ is periodic under $d_i\to d_i+1$, we
can expand it as
\begin{align}
  I_D=\sum_{w^1,\dots,w^D\in {\mathbb Z}} C({w^i})e^{2\pi i w^jd_j},\qquad {\rm where}
  \qquad C({w^i})=\left(\prod_{i=1}^D\int_0^1d\tilde d_i \right)e^{-2\pi i w^j\tilde d_j}I_D.
\end{align}
Introducing $\beta_i=n_i+d_i$, we rewrite $C({w^i})$ as 
\begin{align}
  C({w^i})&=\left(\prod_{i=1}^D\int_{-\infty}^\infty d\beta_i \right)e^{-\pi \beta_j(A^{-1})^{jk}\beta_k}e^{-2\pi i w^l\beta_l}=\left(\prod_{i=1}^D\int_{-\infty}^\infty
            d\tilde \beta_i \right)
           e^{-\pi \tilde \beta_j(A^{-1})^{jk}\tilde \beta_k}e^{-\pi  w^{\bar j}A_{\bar j \bar k} w^{\bar k}},
\end{align}
where $\tilde \beta_j=\beta_j+iA_{jk}w^k$.

Since $A$ is symmetric, $A^{-1}$ is diagonalized by an orthogonal
matrix $O$ $(OO^T=1)$ and is written as
\begin{align}
  (A^{-1})^{ij}=(O^T\hat A^{-1}O)^{ij},
  \qquad (\hat A^{-1})^{kl}=(a^{-1})^k\delta^{kl}, 
\end{align}
where $\hat A^{-1}$ is the diagonal matrix and $(a^{-1})^k$ is a
$k$-th eigenvalue of $\hat A^{-1}$. Defining
$z_i=O_i{}^j\tilde \beta_j$, we obtain
\begin{align}
  C({w^i})&=\left(\prod_{i=1}^D\int_{{\cal C}_i}dz_i e^{-\pi (a^{-1})^i(z_i)^2}\right)e^{-\pi  w^jA_{jk} w^k},
\end{align}
where ${\cal C}_i$ denotes a path in the complex plane defined by
${\rm Re}(z_i)=(-\infty,\infty)$ with a fixed ${\rm Im}(z_i)
=O_i{}^jA_{jk}\omega^k$. With
the help of the relation
\begin{align}
  \int_{-\infty}^{\infty} dx e^{-\pi c (x+iy)^2}
  =\int_{-\infty}^{\infty} dx e^{-\pi c x^2}={1\over \sqrt{c}}, 
\end{align}
we obtain
\begin{align}
  C({w^i})&=\left(\prod_{i=1}^D{1\over \sqrt{(a^{-1})^i}}\right)e^{-\pi  w^jA_{jk} w^k}
           =\sqrt{{\rm det}A}e^{-\pi  w^iA_{ij} w^j}.
\end{align}
Thus, the following relation holds:
\begin{align}
  \sum_{n_1,\dots,n_D\in {\mathbb Z}}e^{-\pi(n_i+d_i)(A^{-1})^{ij}(n_j+d_j)}
  =\sqrt{{\rm det}A}
  \sum_{w^1,\dots,w^D\in {\mathbb Z}} 
  e^{-\pi  w^iA_{ij} w^j}
  e^{2\pi i w^kd_k},
  \label{poissonDdim}
\end{align}
which is the Poisson resummation formula used in
Appendix~\ref{Sec:veffderive}. The above is naturally rewritten by the
vectors and the metric that are defined by $\bm W=w^i\bm E_i$,
$\bm n=n_i\tilde{\bm E}^i$, $\bm d=d_i\tilde{\bm E}^i$,
$\bm E_i\cdot \tilde{\bm E}^j=\delta_i^j$, and
$\bm E_i\cdot \bm E_j=A_{ij}$ as
\begin{align}
  \sum_{n_1,\dots,n_D\in {\mathbb Z}}e^{-\pi|\bm n+\bm d|^2}
  =\sqrt{{\rm det}A}
  \sum_{w^1,\dots,w^D\in {\mathbb Z}} 
  e^{-\pi  |\bm W|^2}
  e^{2\pi i \bm W\cdot \bm d}.
\end{align}

%
\section{Fermion fields in six dimensions}
\label{Sec:fermion6d}
%
We here summarize the notations related to fermion fields in 6D
theories. First, we work with the metric
$\eta_{MN}={\rm diag}(1,-1,-1,-1,-1,-1)$. The 4D gamma matrices
$\gamma^\mu$ can be written by
\begin{align}\label{4dgam1}
  \gamma^\mu=
  \begin{pmatrix}
      0&\sigma^\mu
      \\\bar \sigma^\mu
      & 0
  \end{pmatrix},\qquad
        \sigma^\mu=(I_{2},\sigma^1,\sigma^2,\sigma^3), \qquad 
        \bar \sigma^\mu=(I_{2},-\sigma^1,-\sigma^2,-\sigma^3), 
\end{align}
where $I_2$ is the $2\times 2$ identity matrix and $\sigma^i$
$(i=1,2,3)$ are the Pauli matrices. A 4D Dirac fermion $\psi$ is
denoted by $\psi=\psi_{L}+\psi_R$, where 4D chirality is defined by
\begin{align}
  \gamma_5=i\gamma^0\gamma^1\gamma^2\gamma^3
  =
  \begin{pmatrix}
      -I_2&0\\0&I_2
  \end{pmatrix}
  ,\qquad \gamma_5\psi_L=-\psi_L, \qquad
               \gamma_5\psi_R=\psi_R. 
\end{align}
Thus, the 4D Weyl fermions are written by
\begin{align}
               \psi_L=
               \begin{pmatrix}
                   \xi_L\\ 0
               \end{pmatrix},
  \qquad 
               \psi_R=
               \begin{pmatrix}
                0\\   \bar \eta_R
            \end{pmatrix},
  \label{twocomponentspinor1}
\end{align}
where $\xi_L$ and $\bar \eta_R$ are two-component spinors.

The 6D gamma matrices can be defined by the 4D gamma matrices in
Eq.~\eqref{4dgam1} as
\begin{align}
  \Gamma^\mu=\sigma^3\otimes\gamma^\mu =
  \begin{pmatrix}
      \gamma^\mu &0\\
      0&-\gamma^\mu
  \end{pmatrix},\quad
  \Gamma^5=i\sigma^1\otimes I_4=
         \begin{pmatrix}
             0&iI_4\\ iI_4&0
         \end{pmatrix},\quad 
 \Gamma^6= i\sigma^2\otimes I_4=
         \begin{pmatrix}
             0&I_4\\ -I_4&0
         \end{pmatrix},
\end{align}
so that they satisfy the 6D Clifford algebra: 
\begin{align}
  \{\Gamma^M,\Gamma^N\}=2\eta^{MN}.
\end{align}
To study the 6D chirality, it is useful to define 
\begin{align}
  \Gamma_7=\Gamma^0\Gamma^1\Gamma^2\Gamma^3\Gamma^5\Gamma^6=-\sigma^3\otimes \gamma_5=
  \begin{pmatrix}
      -\gamma_5&0\\0&\gamma_5
  \end{pmatrix}
                      ={\rm diag}(1,1,-1,-1,-1,-1,1,1), 
\end{align}
which satisfies $\{\Gamma_7,\Gamma^M\}=0$.  The 6D Dirac fermion
$\Psi$ is decomposed into a sum of the 6D Weyl fermions $\Psi_{\pm}$,
which are eigenstates of $\Gamma_7$, as
\begin{align}
  \Psi=\Psi^++\Psi^-, \qquad 
  \Gamma_7\Psi^{\pm}=\pm\Psi^{\pm}.
\end{align}
Thus, we can write
\begin{align}
  \Psi^\pm\equiv
  {1\pm \Gamma_7\over 2}\Psi
=
  \begin{pmatrix}
      {1\mp \gamma_5\over 2}&0
      \\0&{1\pm \gamma_5\over 2}
  \end{pmatrix}\Psi,
           \qquad
          \Psi^+=
           \begin{pmatrix}
               \psi^+_{L}\\\psi^+_{R}
           \end{pmatrix}, \quad
          \Psi^-=
           \begin{pmatrix}
               \psi^-_{R}\\\psi^-_{L}
           \end{pmatrix}.
\end{align}
A 6D Weyl fermion $\Psi^\pm$ involves a vector-like pair of 4D Weyl
fermions. By using the two-component spinor notation in
Eq.~\eqref{twocomponentspinor1}, we can also write
\begin{align}
 \Psi=\Psi^++\Psi^-=
  \begin{pmatrix}
      \psi^+_L+\psi^-_R\\
      \psi^+_R+\psi^-_L
  \end{pmatrix}
=
  \begin{pmatrix}
      \xi^+_{L}\\\bar \eta^-_{R}\\\xi^-_{L}\\\bar \eta^+_{R}
  \end{pmatrix}.
\end{align}

Let us study fermion bilinears.  We define
$\overline \Psi\equiv \Psi^\dag \Gamma^0$ and find
\begin{align}
  \overline{\Psi^+}&=(\overline{\psi^+_{L}},-\overline{\psi^+_{R}}), \qquad
  \overline{\Psi^-}=(\overline{\psi^-_{R}},-\overline{\psi^-_{L}}).
\end{align}
The fermion bilinears without derivatives are given by 
\begin{align}
&  \overline \Psi \Psi=\overline{\Psi^-}\Psi^++
  \overline{\Psi^+}\Psi^- ,
\end{align}
which are written in terms of the 4D Weyl fermions as
\begin{align}
\overline{\Psi^-}\Psi^+=  \overline{\psi^-_{R}} \psi^+_{L}
-  \overline{\psi^-_{L}} \psi^+_{R}, \qquad 
\overline{\Psi^+}\Psi^- = \overline{\psi^+_{L}} \psi^-_{R}
-  \overline{\psi^+_{R}} \psi^-_{L}.
\end{align}
To obtain the kinetic terms for fermion fields, we use the fermion
bilinears with a derivative:
\begin{align}\label{6Dferkin1}
  \overline \Psi \Gamma^M\der_M\Psi=\overline{\Psi^+}\Gamma^M\der_M\Psi^++
                \overline{\Psi^-}\Gamma^M\der_M\Psi^- ,
\end{align}
where 
\begin{align}\label{gamdel1}
 \Gamma^M\der_M=
    \begin{pmatrix}
        \gamma^\mu\der_\mu&i(\der_5-i\der_6)\\
        i(\der_5+i\der_6)&-\gamma^\mu\der_\mu
    \end{pmatrix}.
\end{align}
Thus, by using the 4D Weyl fermions, we can rewrite the
above as 
\begin{align}
  &\overline{\Psi^+}\Gamma^M\der_M\Psi^+
    =\overline{\psi^+_{L}}\gamma^\mu\der_\mu\psi^+_{L}
    +   \overline{\psi^+_{R}}\gamma^\mu\der_\mu\psi^+_{R}
    +\overline{\psi^+_{L}}i(\der_5-i\der_6)\psi^+_{R}
    -\overline{\psi^+_{R}}i(\der_5+i\der_6)\psi^+_{L},
  \\ 
&  \overline{\Psi^-}\Gamma^M\der_M\Psi^-                        
    =\overline{\psi^-_{L}}\gamma^\mu\der_\mu\psi^-_{L}
    +   \overline{\psi^-_{R}}\gamma^\mu\der_\mu\psi^-_{R}
    +\overline{\psi^-_{R}}i(\der_5-i\der_6)\psi^-_{L}
    -\overline{\psi^-_{L}}i(\der_5+i\der_6)\psi^-_{R}.
\end{align}
The mixing terms between $\psi^\pm_{ L}$ and $\psi^\pm_{ R}$ include
the derivatives on the extra-dimensional coordinates.

To deal with the gamma matrices and study fermion fields on
$M^4\times T^2/{\mathbb Z}_3$, it is useful to introduce the oblique
coordinates discussed in Sec.~\ref{sec_t2z3}.  With the oblique
coordinates $y^1$ and $y^2$ found in Eq.~\eqref{vecydef1}, we
naturally define the new gamma matrices from $\Gamma^5$ and $\Gamma^6$
as
\begin{align}
  \Gamma^{y^1}&\equiv \Gamma^5+{1\over \sqrt{3}}\Gamma^6={2\over \sqrt{3}}
  \begin{pmatrix}
      0&-\bar \omega\\\omega&0
  \end{pmatrix}\otimes I_4,\qquad
  \Gamma^{y^2}\equiv {2\over \sqrt{3}}\Gamma^6={2\over \sqrt{3}}
  \begin{pmatrix}
      0&1\\-1&0
  \end{pmatrix}\otimes I_4, 
\end{align}
As expected, they satisfy
\begin{align}
  \{\Gamma^{y^i},\Gamma^{y^j}\}=-2g^{ij}, \qquad 
  \{\Gamma^{\mu},\Gamma^{y^i}\}=0,
\end{align}
where $g^{ij}$ is the metric in Eq.~\eqref{invmetric1}.  It is also
natural to define $\Gamma_{y^i}\equiv -g_{ij}\Gamma^{y^j}$, which are
explicitly written as
\begin{align}
 \Gamma_{y^1}=
-i    \begin{pmatrix}
        0&1\\1&0
    \end{pmatrix}\otimes I_4,\qquad 
\Gamma_{y^2}=
-i    \begin{pmatrix}
        0&\bar \omega\\\omega&0
    \end{pmatrix}\otimes I_4. 
\end{align}
We also introduce the useful notations:
\begin{align}
  \Gamma_{y^3}=-\Gamma_{y^1}-\Gamma_{y^2}=
  -i  \begin{pmatrix}
        0&\omega\\\bar \omega&0
    \end{pmatrix}\otimes I_4 ,\qquad \Gamma_{y^{i+3}}=\Gamma_{y^i}.
\end{align}
Then, Eq.~\eqref{gamdel1} is rewritten as
\begin{align}
  \Gamma^M\der_M&=
  \Gamma^\mu\der_\mu-g^{ij}\Gamma_{y^i}\der_{y^j}=
                  \Gamma^\mu\der_\mu-{2\over 3}\sum_{\ell=1}^3\Gamma_{y^\ell}\der_{y^\ell}\\
  &=
    \begin{pmatrix}
        \gamma^\mu\der_\mu & i{2\over 3}(\der_{y^1}+\bar \omega\der_{y^2}+\omega\der_{y^3})\\
        i{2\over 3}(\der_{y^1}+\omega\der_{y^2}+\bar \omega\der_{y^3})
        &-\gamma^\mu\der_\mu
    \end{pmatrix}.
          \label{ferder61}
\end{align}

Let us discuss the BC on $T^2/{\mathbb Z}_3$ related to the
${\mathbb Z}_3$ transformation $y^i\to \hat{\cal S}_0[y^i]$ for
fermion fields. The ${\mathbb Z}_3$ transformation generated by
$\hat{\cal S}_0$ is a $SO(2)\cong U(1)$ rotation with the angle
$2\pi/3$ on a two-dimensional Euclidean space, under which the
derivative $\der_{y^i}$ transforms to $\der_{y^{i-1}}$ as in
Eq.~\eqref{kis0trans}.  It is convenient to define a matrix
$\tilde S_\Psi $ that satisfies
\begin{align}
  \tilde S_\Psi^\dag \Gamma_{y^i}\tilde S_\Psi =\Gamma_{y^{i-1}},
  \qquad [\Gamma^\mu,\tilde S_\Psi ]=0, \qquad \tilde S_\Psi^\dag \tilde S_\Psi =I_2\otimes I_4.
\end{align}
One of the possible choices is
\begin{align}\label{tildeGam_def}
\tilde S_\Psi \equiv 
-  \begin{pmatrix}
      \omega&0\\0&\bar \omega
  \end{pmatrix}\otimes I_4.
\end{align}
Using $\tilde S_\Psi $, we can define the transformation law of the 6D
Dirac fermion $\Psi$ under the $SO(2)$ rotation with the angle
$2\pi/3$ as $\Psi\to \tilde S_\Psi \Psi$ so that the $2\pi$ rotation
gives $\Psi\to -\Psi$.  One sees that the $2\pi/3$ rotation keeps the
fermion bilinear in Eq.~\eqref{6Dferkin1} invariant as required from
the 6D Lorentz invariance, with the help of the relation
\begin{align}
  \tilde S_\Psi^\dag   \Gamma^0\Gamma^M\der_M \tilde S_\Psi 
  &=
    \tilde S_\Psi^\dag \Gamma^0
    \left(    \Gamma^\mu\der_\mu-{2\over 3}\sum_{\ell=1}^3\Gamma_{y^\ell}\der_{y^{\ell-1}}\right)
    \tilde S_\Psi \\
  &=\Gamma^0\left(    \Gamma^\mu\der_\mu-{2\over 3}\sum_{\ell=1}^3\Gamma_{y^{\ell-1}}\der_{y^{\ell-1}}\right)
    =\Gamma^0\Gamma^M\der_M.
\end{align}

We can define the BC for the 6D Dirac fermion 
$\Psi(x^\mu,y^i)$ as
\begin{align}
  \Psi(x^\mu,\hat{\cal S}_0[y^i])=-\omega^{p_s}\tilde S_\Psi \Psi(x^\mu,y^i),
  \label{ferBCgendef1}
\end{align}
where $p_s\in \{0,\pm 1\}$ can be chosen by hand, and the overall
minus sign on the right-hand side originates from the fermion number
operator. We note that the BC in Eq.~\eqref{ferBCgendef1} is
consistent with $\hat {\cal S}_0^3=\hat {\cal I}$ as required. Using
the 4D Weyl fermions, we rewrite the BC as follows:
\begin{align}
&  \psi^\pm_{ L}(x^\mu,\hat{\cal S}_0[y^i])=\omega^{p_s\pm 1}\psi^\pm_{ L}(x^\mu,y^i),\qquad 
  \psi^\pm_{ R}(x^\mu,\hat{\cal S}_0[y^i])=\omega^{p_s\mp 1}\psi^\pm_{ R}(x^\mu,y^i), 
\end{align}
which allows us to leave a 4D chiral fermion spectrum as the zero mode
from a 6D Weyl fermion $\Psi^\pm$.

\bigskip\bigskip

\end{document}